# 2D Layered Heterojunctions for Photoelectrocatalysis

Mengjiao Wang*, Michal Langer, Roberto Altieri, Matteo Crisci, Silvio Osella* and Teresa Gatti*

*KEYWORDS: 2D layered nanostructures, heterojunctions, photoelectrocatalysis, light energy conversion*

**ABSTRACT:** Two-dimensional (2D) layered nanomaterials heterostructures, arising from the combination of 2D materials with other low-dimensional species, feature large surface area to volume ratio, which provides a high density of active sites for catalytic applications and in particular for (photo)electrocatalysis (PEC). Meanwhile, their unique electronic band structure and high electrical conductivity enable efficient charge transfer (CT) between the active material and the substrate, which is essential for catalytic activity. In recent years, researchers have demonstrated the potential of a range of 2D material interfaces, such as graphene, graphitic carbon nitride (g-C3N4), metal chalcogenides (MCs), and MXenes, for (photo)electrocatalytic applications. For instance, MCs such as $MoS_2$ and $WS_2$ have shown excellent catalytic activity for hydrogen evolution, while graphene and MXenes have been used for the reduction of carbon dioxide to higher value chemicals. However, despite their great potential, there are still major challenges that need to be addressed in order to fully realize the potential of 2D materials for PEC. For example, their stability under harsh reaction conditions, as well as their scalability for large-scale production are important factors to be considered. Generating heterojunctions (HJs) by combining 2D layered structures with other nanomaterials is a promising method to improve the photoelectrocatalytic properties of the former. In this review, we inspect thoroughly the recent literature, to demonstrate the significant potential that arises from utilizing 2D layered heterostructures in PEC processes across a broad spectrum of applications, from energy conversion and storage to environmental remediation. With the ongoing research and development, it is likely that the potential of these materials will be fully expressed in the near future.

## 1. Introduction

Electrocatalysis constitutes one of the most important branches in the modern chemical industry. It enables chemical reactions that generate few by-products and require low amounts of starting materials, leading to cleaner and more sustainable chemical processes. In this field, electric power is converted into an energy source to trigger a variety of chemical reactions, such as water splitting,[1] nitrogen reduction,[2] $CO_2$ reduction,[3] treatment of pollutants[4] and much more. However, to replace the traditional thermal catalysis with an eco-friendly electrocatalysis, the efficiency and the cost of the whole electrocatalytic system still remain a problem. For example, in the field of $H_2$ production, the $H_2$ generated through water splitting accounts only ~ 3.9% of the total output due to its double cost compared to the thermal catalysis method, which still relies on fossil fuels as the primary $H_2$ source.[5] There are two main strategies to decrease the cost and improve the catalytic system: one is to make use of alternative clean and cheap energy sources to promote the chemical reactions, and another one is to develop more efficient catalysts at lower costs. In this context, the introduction of light as an energy source and the development of cheap catalysts have become interesting directions. After Fujishima and Honda firstly split water by resorting to a combination of electricity and solar light, scientists found that specific kinds of electrocatalysts are able to convert photo energy into the energy for chemical reactions.[6] While relying entirely on solar energy for the catalytic reactions represents the most eco-friendly approach and eliminates the need for electricity consumption, the current efficiency of pure photocatalysis falls short of meeting the practical application requirements. Therefore, photoelectrocatalysis (PEC) turns out as a more versatile strategy that combines a clean energy source while also ensuring high reaction efficiency.

A proper photoelectrocatalyst is an essential component in a PEC reaction, as it should be able to absorb light and use the energy from photons to drive the chemical reaction, while also serving as a classical catalyst to increase the reaction speed. As shown in Figure 1, in PEC, a photoelectrode is used to generate electron-hole pairs upon illumination. Specifically, when photons with sufficient energy are absorbed by the photoelectrocatalyst on a photocathode, they can excite electrons inside the catalyst and drive specific reduction reactions with the excited electrons. Similarly, a photoanode is able to accumulate photo-generated holes and catalyze oxidation reactions. Based on the above mechanism, the choice of photoelectrocatalyst is crucial in determining the efficiency and selectivity of the PEC reaction. Ideally, a photoelectrocatalyst requires the following criteria: i) a suitable bandgap (smaller than the energy of the incident light) to have a high absorption coefficient for the relevant wavelengths of light; ii) a suitable band alignment to ensure the PEC, meaning that the conduction band (CB) should be more negative than the reduction potential of the desired reaction, while the valence band should be more positive than the oxidation potential of the reaction. This ensures that the generated electrons and holes have enough energy to participate in the specific reaction; iii) stability under the reaction conditions, including light exposure, strong pH, and high bias. Stability is critical to ensure the material maintains its photoactivity and catalytic activity over time; iv) high mobility of charge carriers to facilitate a charge transfer (CT) to the surface of the catalyst and improve the catalytic activity.

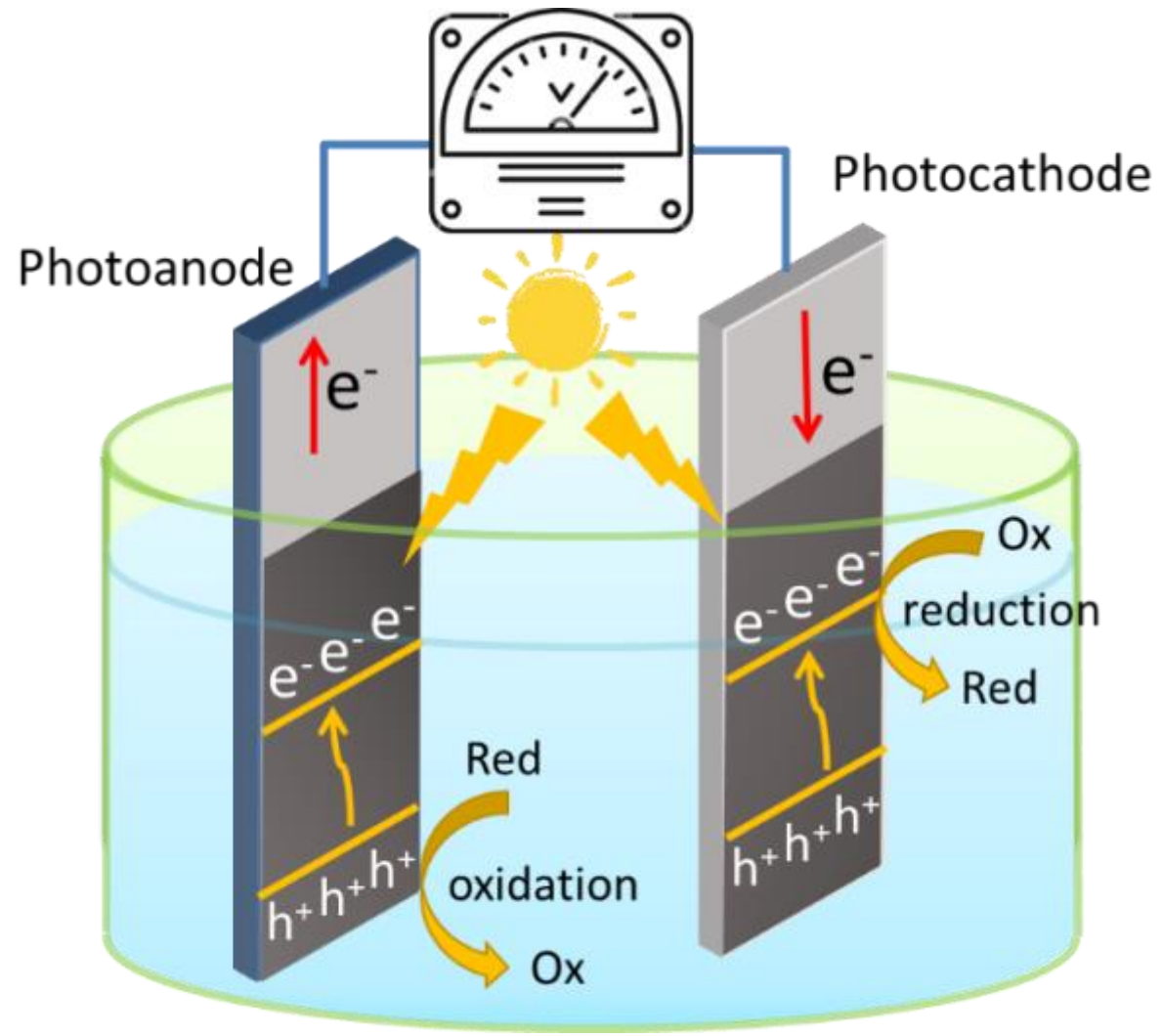

**Figure 1.** General mechanism of PEC reactions.

Two-dimensional (2D) materials are characterized by strong in-plane covalent bonds and weak van der Waals (vdWs) interactions between layers, resulting in atomically thin materials with unique electric and optical properties. The most well-known 2D materials include graphene, graphitic carbon nitride (g-$C_3N_4$), metal chalcogenides (MCs), layered double hydroxides (LDHs), MXenes, bismuth oxyhalides (BiOX, X= Cl, Br or I), hexagonal boron nitride (h-BN), $Pd_3(PS_4)_2$, transition metal oxides, 2D metal organic frameworks (MOFs), Xenes, metal phosphorous trichalcogenides ($MPCh_3$), and other materials with 2D structure on atomic level (Figure 2).[7–12] 2D materials have been extensively studied in PEC, where they serve various roles, including functioning as photoelectrocatalysts, co-catalysts for bandgap tuning, CT mediators or supporting matrices for catalysts.[13–16] Usually, 2D materials are mainly selected for PEC due to the following properties:

i) Large surface area. As a catalyst, high surface area of 2D materials can increase the density of active sites (surface defects, unsaturated atoms, and/or active edges et al.) and improve the efficiency of the catalytic reaction. As a supporting matrix, the large surface area enables a wide distribution of the real catalysts. Meanwhile, interior atoms could be brought closer to the surface in thinner atomic layers, thereby facilitating the contact between the catalysts and reactants.

ii) Strong interaction with light. Due to their thin nature, 2D materials can strongly interact with light, leading to efficient generation of charge carriers under illumination.

iii) Good CT properties. 2D materials have excellent CT properties, which allow for efficient transport of charge carriers to and from the catalyst. This can improve the overall efficiency of the photoelectrocatalytic reaction.

iv) Tunable properties. The properties of 2D materials can be tuned by adjusting their composition, thickness, and surface chemistry. This allows for precise control over their electronic and optical properties, which can be optimized for specific applications.

v) Chemical stability. Many materials are still chemically stable after nanosized into 2D layered structures, even under harsh reaction conditions like high voltage, strong irradiation, and extreme pH. This can improve the stability and longevity of the photoelectrocatalyst.

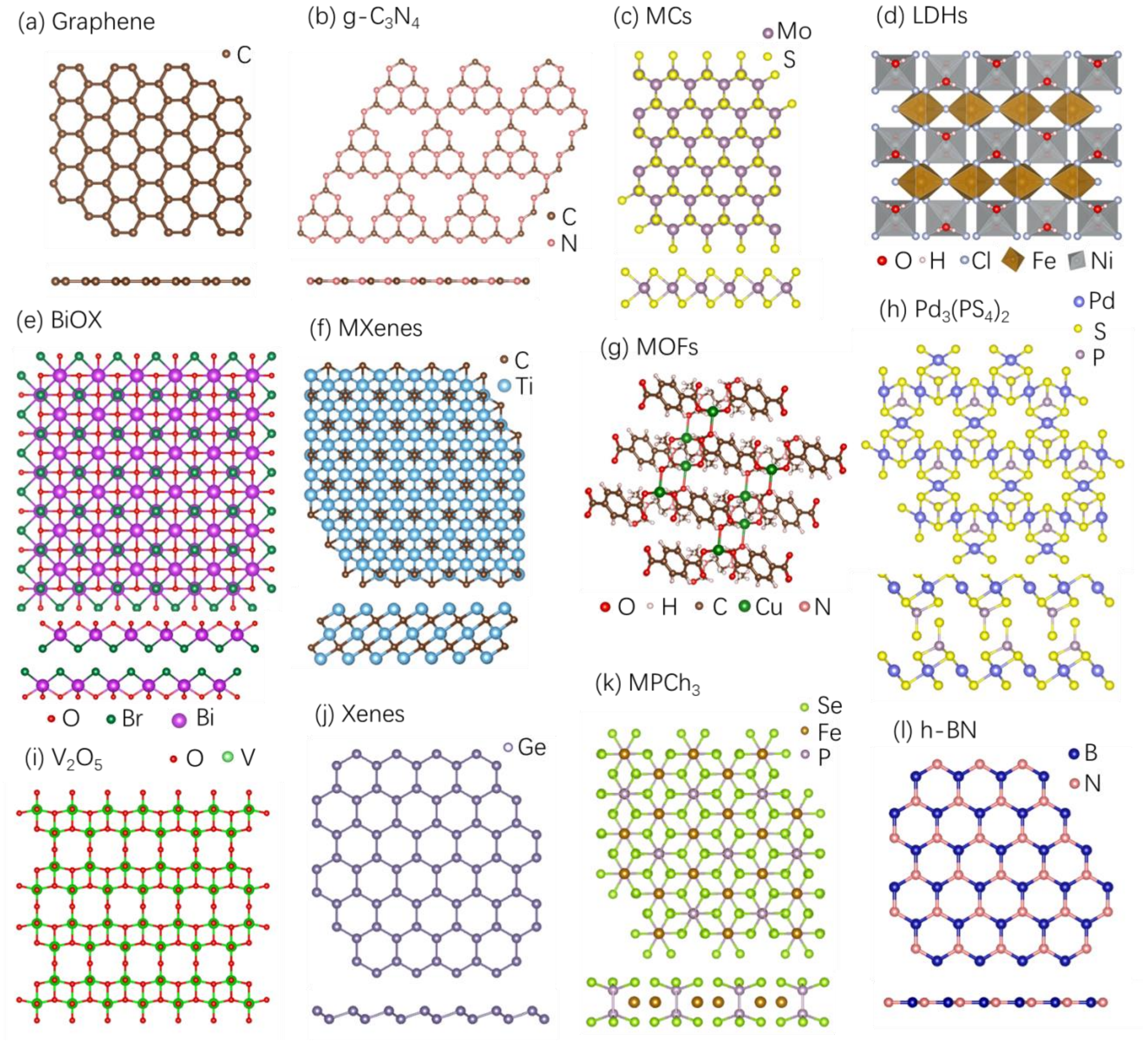

**Figure 2.** For each group of 2D layered materials, the structures of a representative material are shown. (a) Graphene. (b)g-C3N4. (c)MCs, with $MoS_2$ as an example. (d) LDHs, with NiFe-LDH as an example. (e) BiOX, with BiOBr as an example. (f) MXenes, with $Ti_3C_2$ as an example. (g) MOFs, with Cu(II) catena-((μ4-terephthalato)-(N,N-dimethylformamide)) as an example. (h) $Pd_3(PS_4)_2$. (i) $V_2O_5$. (j) Xenes, with germanene as an example. (k) $MPCh_3$, with $FePSe_3$ as an example. (l) h-BN.

Based on this background knowledge, this review discusses up to date 2D material heterojunctions (HJs) applied in PEC. The content focuses on the introduction of different HJs based on 2D materials (Section 2). Then the methods of synthesis of 2D HJs are introduced in Section 3, the computational methods describing (photo)electrocatalysis in Section 4 and the applications of 2D HJs in PEC are summarized in Section 5, including water splitting, organic degradation, $N_2$ reduction reaction, $CO_2$ reduction reaction, antimicrobial reaction, metal ion detoxification reaction and other new reactions under PEC conditions. Finally, we conclude and provide a prospective on the potential development of 2D material HJs for PEC in the near future.

## 2. Heterojunctions based on 2D materials

HJs with 2D materials are interfaces between 2D materials and other nanomaterials. On one side, the large surface area of 2D materials provides enough space to combine with other nanomaterials, and the strong interaction between 2D materials and other materials can lead to enhanced charge transfer, which can be exploited for PEC. On the other side, the band alignment and bandgap of the HJs can be tuned by the 2D compounds. In fact, the 2D semiconductors HJs can interact with the surface of other components and modify the energy levels. This interaction can lead to changes in the bandgap of both the components, which affects the optical and electronic properties of the whole HJs. For example, after the introduction of $MoS_2$, there is a red shift in the absorption band edge of $SnO_2$ from 380 nm to 500 nm, indicating an enhancement in light response; the peak intensity of the PL decreases as well, showing that the electron and hole are effectively separated (Figure 3a and 3b).[17] Combining the granular $SnO_2$ with $MoS_2$ can not only expand the specific surface area of the composite and increase the surface adsorption sites of the reagents, but also allow for full use of the optical performance of 2D $MoS_2$ to increase the light conversion efficiency, which is proved by the increased photocurrent of a $MoS_2/SnO_2$ HJ compared to pure $SnO_2$ (Figure 3c).

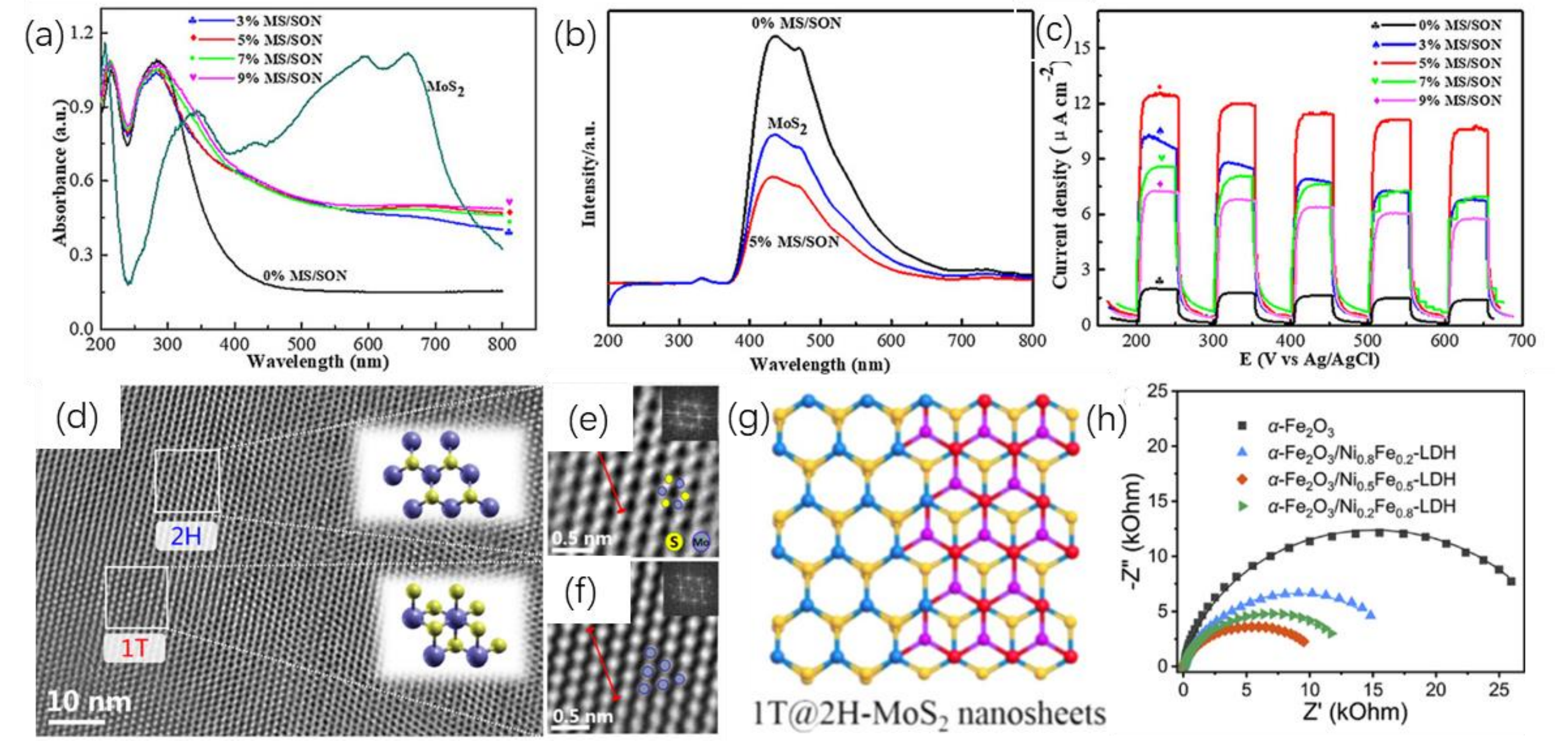

**Figure 3.** (a) UV-vis absorption spectra of $MoS_2/SnO_2$ composites with different $MoS_2$ weight ratios. (b) PL spectra of 0.5 % $MoS_2/SnO_2$ and pure $MoS_2$. (c) Transient photocurrent density versus time of $MoS_2/SnO_2$ composites in $CO_2$ atmosphere.[17] © 2019 Wiley-VCH Verlag GmbH& Co.KGaA, Weinheim. (d) High-resolution transmission electron microscopy (HRTEM) image of the lateral HJ of $MoS_2$. Inset: Schematic structures of the unit cells of 1T and 2H phase. (e, f) Two filtered images of the region enclosed by the white square of (d) to show the reversed contrast. (g) Schematic of the top view of 1T/2H $MoS_2$ HJ.[18] © 2016 American Chemical Society. (h) Nyquist plots of electrochemical impedance spectroscopy measured under AM 1.5G (100 $mW/cm^2$) illumination.[19] © 2018 Elsevier B.V.

The following 2D materials have been studied to construct the HJs for PEC: graphene, g-$C_3N_4$, metal chalcogenides (MCs), layered double hydroxides (LDHs), MXenes, bismuth oxyhalides and 2D MOF material. Novel 2D materials such as h-BN, Xenes, $Pd_3(PS_4)_2$ and $MPCh_3$ have been reported as promising photoelectrocatalysts, but the HJs combined with them have not been reported yet.

Graphene is an isolated 2D material consisting of a single layer of carbon atoms arranged in a hexagonal pattern, forming a honeycomb lattice (Figure 2a). It was first exfoliated in 2004, since when a plenty of applications have been discovered on 2D layered graphene.[20] This unique arrangement allows the electrons in graphene to move freely across its surface, giving it an exceptional electrical conductivity. In the field of PEC, graphene oxide is commonly used as well because the functional groups enhance dispersibility in solvents like water, allowing for easier integration into various matrices or coatings. For graphene oxide (GO), the introduced oxygen-containing groups make it insulate, while reduced graphene oxide (rGO) is more frequently used because the reduction of functional groups can recover the electrical conductivity. Therefore, compared with graphene, rGO are promising candidates to function as co-catalyst together with other photoelectrocatalysts for enhancing the charge separation and charge transfer.

G-$C_3N_4$ is a semiconducting material with a unique 2D layered structure consisting of nitrogen-containing aromatic rings, which is similar to graphene (Figure 2b). The electronic properties of g-$C_3N_4$ depend on its composition and structure, which influence its band structure and the distribution of electronic states. g-$C_3N_4$ is classified as an n-type semiconductor because it has excess electrons as majority charge carriers, which can contribute to electrical conductivity. Apart from its unique electrical property, g-$C_3N_4$ exhibits good stability, unique electronic property, strong light absorption and photocatalytic activity, which makes it a promising material for applications in PEC. Except g-$C_3N_4$, g-$C_3N_5$ is used as photoelectrocatalyst as well. Studies show that g-$C_3N_5$ with extended conjugation via the introduction of extra nitrogen into the triazine units possesses higher thermodynamic stability, better electronic properties, and narrower band gap (2.2 eV) compared with g-$C_3N_4$ (2.7 eV).[21]

2D MCs are a vast class of interesting and novel material. The most popular 2D materials in this group are the transition metal dichalcogenides (TMDCs), which are denoted as $MX_2$ (M: metals; X: S, Se, Te), and with a general layered structure of stacked X-M-X layers (Figure 2c). Some examples are $MoS_2$[22], $WS_2$[23] and $MoSe_2$[24] that are semiconductors in their most stable thermodynamic phase, while other TMDCs present semi-metallic ($WTe_2$, $VS_2$) or metallic properties instead ($TaS_2$, $NbS_2$).[25] Naturally, only the semiconducting materials have been studied for PEC applications and therefore we will take a closer look at the state-of-the-art of mainly sulfides and selenides of Mo and W. Specifically, there are two different phases of MoS2: 1T-MoS2 and 2H-MoS2. 1T-$MoS_2$ has been synthesized by converting 2H-$MoS_2$.[18] One notable distinction is its metallic behavior, meaning it can conduct electricity, while the 2H- phase is typically semiconducting. This metallic nature of 1T-$MoS_2$ makes it particularly interesting as a charge separator and mediator in the HJ photocatalysts. For instance, in the 2H/1T HJ, the electrical coupling and synergistic effect between 2H and 1T phases can greatly facilitate the efficient electron transfer from the active sites of $MoS_2$, which significantly improves the HJ photocatalytic performance (Figure 3d-g).

LDHs are a class of inorganic compounds with a layered structure that consists of positively charged metal hydroxide layers and exchangeable anions between these layers. The LDH structure consists of infinite sheets with brucite-like layers with interlamellar ions (Figure 2d). In brucite-like layer, the metal hydroxide layers are composed of divalent or trivalent metal cations and $OH^-$ species, where metal cations in the sheets are present at the centers of the edge-sharing octahedra constructed by $OH^-$ species. There are different kinds of metal cation combinations in the structures used for PEC, such as Co/Mn[26], Ni/Mn[27], Zn/Cr[28], Ni/Co[29], Co/Fe[30], Co/V[31], Cu/Ti[30] and Ni/Fe[19,32] et al. The interlayer anions are typically negatively charged and can be mono- or polyatomic, such as $CH_3COO^-$, $Cl^-$, $CO_3^{2-}$, $NO_3^-$ and $SO_4^{2-}$. These anions are held in place by electrostatic forces and hydrogen bonding with the metal hydroxide layers. The layers of LDHs are generally held together by weak vdWs forces, making them suitable for intercalation. As a result, they are employed as co-catalysts to enhance PEC efficiency. For example, by tuning the Ni/Fe molar ratio to 1:1, the α-$Fe_2O_3$/NiFe-LDH composite photoanode is able to obtain an optimized photocurrent of almost 150 μA/cm$^2$, which is higher than α-$Fe_2O_3$/NiFe-LDH photoanode with other Ni/Fe ratios and ~3 times higher than pristine α-$Fe_2O_3$ as well.[19] As illustrated in Figure 3h, the α-$Fe_2O_3$/$Ni_{0.5}Fe_{0.5}$-LDH photoanode shows the smallest electron transfer resistance (10.6 kΩ cm$^2$) among these four samples. This reveals the fast CT kinetics that occur in the system.

As a series of 2D layered inorganic semiconductor materials, bismuth oxyhalides (BiOX, X = Cl, Br, and I) have received great research interest in various photo(electro)catalytic applications, due to their suitable band gaps, chemical stability, absence of toxicity, and corrosion resistance. Among the BiOX catalysts (Figure 2e), BiOI is the most efficient visible light harvesting photocatalyst because of its narrow bandgap (1.7 eV).[33] However, the pure BiOI is not efficient enough for PEC, thus it is commonly used as a component in hybrid photoelectrocatalysts. For instance, Pt/BiOI has been reported as good photoelectrocatalysts for various reactions.[34]

2D MXenes are a family of 2D transition metal carbides, nitrides, and carbonitrides that have attracted significant attention in the field of PEC.[35] MXenes have a layered structure consisting of transition metal atoms sandwiched between two or more layers of carbon or nitrogen, resulting in a high surface area and high electrical conductivity (Figure 2f). However, the metallic nature of most kinds of MXenes do not make them capable of having an optical response ability, which greatly hinders their application in PEC as a pure photoelectrocatalyst.[33] MXenes have shown promising potential as a component for HJs due to their nature as a 2D material, including high surface area, tunable electronic properties, and chemical stability.

MOFs are a class of crystalline materials consisting of metal ions or clusters linked together by organic ligands (Figure 2g).[33] They have a highly ordered, porous structure with high surface area and can be synthesized in a wide variety of structures and compositions. The combination of their tunable pore size, high surface area, and chemical functionality makes MOFs useful for a wide range of catalysis applications. Compared with three-dimensional MOFs, 2D MOFs can provide more coordinatively unsaturated metal sites, which is helpful to the mass transfer and electron transfer processes.[11] 2D-MOFs can be considered as having defective, exchangeable coordination positions at the metal nodes. Besides, the bandgap energy ranges from 1.94 to 2.12 eV, depending on the synthesis parameters of the samples, making them very promising candidates for different PEC reactions.

In general, the role of 2D materials in HJs can be divided into three kinds: (i) as main catalysts, (ii) as CT mediators or (iii) as supporting matrices. Usually, 2D materials always have synergetic performance instead of one simple function in PEC, for instance in the *p-n* junction formed with 2D materials, the 2D materials can perform as catalyst but also CT mediators for separating the charge carriers. Further elaboration on this is discussed in the following sections.

### 2.1 2D materials as main catalyst

2D materials with semiconductive property are widely used as the main catalyst providing the active sites for PEC reactions within HJs. Specifically, the 2D materials can separate the photogenerated charges and collect more charges from the electric field due to the tunable band structure, then transfer the charges to the active sites for photoelectrocatalytic reactions. $MoS_2$/indium tin oxide (ITO) HJ is a typical example.[36] During the PEC reaction, the carrier separation occurs by injecting photogenerated electrons from the CB of $MoS_2$ to ITO, whereas photogenerated holes in the valence band of $MoS_2$ cannot diffuse to ITO because of the existence of a *p*-type Schottky barrier. These holes accumulate on the surface of $MoS_2$ and participate in oxidation reactions. Meanwhile, the reverse transport of positive and negative charge carriers at the $MoS_2$/ITO interface retards the recombination of electron/hole pairs and prolongs the carrier lifetime, leading to a superior photocurrent response.

There are different types of HJs formed by involving 2D materials and, depending on the band positions of the semiconductor components, they include type I, type II, Z-scheme and S-scheme HJs (Figure 4a-d). In type I HJs, the photogenerated electrons usually transfer from the more negative CB to the less negative CB for reduction reaction, while the holes transfer from the more positive band to the less positive one for oxidation reaction, in this way separating the photogenerated charge carriers and allowing their exploitation for PEC. For example, at a BiOBr/MoS2 HJ, the photogenerated holes and electrons from 2D BiOBr can transfer to the valence band (VB) and CB of 2D $MoS_2$, respectively, because of the intrinsic band structure of the two materials.[37] However, the original band position is not the only prerequisite to determine the CT route, as defect states or the internal electric field (IEF) can influence the CT route as well. For instance, the defects or oxygen vacancies on BiOBr can trap part of the photogenerated holes, thus only part of the holes from BiOBr can transfer to $MoS_2$, though the VB of $MoS_2$ is less positive.

In a type II HJ, the photogenerated electrons and holes will still transfer to the less negative CB and less positive VB, respectively. Different from type I HJ, the conduction band minimum and the valence band maximum are on different components in the type II HJ, resulting in a staggered band alignment. As a 2D semiconductor with suitable band alignment, g-$C_3N_4$ is frequently used as one of the components to construct type II HJs with other semiconductors. The g-$C_3N_4$/$TiO_2$ HJ is a widely reported HJ used for PEC, since both components are efficient and stable photoelectrocatalysts.[38–41] g-$C_3N_4$ and BiOI form a Type-II HJ at which photogenerated electrons in g-$C_3N_4$ are transferred to BiOI, while photogenerated holes in BiOI are transferred to g-$C_3N_4$.[42] Effective charge separation, synergistic trap filling leading to a lower CT resistance, and enhanced light absorption have been identified as dominant factors for the enhancement of photocurrent when compared to the pristine photoanodes. In addition, S doped g-$C_3N_4$ acts as electron donor when placed in contact with samarium vanadate (SmV). This can create a type II HJ, where reduction reaction occurs at the CB of SmV, while oxidation reaction occurs at the valence band of S doped g-$C_3N_4$[43]. Other 2D materials with semiconducting properties such as α-$Fe_2O_3$/$MoS_2$[44], BiOI/$BiPO_4$[45], $WO_3$/$WS_2$[46] and others are reported as type II HJ for PEC as well, and are shown in Table 1.

**Table 1.** Photoelectrocatalytic HJs containing 2D materials.

| Catalyst | Synthesis | Light source | Photocurrent | Type | application | Ref |
|---|---|---|---|---|---|---|
| **rGO/g-$C_3N_4$/BiVO** | Hydrothermal | 500 W halogen lamp, 100 mW.cm-2 | 14.44 mA/$cm^2$ 1.0 V vs Ag/AgCl | II | HER | 47 |
| **rGO/AgCl** | Ultra-sonication, hydrothermal | 300 W Xenon lamp (λ ≥ 400 nm) | 2.5 μA /$cm^2$ | / | Organic degradation | 48 |
| **rGO/$CeO_2$/$TiO_2$** | Electrochemical method | 500W Xe lamp (>365 nm), 110 mW cm-2 | / | / | Organic degradation | 49 |
| **rGO/Cu** | Hummers' method, electrochemical method | Xe-arc lamp (100 mW cm−2) | ~14 mA/$cm^2$ at -2.4 V vs. Ag/AgCl | / | $CO_2$ reduction reaction ($CO_2$RR) | 50 |
| **Pt/CdS/rGO** | Solvothermal | Xe arc lamp (150 W) | ~1.4 mA/$cm^2$ at -0.3 V. SCE | / | Methanol oxidation reaction (MOR) | 51 |
| **GO/$In_2S_3$/$TiO_2$** | Hydrothermal, electrodeposition | 300 W Xe short arc lamp, 100 mW cm-2 | 0.45 mA/$cm^2$ at 0.8 V vs Ag/AgCl | | HER | 52 |
| **GO/$Ag_3PO_4$/Ni** | Electrochemical method reported | 500W Xe lamp (>420 nm), 110 mW cm-2 | 15 mA/$cm^2$ at 1 V vs. SCE | / | Organic degradation | 14 |
| **Nd:g-$C_3N_4$/BiOI** | Polycondensation, hydrothermal | Air Mass, AM 1.5 G | 0.03 mA/$cm^2$ at 0.2 V vs RHE | II | HER | 53 |
| **g-$C_3N_4$/$Ag_3PO_4$** | Polycondensation, in situ precipitation | visible lamp , λ> 400 nm | 13 μA/$cm^2$ | II | organic degradation | 54 |
| **g-$C_3N_4$/$TiO_2$** | Polycondensation, sonication | AM 1.5G (100 mW cm−2) | 142.7 μA/$cm^2$ at 1.23 V vs. RHE | II | OER | 38 |
| **SmV/g-$C_3N_4$** | Polycondensation, hydrothermal | 400 W Xe light | ~0.13 mA/$cm^2$ | II | HER, | 43 |
| **BiOI/g-$C_3N_4$** | Hydrothermal | 300 W Xenon lamp | 0.1 μA/$cm^2$ | S-scheme | Water splitting | 55 |
| **$BiVO_4$/GQD/g-$C_3N_4$** | Hummer's method, hydrothermal, thermal polycondensation | 500 W halogen lamp | 19.2 mA/$cm^2$ at 1.0 V vs Ag/AgCl | II | HER | 56 |
| **$Co_3O_4$/g-$C_3N_4$** | Hydrothermal | Xenon lamp | 0.01 μA/$cm^2$ | S-scheme | Organic degradation | [13] |
| **P/g-$C_3N_4$** | chemical vapor deposition (CVD) | 300 W Xenon lamp | 202 μA/$cm^2$ at −1 V vs. Ag/AgCl | / | HER | [57] |
| **g-$C_3N_4$/$TiO_2$** | thermal condensation | 300 W Xe lamp (200 ~800 nm) | 0.63 mA/$cm^2$ at 0.0 V vs. Ag/AgCl | II | Organic degradation | [40] |
| **CQDs/g-$C_3N_4$** | thermal polycondensation, LPE | 500WXe lamp with a UV cut-off filter (>420 nm) | ~0.025 mA/$cm^2$ | / | Organic degradation | [58] |
| **C/g-$C_3N_4$** | Pyrolysis, thermal condensation | white light LED lamp (50 W) | ~19 μA cm−2 at 1.23 V vs RHE | / | Organic degradation | [59] |
| **g-$C_3N_4$/$TiO_2$** | thermal condensation, wet chemical synthesis | 100 mW cm-2 (AM 1.5 G) | 72.3 μA/$cm^2$at 1.23 V vs RHE | II | OER | [39] |
| **g-$C_3N_4$/$Sn_3O_4$/Ni** | | LED | | Z-scheme | Metal ions reduction | [60] |
| **g- $C_3N_4$ /α-$Fe_2O_3$** | Hydrothermal, thermal condensation | 500W Xe lamp (>420 nm), 100 mW cm-2 | 4.97 mA/$cm^2$ at 1.5 V vs. Ag/AgCl | II | Organic degradation | [61] |
| **g-$C_3N_5$/BiOBr** | Hydrothermal, thermal condensation | Xe lamp (>420 nm), 110 mW cm-2 | −1.2 μA/$cm^2$ | II | N2RR | [21] |
| **$BiVO_4$/$SnS_2$** | Hydrothermal | solar simulator, 100 mW/cm2 | 0.21 mA/$cm^2$ at 1.23 V vs RHE | Z-scheme | HER | [62] |
| **Au-WS2** | Hydrothermal | 300 W Xenon lamp | 23 mA/$cm^2$ at 1 V vs. SCE | / | Pollutant degradation | [63] |
| **GaTe/ZnO** | CVD | 300 W Xe lamp (λ > 420 nm), 100 mW cm–2 | −2.5 mA/$cm^2$ at −0.39 V vs RHE | II | HER | [64] |
| **Ni/NiFe-LDH/$Co_3O_4$** | hydrothermal | Xenon lamp | 0.16 μA/$cm^2$ at 0.7 V vs. RHE | II | Organic degradation | [32] |
| **α-$Fe_2O_3$/NiFe-LDH** | Hydrothermal, electrodeposition | | 150 μA/$cm^2$ at 1.23 V vs RHE | | OER | [19] |
| **$Ti_3C_2T_x$/$Bi_{12}TiO_{20}$** | Hydrothermal | xenon lamp | / | / | Organic degradation | [65] |
| **Co-$Ti_3C_2$** | Hydrothermal | solar simulator, 100 mW cm−2 | 2.99 mA/$cm^2$ at 1.23 V vs RHE | / | OER | [66] |
| **$TiO_2$/$Ti_3CN$ MXene** | Hydrothermal | | | / | $CO_2$ RR | [67] |
| **Pd/N-$TiO_2$/$Ti_3C_2$** | Hydrothermal | 300 W Xenon lamp with 200 mW/cm2 | 1 mA/$cm^2$ at -0.7 V | / | $CO_2$RR | [68] |
| **BiOBr/$TiO_2$** | Chemical bath deposition | 300W xenon lamp (λ ≥ 420 nm) | 0.033 mA/$cm^2$ | II | Organic degradation | [69] |
| **$WO_3$/$BiVO_4$/ $NiCo_2O_x$** | Thermal calcination, chemical bath deposition | AM 1.5G illumination | 2.85 mA/$cm^2$ at 0.7 V vs. RHE | | OER | [70] |
| **Pt-$Bi_2WO_6$/ $La_2Ti_2O_7$** | Hydrothermal | 500 W Xenon lamp 186 mW/cm2 with a UV cut-off filter (>420 nm) | ~0.5 $mA/cm^2$ at −0.25 V vs. SCE | | MOR | 71 |
| **BiOI/$BiPO_4$** | electrodeposition | xenon lamp | 60 µ$A/cm^2$ | II | Organic degradation | 45 |
| **Pt/BiOI** | hydrothermal | 500 W Xenon lamp with a UV cut-off filter (>420 nm) | 70.3 $A/cm^2$ at −0.25 V | / | ethanol oxidation | 34 |
| **$TiO_2$/$P_3$HT** | Hydrothermal, dip-coating | / | 0.68 $mA/cm^2$ at 1.0 V | II | Organic degradation | 72 |
| **Pt/$La_2Ti_2O_7$** | Hydrothermal, photo-reduction | 150 W Xe arc lamp | ~0.15 $mA/cm^2$ at −0.2 V vs. SCE | / | Methanol oxidation | 73 |
| **CuBDC/Cu** | solvothermal method | 300W Xe lamp (>420 nm), 100 mW cm-2 | 0.6 $mA/cm^2$ at -0.3 V vs. RHE. | / | CO2RR | 11 |

In addition to the CT-route in type II HJ, the Z-scheme PEC mechanism can be exploited for specific catalytic reactions. In a Z-scheme HJ, the less energetic electrons are transferred from the CB of one semiconductor to the valence band of another semiconductor through an intermediate redox mediator, which acts as a relay system (Figure 4c). This process results in the generation of electron-hole pairs in both semiconductors, allowing for simultaneous oxidation and reduction reactions to occur on their surfaces. However, it is not mandatory to have an intermediate in a Z-scheme HJ: direct Z-scheme catalysts have been developed as well. For example in the PEC of $H_2O_2$ production by $MoS_2$/$CoMoS_4$, the original band structure of $MoS_2$/$CoMoS_4$ is not able to catalyze the $O_2$ reduction to $H_2O_2$ in theory(Figure 4e).[74] This is because the E ($O_2 \rightarrow \bullet O_2^-$) (–0.33 eV vs. NHE) is more negative than the CB of $CoMoS_4$ (–0.19 eV vs. NHE). In reality, instead of the traditional way of CT, the photogenerated charges follow the Z-scheme mechanism of charge recombination inside the HJ. Specifically, in the tight solid–solid contact HJ interface between $MoS_2$ and $CoMoS_4$, the CB of $CoMoS_4$ is very close to the VB of $MoS_2$, providing a shorter electron transfer path, and the photo-generated electrons at the CB of $CoMoS_4$ could quickly migrate to the VB of $MoS_2$ and immediately combine with the holes there. Thereby, the longevity of the remaining electron (CB, $MoS_2$) / hole (VB, $CoMoS_4$) pairs is greatly prolonged, giving the HJ excellent oxidation and reduction activity. A 2D/2D $BiVO_4$/$SnS_2$ nanocomposite can form a Z-scheme HJ by coupling both $BiVO_4$ and $SnS_2$.[62] Inside this HJ, the electron transfer from $BiVO_4$ to $SnS_2$ is permitted, leaving holes on VB of $BiVO_4$ and electrons on the CB of $SnS_2$, making this HJ active in PEC reactions.

The formation of HJs from two different semiconductors can enhance the electron transfer capability through establishing IEF, since one material has a higher electron affinity and lower ionization energy compared to the other. This is the S-scheme HJ (Figure 4d). Typically, two types of semiconductors in the S-scheme can align the Fermi energy after contact, which results in a band bending forcing the less energetic photogenerated electron in the CB of the oxidation catalyst and hole in the VB of reduction catalyst at the interfacial region. Different from Z-scheme HJs, in which the two materials have similar electron affinities and ionization energies, the IEF, band bending, and Coulombic attraction act as the driving forces of the charge recombination inside the S-scheme HJ. For example, the combination of 2D BiOI nanoflakes and g-$C_3N_4$ in a HJ forms a *p-n* junction and builds an IEF on the interface of the two materials, with more negative charges on the BiOI surface and more positive charges on the g-$C_3N_4$ surface (Figure 4f).[55] Under illumination, the photogenerated electrons transfer from BiOI to g-$C_3N_4$ under the influence of the IEF, while the photogenerated holes transfer on the opposite direction. The rapid recombination of photogenerated charges is prevented by the transfer. In this way, the accumulated electrons on g-$C_3N_4$ can catalyse a reduction reaction, meanwhile BiOI can provide reaction sites for an oxidation reaction.

It is worth to mention that in some S-scheme HJs, there can be several competitive CT processes and the pathway based on the S-scheme mechanism might not be dominant.[75] As shown in Figure 4g, in the $MoS_2$/$WSe_2$ HJ, photogenerated electrons and holes are formed in both $MoS_2$ and $WSe_2$ (process 1). Because of the carrier concentration gradient, an IEF forms at the interface between $MoS_2$ and $WSe_2$. This IEF accelerates the electrons to the CB of $MoS_2$ the hole to the VB of $WSe_2$ (process 2). Though the charge recombination process 3 and 4occurs simultaneously *via* the interface transitions and inside the crystals, the time scale of process 3 is at a nanosecond (ns), which is 3 or 4 orders of magnitude slower than CT process 2. Thus, the CT is still dominate to the separation of the photoinduced charge pairs. As a result, the recombination process is suppressed by the consumption and inverse transfer of the electrons and holes, inducing a more significant photocurrent than individual $MoS_2$ and $WSe_2$. Consequently, the photocurrent of the HJ is 1.79 times larger than that of pristine $MoS_2$ and 1.94 times larger than that of pristine $WSe_2$.

Except for the intrinsic band structures of 2D semiconductors, the defect states can influence the band engineering of the HJs as well. For example, $Al_2O_3$ can construct a type I HJ with 2D $SnS_2$.[76] While the defect state is introduced into $Al_2O_3$, instead of recombing with the holes inside $SnS_2$, the photogenerated electrons can transfer to the defect state of $Al_2O_3$ and move to the Pt electrode for PEC. Meanwhile, the holes on the VB of $SnS_2$ are free for further PEC.

Additionally, the morphology of the two nanomaterials involved in the HJ can influence the CT and the PEC efficiency. As mentioned before, 2D materials have a high surface area to provide more active sites, and thin thickness for short CT length to move to the catalytic reaction point. $MoS_2/WS_2$ HJs are widely used in different kinds of photoelectrocatalytic reactions. Sherrel et al. prepared 100 μm large flakes samples to enable minimization of atomic defects and nanosheets edge density, while the atomically sharp and clean interfaces between the flakes are responsible for reducing charge carrier recombination. The PEC performance turns out better than the $MoS_2/WS_2$ HJ with smaller flakes, which have higher density of defects.[77] Other HJs containing different 2D layered materials can also be formed, such as, g-$C_3N_4$/nitrogen-doped graphene/$MoS_2$[78], g-$C_3N_4$/BiOBr[21] $MoSe_2$/g-$C_3N_4$[79], $VS_x$/graphene[80], etc. These HJs provide a broad optical window for effective light harvesting, short diffusion distance for excellent charge transport, as well as a large contact area for fast interfacial charge separation and PEC reactions. The measured photocurrent density of the HJs is generally enhanced compared to the individual components, which is proof of the superior ability of HJs to generate and transfer photoexcited charge carriers.

The combination between 2D semiconductors and another metallic counterpart can benefit the charge separation as well. For instance, the $Sn_xMo_{1-x}S_2/MoS_2$ HJ is obtained with atomically sharp interface between the metallic $Sn_xMo_{1-x}S_2$ and semiconducting $MoS_2$.[81] The electrons generated by photon energy can rapidly transfer to the mixed sulphide and then to the counter electrode, leaving a rich amount of holes for PEC oxidation reactions.

In the PEC processes, the CT pathway is usually determined by the combination of several confirmatory experiments. For example, free radical capture experiments were performed in the PEC of organic degradation by $Co_3O_4$/g-$C_3N_4$ to detect the active species, indicating that •OH and $•O_2^-$ play a dominant role during the degradation reaction.[13] Based on the analysis of band position from theoretical calculations and reaction energy of the active species free radical capture experiments, it is clear that the more energetic electrons on the CB of $Co_3O_4$ and holes on the VB of g-$C_3N_4$ participate in the formation of •OH and $•O_2^-$, while the holes on $Co_3O_4$ and electrons on g-$C_3N_4$ recombine on the interface of the HJ.

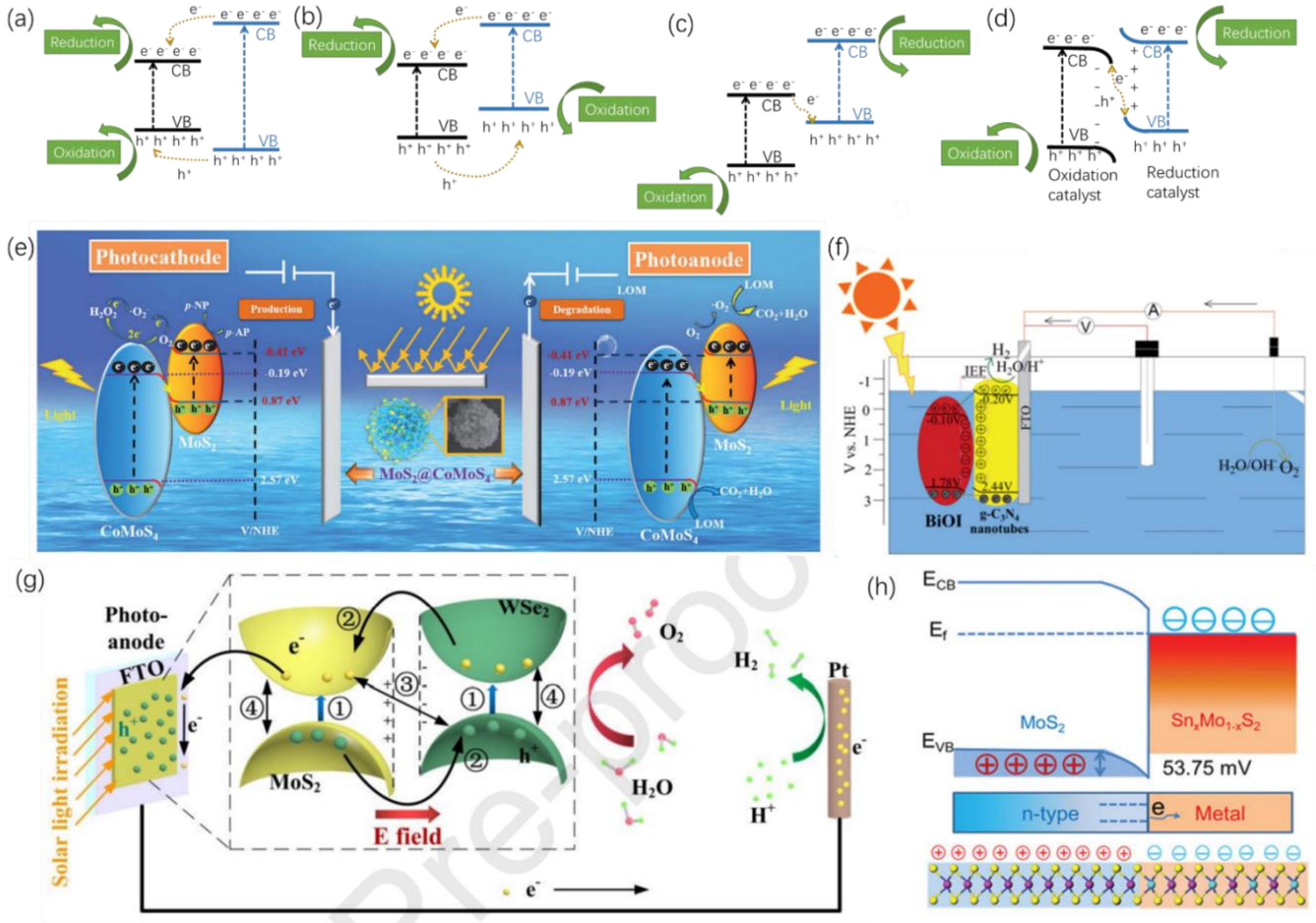

**Figure 4.** Scheme of (a) type I HJ, (b) type II HJ, (c) Z-scheme HJ and (d) S-scheme HJ. (e) Possible PEC mechanism of MoS2/CoMoS4.[74] © the Royal Society of Chemistry 2020. (f) Mechanism of BiOI/g-C3N4 over photoelectrode under light irradiation.[55] © 2022 Elsevier B.V. (g) Schematic diagram of the charge carrier separation and transfer processes of MoS2/WSe2 photoanode in an electrolyte under light illumination.[75] © 2019 Elsevier B.V. (h) Schematic of the band profile for SnxMo1−xS2/MoS2 heterostructure according to Kelvin probe force microscopy (KPFM) characterization. Bottom: schematic of the heterostructure's cross section. Purple, cyan, and yellow balls represent Mo, Sn, and S atoms, respectively. Red circles with plus symbol and blue circles with minus symbol represent holes and electrons, respectively.[81] © 2020 Wiley-VCH GmbH.

## 2.2 2D materials as charge transfer mediators

As CT mediators, 2D materials do not behave as the direct catalyst in a HJ, but they act as support for charge separation and transfer. Charge transport must take place between the components of the HJs during PEC. By adding a 2D material, the CT pathway can be completely changed. As discussed in Session 2.1.1, graphene and its derivatives are widely used as CT mediators because of their high electrical conductivity. Therefore, they are more frequently used for CT rather than for charge generation, and other semiconductors are usually needed in the HJ to absorb and convert the light energy. For example, $Cu_2O$ is largely used for PEC of $CO_2$RR.[82] With the addition of 2D graphene, PEC measurements demonstrated better performance, which is rooted in a suppressed charge carrier recombination. RGO can combine with other semiconductors in forming HJ such as $BiVO_4$/g-$C_3N_4$ to fabricate more complex and efficient photoelectrocatalyst.[47,56] Samsudin et al. loaded rGO on $BiVO_4$/g-$C_3N_4$ to fabricate a new HJ. Upon its integration in the HJ, rGO has become an intermediate material for charge migration in the HJ (Figure 5a). The electrons at the CB of g-$C_3N_4$ smoothly migrate towards the CB of $BiVO_4$ through rGO due to the electrostatic attraction between g-$C_3N_4$ and rGO. In addition, the observed smooth electron migration can be attributed to the lower Fermi level of rGO, which is higher in energy (–0.08 eV vs. NHE) compared to the calculated CB of g-$C_3N_4$ (–1.30 eV vs. NHE). Thereby, the photo-induced charge carriers generated *via* the rGO/g-$C_3N_4$/$BiVO_4$ photocatalyst will flow to the counter electrode via an external circuit for PEC. Similarly, the incorporation of 2D rGO within semiconducting nanoparticles such as AgCl and CdS[51], plays a vital role in charge separation[48]. In these combinations, rGO can attract and transfer one type of charge carriers, leaving the other type on the semiconductors, thus enhancing the charge separation and transfer. For example, the photo-excited electrons from AgCl quantum dots are transferred to the surface of rGO thanks to the favourable energy level between the CB of AgCl and the work function of rGO, and then flowed to the collecting electrodes, leaving the holes on the VB of AgCl QDs, suitable for oxidation reaction. rGO is used in the $CeO_2$-$TiO_2$ system for decreasing the recombination rate of charge pairs.[49] It was found that an optimized amount of rGO results in an improvement of degradation efficiency, while an excess loading reduces the PEC efficiency, as a too large amount of rGO coating blocks light absorption from the binary $CeO_2$-$TiO_2$ HJ.

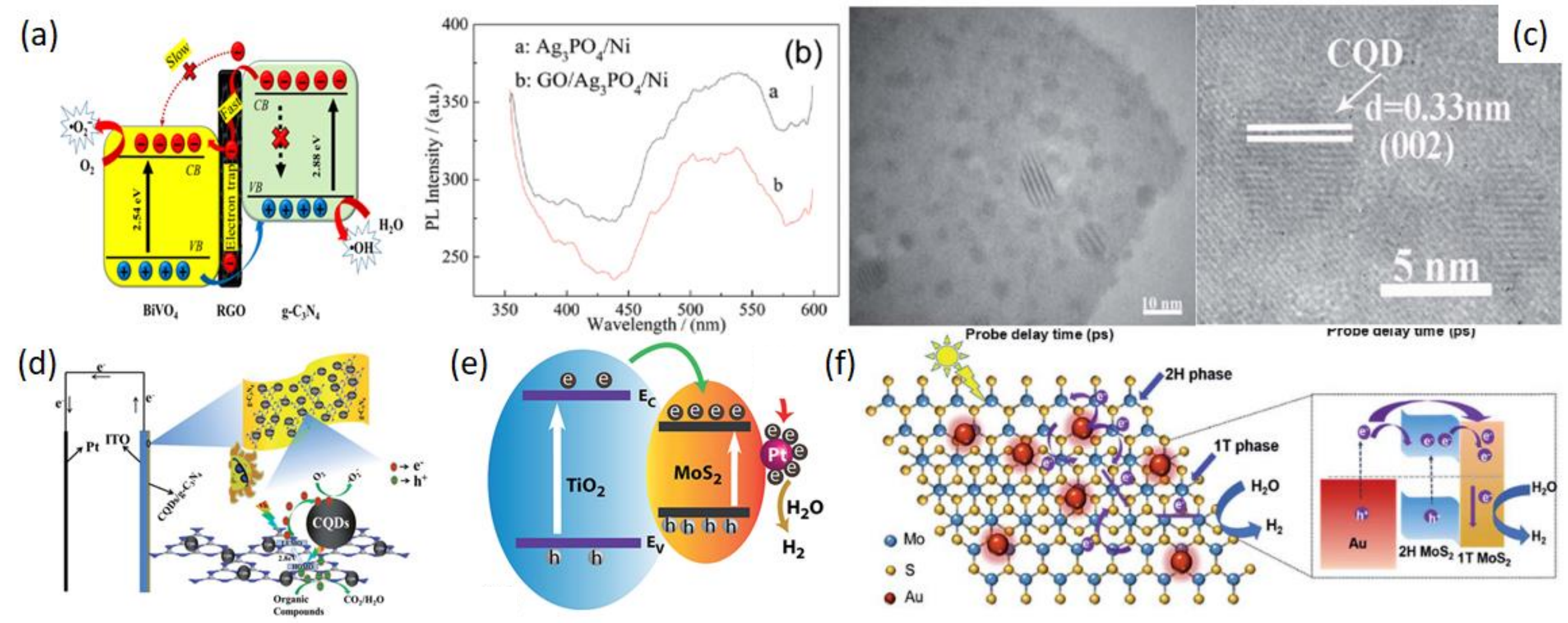

**Figure 5.** (a) Possible schematic illustration of CTinside the g-$C_3N_4$/$BiVO_4$ decorated with rGO.[47] © 2020 Elsevier B.V. (b) Room-temperature PL spectra of $Ag_3PO_4$/Ni and GO/$Ag_3PO_4$/Ni thin films.[14] © The Royal Society of Chemistry and the Centre National de la Recherche Scientifique 2020. (c) Transmission electron microscopy (TEM) images of the CQDs/g-$C_3N_4$. (d) Mechanism of the photoelectrocatalytic process for the CQDs/g-$C_3N_4$ under visible light irradiation.[58] This article is licensed under a Creative Commons Attribution-NonCommercial 3.0 Unported Licence. (e) Schematic illustration of CTat the multiple interfaces and the mechanism of enhanced visible light PEC activity of Pt NPs decorated on $TiO_2$/$MoS_2$.[83] © The Royal Society of Chemistry 2017. (f) A schematic illustration of PEC hydrogen evolution over the 1T-2H MoS2/Au HJ.[84] The Royal Society of Chemistry 2019.

Moreover, addition of rGO sheets can inhibit the recombination of charge carriers in a HJ such as $Ag_3PO_4$/Ni, due to the excellent conductivity of rGO.[14] As shown in Figure 5b, the pure $Ag_3PO_4$/Ni has a broad optically active recombination at 475–510 nm by the photogenerated holes with electrons around surface oxygen vacancies. While the quenching of the PL intensity of the rGO/$Ag_3PO_4$/Ni composite film indicates that the optically active recombination of charge carriers is suppressed by rGO. Interfacial charges can be transferred from the CB of $Ag_3PO_4$ to rGO, thus effectively separating photogenerated charge carriers. This helps to prolong the charge carrier lifetime and should lead to improved PEC activity. This better PEC performance is proved by faster rhodamine B (RhB) degradation than in the pristine $Ag_3PO_4$/Ni. Alternatives of rGO such as g-$C_3N_4$ work with the same mechanism in conjunction with $Ag_3PO_4$.

2D MXenes are another large group of CT-mediators researched in recent years. For example, the *in situ*-growth $TiO_2$/$Ti_3CN$ HJ demonstrates excellent light absorption abilities, possess large specific surface area and contains $Ti^{3+}$ species and oxygen vacancies, which are beneficial for the generation and transport of electron/hole pairs.[35] In $Ti_3C_2T_x/Bi_{12}TiO_{20}$ HJ, it was proved that $Ti_3C_2T_x$, as an useful co-catalyst, creates an IEF at the contact interface with $Bi_{12}TiO_{20}$ and also an external one, which are both responsible for the enhanced PEC performance.[65]

$MoS_2$ can perform as a CT-mediator as well. In $MoS_2/La_2Zr_2O_7$, the 2D $MoS_2$ surface presents a large active area for light absorption and photoelectrons generation, while rich oxygen vacancies in $La_2Zr_2O_7$ with a stable pyrochlore structure trap photogenerated electrons from $MoS_2$ and make them available for PEC.[85] Besides, the $MoS_2/La_2Zr_2O_7$ HJ distributes a localized charge density in the photoelectrocatalyst, which increases the adsorption sites of the reactants. All these factors drive efficient PEC reactions. To further improve the catalytic property, HJs with vertically aligned $MoS_2$ could be induced. For example, Fe doped $Ni_3S_2$ can be loaded on the edges of the $MoS_2$ to build an active interface which is more favorable for adsorption and desorption of the reactants and products.[86]

In some HJs, the 2D materials with semiconducting properties can act as charge generators and mediators, while the other component can attract the photogenerated charges from the 2D materials and use them for PEC. As discussed in section 2.1, g-$C_3N_4$ has drawn attention as a promising new layered material with a bandgap of ~2.7 eV. When combined with carbon quantum dots (CQDs), the contact area between g-$C_3N_4$ nanosheets and CQDs may be greatly increased due to the small size of the CQDs (Figure 5c).[58] Therefore, nanoscale junctions under close contact may be established. Photogenerated charges can transfer rapidly to the surface and further travel to the CQDs for PEC reactions (Figure 5d). Similarly, g-$C_3N_4$ can combine with other materials to form more efficient HJs such as $La_2O_3$/g-$C_3N_4$, CoO/g-$C_3N_4$, and $La_2O_3$/CoO/g-$C_3N_4$[87], in which g-$C_3N_4$ mainly absorbs photons and converts them into active charges to be transferred to the other parts of the junction for PEC.

Metal oxides like $WO_3$ can act as charge transport mediators.[70] For example, in the HJ $WO_3/BiVO_4$, 2D $WO_3$ is used as the electron transport layer, providing direct and fast electron transport pathways in the photoinduced charge transport process to suppress the back reaction.

In some situations, the 2D layered catalyst works as a charge separator and transfer mediator on one photoelectrode, while the needed charge carriers are transferred to the counter electrode for the PEC reaction. In water splitting, the catalyst actually catalyze oxygen evolution reaction (OER) directly and enhance the hydrogen evolution reaction (HER) indirectly because these two half reactions complement each other. Since the catalytic property is finally determined by detecting the production of $H_2$, the HJs are classified as catalysts for HER. The NiFe-LDH/$Co_3O_4$ HJ is such a catalyst for HER, in which the active electrons are generated on the NiFe-LDH/$Co_3O_4$ photoanode and move to the counter Pt electrode for the reduction reaction.[32] Another example is photoelectrocatalytic reduction of U(VI) of g-$C_3N_4/Sn_3O_4$/Ni.[60] In this system, the charge separation is guaranteed by the outside electric field and the S-scheme charge recombination pathway. Under the influence of a bias voltage, the photogenerated electrons on the CB of g-$C_3N_4$ transfer to the cathode, while the photogenerated electrons on the CB of $Sn_3O_4$ can neutralize the photogenerated holes on the VB of g-$C_3N_4$ and prevent the charge recombination inside the g-$C_3N_4$. Besides, photoelectrocatalysts such as h-$MoO_3$/1T-$MoS_2$[88], Nd-doped g-$C_3N_4$/BiOI[53], $Co_3O_4$/$NiMNO_3$[89], $TiO_2$/g-$C_3N_4$[40] and B-phase $TiO_2/MoS_2$[90] are charge generators and separators, providing active species for the corresponding PEC reactions.

The previous tremendous research work has proved the excellent property of HJs with 2D materials and other semiconductors as charge separators. Indeed, the combination between metals and 2D materials can be a promising strategy for efficient photoelectrocatalysts. It is also known that noble metals are perfect catalysts for a lot of chemical reactions, but their cost limits the application towards industrial level. To reduce the cost and ensure the quality of the catalyst at the same time, scientists attempt to combine the "best" charge carrier separator, namely 2D materials, with the "fastest" reaction active sites, namely, noble metals, for specific PEC. With this design of catalyst, the small amount of noble metal usually functions as the reaction active site, while the 2D materials are charge separators to convert light and electric energy and transfer the charge carriers to the active sites. Paul et al. introduced Pt dots to further improve the catalytic property of a n-doped $MoS_2/TiO_2$ (B-phase) HJ. Instead of being the real catalyst, the HJ works as a CT mediator to concentrate the electrons and transfer them to the Pt dots for HER (Figure 5e).[83] A similar strategy is applied in the 2D Au/$WS_2$ photoanode system, in which Au nanodots were successfully incorporated into 2D $WS_2$ nanosheets in order to collect active species, adsorb and catalyze toxic chemicals degradation, while $WS_2$ is in charge of generating charge carriers with its photoelectric property.[63] Hu et al. designed a $La_2Ti_2O_7$/Pt photoanode by following this strategy of building HJs.[73] The 2D perovskite $La_2Ti_2O_7$ is used as light converter and support for the decoration of Pt nanoparticles, while Pt acts as active site for photoelectrocatalytic methanol oxidation. Moreover, the loading of small amount of Pt, Ag, Rh or Au on B-doped g-$C_3N_4$ 2D layered materials can facilitate the PEC reduction reaction as well, since the noble metal behave as a fast catalytic reaction center, while the 2D layered g-$C_3N_4$ is in charge of electron-hole separation and migration.[91,15] In all of these above examples for PEC, a three-electrode system and external electrochemical workstation are applied for the test, indicating that this is still in the early stages of research. To push these cheap catalysts to the real application at industrial level, an artificial photosynthetic cell is needed for the characterization of the catalysts. Xu et al. prepared Pd/N-$TiO_2/Ti_3C_2$ as photocathode, in which N-$TiO_2/Ti_3C_2$ is responsible for generating and transferring active electrons to Pd for PEC, while $BiVO_4$ is chosen to replace Pt as the anode in the artificial photosynthesis cell.[68] Similar examples are listed in Table 1.

The combination of 2D semiconductors and materials with metallic properties can accelerate the CT at the interface of the HJ as well. Zheng et al. synthesized a 1T-2H $MoS_2$/Au HJ[84]. Under illumination, both the plasmonic Au and semiconducting 2H-$MoS_2$ act as light absorbers (Figure 5f). The hot electrons from Au with high enough energies can overcome the Schottky barrier and be injected into the 2H-$MoS_2$ CB. Besides, the Schottky barrier existing at the interface of Au and 2H-$MoS_2$ can efficiently prevent the injected hot electrons to recombine. Moreover, the intimate lateral 1T-2H $MoS_2$ heterostructure improves the intralayer CT from the CB of 2H $MoS_2$ to 1T-$MoS_2$. Additionally, the metallic 1T-$MoS_2$ enhances the electronic conductivity as well as provides abundant catalytically active sites for PEC water splitting.

### 2.3. 2D materials as supporting matrix

Thanks to the large surface area, 2D materials are frequently chosen as supporting matrix for other photoelectrocatalysts. By growing smaller nanosized photoelectrocatalysts on the 2D material matrix, the aggregation of the photoelectrocatalysts is prevented and it can be more uniformly distributed than without the 2D material matrix. For instance, nanosized pure lamellar structure of nickel boron oxide (Ni-$B_i$) suffer from low productivity and aggregation by direct synthesis, while on graphene matrix the ultrathin 2D Ni-Bi arrays vertically aligns and thus maximizes the exposure of the active sites for PEC (Figure 6a and 6b).[92] As shown in Figure 6c, the photocurrent density of Ni-Bi/graphene displays a distinct enhancement compared to the pristine Ni-$B_i$, suggesting an improved charge separation and transfer at the Ni-Bi/graphene HJ.

Peng et al. reported a novel photoelectrocatalytic system by growing carbon nitride (CN) on the surface of carbon paper under in situ crystallization by using dicyandiamide and calcination step afterwards[59]. As shown in Figure 6d, the synthesized CN by dicyandiamide or melamine exhibits a vertically aligned morphology with a homogeneous layer distribution and excellent contact with the carbon paper. It is notable that electrical conductivity of both CN/C samples is improved compared to that of carbon paper, and comparable to the commercial carbon paper. The good electrical conductivities of self-standing CN/C pave the way toward their application as freestanding photoanodes in PEC.

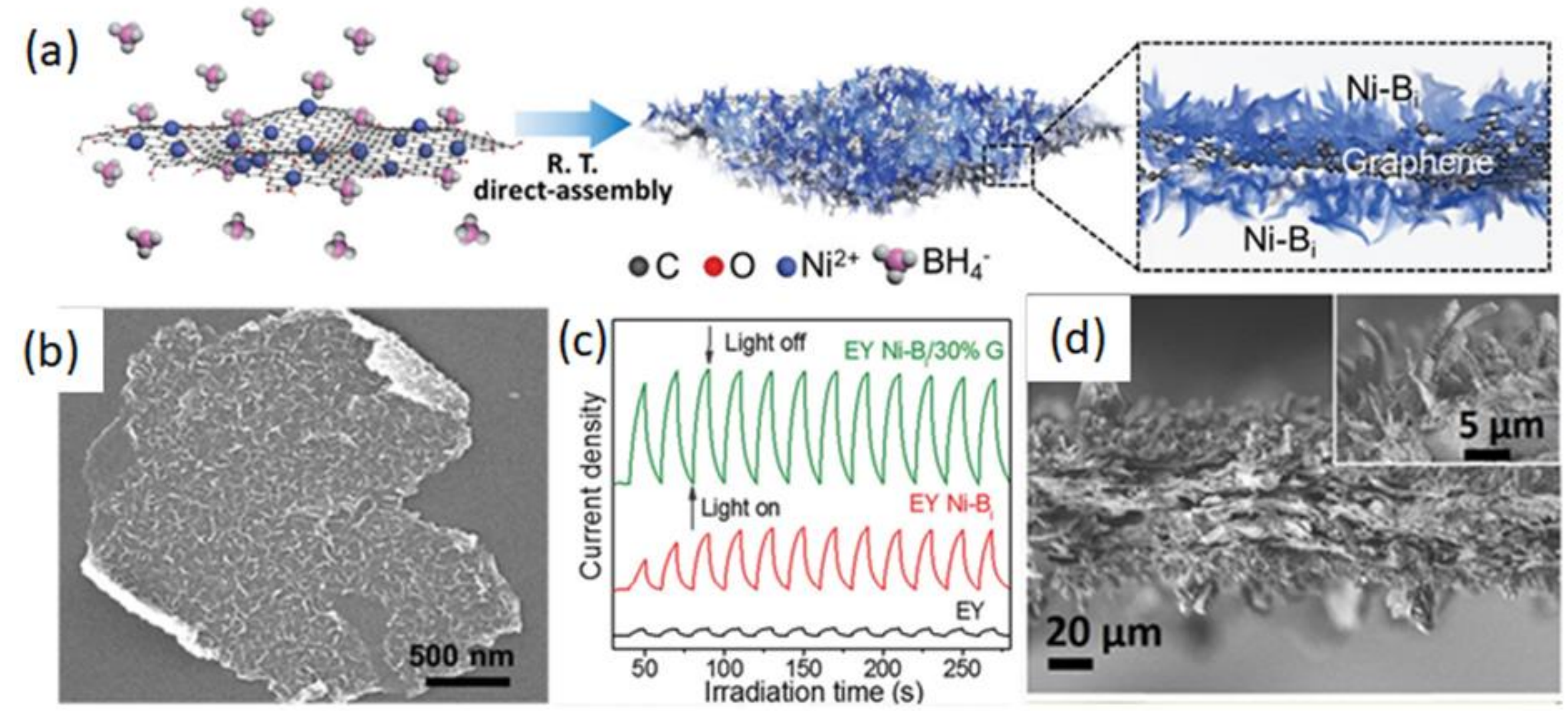

**Figure 6.** (a) Schematic illustration of the synthesis of the 2D-on-2D trilayered nickel boron oxide/graphene (Ni–Bi/G) sandwich heterostructure. (b) Field emission scanning electron microscopy images of Ni–Bi/G.(c) Transient photocurrent response bare EY, EY-(Ni–Bi) and EY-(Ni–Bi/G) Eosin Y is used (EY) as a photosensitizer.[92] © The Royal Society of Chemistry 2017 (d) cross-sectional SEM images of CN/C.[59] © 2019, American Chemical Society.

## 3. Synthetic approaches

Various methods are employed in the synthesis of 2D materials and the HJs containing them. These approaches can generally be categorized into two main groups: *top-down* and *bottom up* methods. i) *Top-down* methods focus on the dimensional reduction of the bulk counterpart to obtain the final nanomaterial. Among all kinds of top down methods, electrochemical exfoliation, sonication exfoliation and acid etching are common methods used in the fabrication of photoelectrocatalyst (Figure 7).[93,94] ii) *Bottom up* methods typically involve synthesizing the material beginning with the atoms or small molecular components, and the example of this strategy are CVD, hydrothermal synthesis, electrodeposition and polycondensation. The synthesis of the HJs generally includes one or more steps of synthesis method, and the multi-step synthesis combines various methods to build HJs with close interactions. The synthesis of HJs can be divided into *in-situ* and *ex-situ* strategies: i) in the in-situ synthesis the HJs are formed *in-situ* when one or more components are synthesized simultaneously. In this strategy, the bottom-up method is always used for crystallizing the last components and form HJs eventually. ii) in the ex-situ synthesis, the components are synthesized independently and combined in the last step with a separate synthesis method. In this section, we will introduce mainly the combination of the synthesis methods on 2D HJs with the above-mentioned classification.

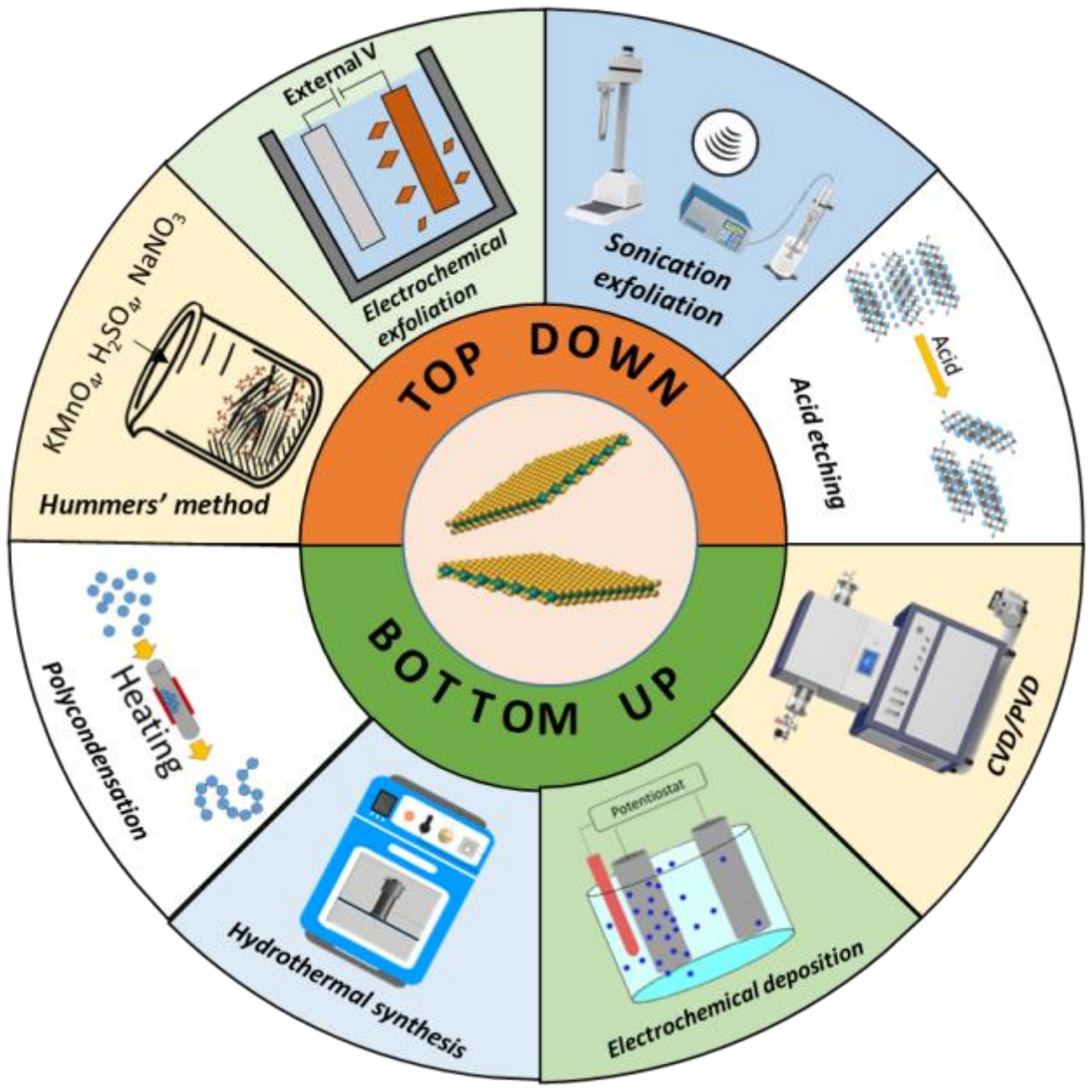

**Figure 7.** Commonly used top–down and bottom–up approaches for the synthesis of HJs containing 2D photoelectrocatalysts.

### 3.1 *In-situ* synthesis

In the *in-situ* synthesis, the formation of HJ is based on the direct and homogenous growth of one or more components. Usually, the more stable components are synthesized, followed by the addition of the less stable components to form the HJs. In this case, the 2D materials are usually firstly fabricated because of their better stability compared to other nanomaterial components and their ability as a platform to load and separate other nanomaterials.

For some popular 2D materials in the HJs, specific synthesis methods have been studied and developed on a relatively mature level, and they are synthesized individually as a platform for nucleation of the other components in the next synthesis steps. As a component of photoelectrocatalyst HJs, g-$C_3N_4$ is synthesized mainly by thermal polycondensation method as a typical 2D platform for the HJ formation, and the in-situ formation of smaller nanoparticles on the surface of g-$C_3N_4$ can prevent the aggregation and result in more efficient interactions between two components. The precursors for g-$C_3N_4$ consist of urea, melamine, thiourea, and dicyandiamide, which are carbon and nitrogen-rich organics. For instance, by heating acid treated urea powder under 550 ◦C for 2 h, Zhu et al synthesized ultrathin g-C3N4 nanosheets.[15] Then Pt nanoparticles are loaded uniformly on the 2D g-$C_3N_4$ by a simple reflux method, as shown in Figure 8a and b. Similarly, $Cu_2O$ nanoparticles[95] and CoMn-LDH[26] can attach on g-$C_3N_4$ by a solution phase method. Besides, hydrothermal method is commonly used for in-situ crystallization of other components on g-$C_3N_4$ to form photoelectrocatalytic HJs, and typical examples include $TiO_2$/g-$C_3N_4$,[41] $BiVO_4$/g-$C_3N_4$,[56] BiOCl/g-$C_3N_4$,[96] BiOl/g-$C_3N_4$,[55] $La_2O_3$/g-$C_3N_4$, $Co_3O_4$/g-$C_3N_4$,[87] ZnTe/g-$C_3N_4$,[97] NiMn-LDH/g-$C_3N_4$,[27] CuTi-LDH/g-$C_3N_4$,[98] $Sn_3O_4$/g-$C_3N_4$,[60] $MoSe_2$/g-$C_3N_4$,[79] SmV/g-$C_3N_4$,[43] Pt/g-$C_3N_4$,[16] $MoS_2$/g-$C_3N_4$.[99] Except hydrothermal synthesis, other methods such as CVD,[57] solvothermal method[16] can be used to grow HJs on the surface of g-$C_3N_4$ as well. For most nanomaterials with high surface energy, the risk of sintering and phase changing is high in a calcination step, thus the thermal polycondensation method for g-$C_3N_4$ is mostly in the first step. However, there are exceptions for the very stable nanomaterials connected to g-$C_3N_4$. For example, in α-$Fe_2O_3$/g-$C_3N_4$ HJ, α-$Fe_2O_3$ is made by hydrothermal method first, then goes through the calcination steps under 550 °C which is the temperature to form g-$C_3N_4$ (Figure 8c).[100] Similarly, TiO2 nanotube arrays can be synthesized by electrochemical anodization and used in the further step for $TiO_2$/g-$C_3N_4$ HJ by polycondensation as well.[40] It is possible to synthesize other types of carbon nitride such as g-$C_3N_5$ and CN by polycondensation with 3-amino-1,2,4-triazole and dicyandiamide as a precursor.[59, 101] Further steps are then necessary to form the HJs. Apart from polycondensation method, CVD has been used for carbon nitrides and the HJs as well. For example, by using methylamine-borane as a precursor, borocarbonitride layers are deposited on $TiO_2$ nanoribbons for photoelectrocatalytic water splitting.[102] Murugan et al. also reported a simple one-step soft template synthesis to prepare g-$C_3N_4$/$TiO_2$, which can be used as a photoanode for OER.[39]

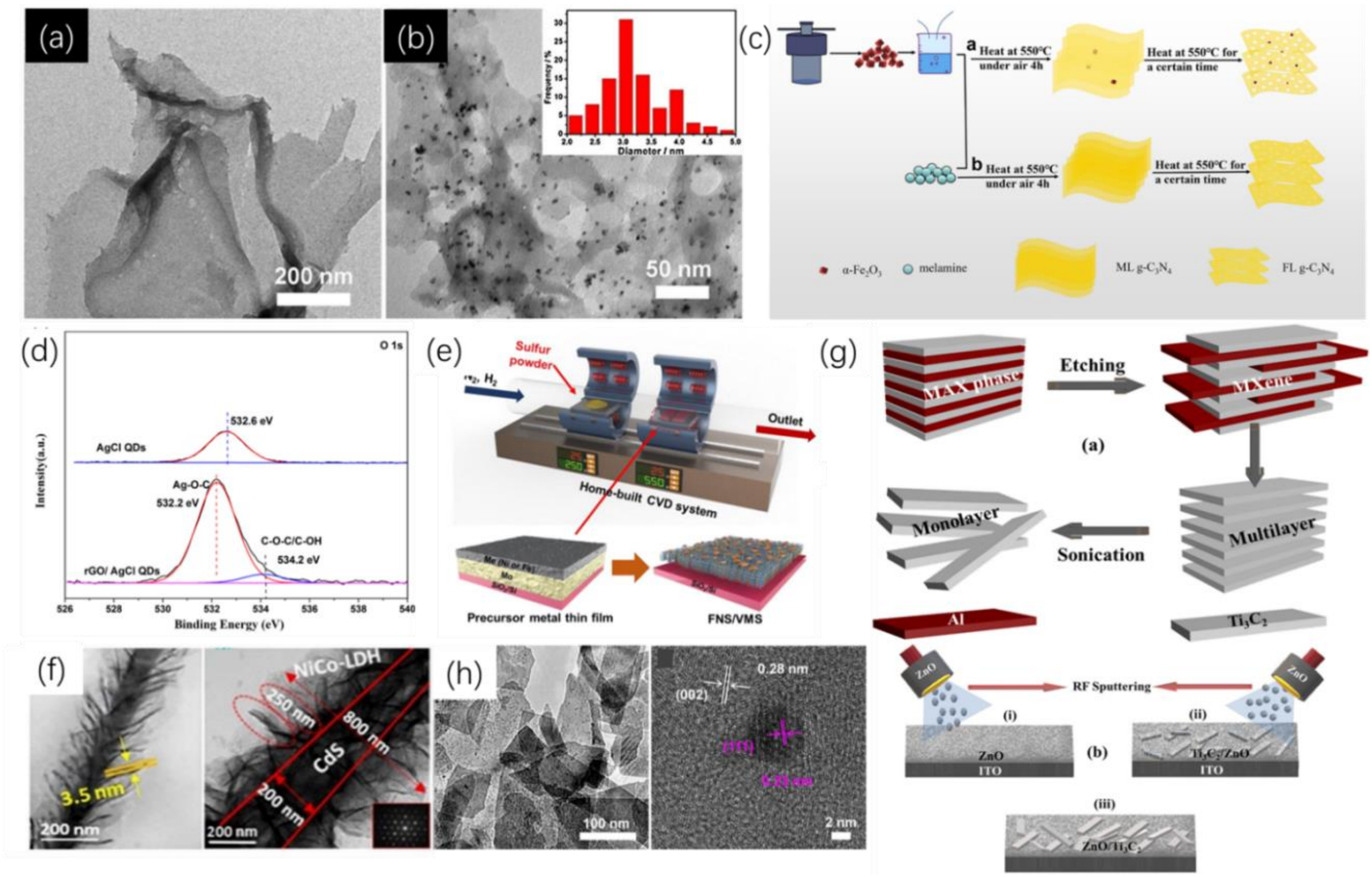

**Figure 8.** TEM images of pure ultrathin g-$C_3N_4$ nanosheets (a) and Pt/g-$C_3N_4$ nanocomposites (b). The inset of Figure 8b is the size distribution of Pt nanoparticles in Pt/g-$C_3N_4$ nanocomposites.[15] © 2016 Elsevier B.V. (c) Schematic illustration of the synthetic route used to prepare the g-$C_3N_4$.[100] © 2020 The Chemical Society Located in Taipei & Wiley-VCH Verlag GmbH & Co. KGaA, Weinheim. (d) High resolution XPS spectra of O.[48] © 2019 The American Ceramic Society. (e) Schematic of experimental procedures and characterization of synthesized metal sulfide decorated $MoS_2$. © 2021 Elsevier B.V. (f) HRTEM images of the CdS@NiCo-LDH.[29] © 2018 Published by Elsevier B.V. (g) Schematic representation of the $Ti_3AlC_2$ MAX phase transformation into multilayer and monolayer Ti3C2 MXene and magnetron sputtering of (i) ZnO (ii) ZnO on monolayer $Ti_3C_2$ ($Ti_3C_2$/ZnO) and (iii) $Ti_3C_2$ on ZnO (ZnO/$Ti_3C_2$).[103] © 2021 Elsevier B.V. (h) TEM and HRTEM images of Pt/$La_2Ti_2O_7$.[73] © 2017 Taiwan Institute of Chemical Engineers. Published by Elsevier B.V.

RGO is generally synthesized by exfoliating graphite power via Hummers' method. Then other components will crystallize on the surface of rGO, such as Cu nanoparticles by electrochemical reduction process,[50] $Ag_3PO_4$ by electrodeposition,[14] Pt/CdS combination by hydrothermal method.[51] Graphene can be fabricated via CVD in which methane is used as the carbon source, and other components are deposited on the surface of graphene through different methods. For instance, $Cu_2O$ can combine graphene by electrodeposition. Besides, hydrothermal method has been applied to reduce GO and construct a stable interaction in HJs containing rGO as well. For example, to prepare rGO/AgCl HJ, Kadeer et al. first mixed GO and AgCl quantum dots uniformly, then 2D GO was reduced and form Ag–O–C bonds simultaneously, which was confirmed by X-ray photoelectron spectroscopy (XPS) spectra of O (Figure 8d).[48]

For the HJs containing TMDCs, they are usually synthesized by hydrothermal method which is used as the step of formation of HJ. For example, to synthesize BiOBr/$MoS_2$/GO HJ, $(NH_4)_2MoO_4 \cdot 4H_2O$ and $(H_2N)_2S$ are chosen as precursors of Mo and S. Then the as-prepared GO is added in the precursor as a nucleation point for MoS2.[104] Similar examples include $MoS_2$/$TiO_2$,[90] $BiVO_4$/$SnS_2$,[62] $MoS_2$/$La_2Zr_2O_7$,[85] $Al_2O_3$/$SnS_2$,[76] $MoS_2$/$CoMoS_4$.[74] Moreover, hydrothermal synthesis can be used as a single step to form HJs. For instance, to synthesize $MoS_2$/$SnO_2$ HJ, $Na_2MoO_4$, thiourea and $SnCl_4 \cdot 5H_2O$ are used as Mo, S and Sn precursors and mixed in the Teflon-lined autoclave for a one-step synthesis. It is easy to control the ratio between components by simply modifying the amount of the precursors. $MoS_2$ layered materials can be obtained by liquid phase exfoliation and applied as a platform for nucleation of other components such as Au.[84] Besides, it is possible to synthesize layered TMDCs through vulcanization or selenization of the corresponding oxides in the HJs, for example, from $MoO_x$/Si to $MoS_2$/Si and $MoSe_2$/Si.[105] It has to be mentioned that environmental friendly chemicals have been used to replace precursors with health risk. In the hydrothermal synthesis of Au-$WS_2$ HJ, organosulfur from rotten garlic powder is used as a S precursor alternative of $CH_2N_2S$, providing a possibility for "green synthesis".[63] CVD is a commonly used method for constructing HJs containing MCs as well. For example, GaTe/ZnO is synthesized by a 2-step of CVD: the first step is to grow ZnO nanowire, while the second step is to grow GaTe nanosheets on ZnO.[64] On the other hand, $MoS_2$/$WS_2$ and Fe-doped $Ni_3S_2$/$WS_2$ can be synthesized by one step CVD considering the possibility of both components of changing from oxide hydrate or metal precursors to sulfides simultaneously (Figure 8e).[77, 86] Furthermore, $MoS_2$/Si,[106] $Sn_xMo_{1-x}S_2/MoS_2$,[81] $MoS_2/WO_3$[107] HJs can be synthesized by decorating $MoS_2$ via CVD. Zheng et al. reported a chemical bath method to coat $MoS_2$ on $BiVO_4$ photoanode. In principle, the $(NH_4)_2MoS_4$ precursor remains on $BiVO_4$ by a merging step, then a calcination is needed to obtain stable $MoS_2/BiVO_4$ HJ.[108]

LDHs can be easily synthesized by different kinds of wet chemical methods. Usually, the LDHs are easily formed on other components to produce HJs because of their flexibility in synthesis. For instance, CoFe LDH crystallizes easily on the surface of $BiVO_4$ in the ethanol solution of $Co(NO_3)_2$ and $Fe(NO_3)_3$, thus forming HJs photoelectrocatalytic water splitting.[30] Pirkarami et al. grow NiCo LDH microsheets on the surface of CdS by hydrothermal method by using $Co(NO_3)_2$ and $Ni(NO_3)_2$ as precursors (Figure 8f).[29] Zhu et al. grow NiFe LDH on $\alpha$-$Fe_2O_3$ particles via a fast electrodeposition method. Besides, LDHs can be used as nucleation points for other components. For example, $Co_3O_4$ is synthesized with NiFe-LDH as nucleation point to form a p-n HJ for photoelectrocatalytic removal of contaminants.[32]

2D layered MXene is mostly obtained by acid etching method, then used for combining other nanomaterials to form HJs. As a popular MXene used in PEC, $Ti_3C_2$ is obtained from $Ti_3AlC_2$ by selectively etching Al layers inside $Ti_3AlC_2$ structure (Figure 8g). Then the layered $Ti_3C_2$ is used for crystallization of other nanomaterials, such as Ti-ZnO[109] and ZnO[103] by magnetron sputtering, $Bi1_2TiO_{20}$[65] and $TiO_2$[68] by hydrothermal method.

BiOI is synthesized by hydrothermal method, followed by a second step of hydrothermal to produce Pt/BiOI as a photoelectrocatalyst for ethanol oxidation reaction.[34] Moreover, it can be deposited on $BiPO_4$ by electrochemical method for photoelectrocatalytic degradation of tetracycline.[45]

Except the above mentioned popular 2D layered materials and the HJs based on them, additional new materials have been reported as active for PEC, and the corresponding synthesis method is newly designed as well. Ultrathin $La_2Ti_2O_7$ nanosheets can be synthesized via facile hydrothermal method with $La(NO_3)_3$ and $Ti(SO_4)_2$ are precursors, while photo-reduction method was applied to load Pt nanoparticles to form a HJ for methanol oxidation reaction (Figure 8h).[73] $Bi_2WO_6/La_2Ti_2O_7$ can be obtained by a simple hydrothermal method as well.[71]

### 3.2 *Ex-situ* synthesis

In the *ex-situ* methodologies, the components for the HJs are synthesized or prepared firstly, then a process is formed to anchor the existing components to each other via covalent or noncovalent interactions. In general, top-down synthesis is applied for producing 2D layered materials, which are combined with other components in the last synthesis step. For example, Wang et al. synthesized MoS2/BiOBr HJs via a two-step liquid phase exfoliation. The two components are combined with vdWs interaction, which is proved by the shift of the MoS2 characteristic peaks in the Raman spectra (Figure 9a).[37] Si et al. managed to obtain $MoS_2/WSe_2$ by liquid phase exfoliation as well with isopropanol/H2O solvent.[75]

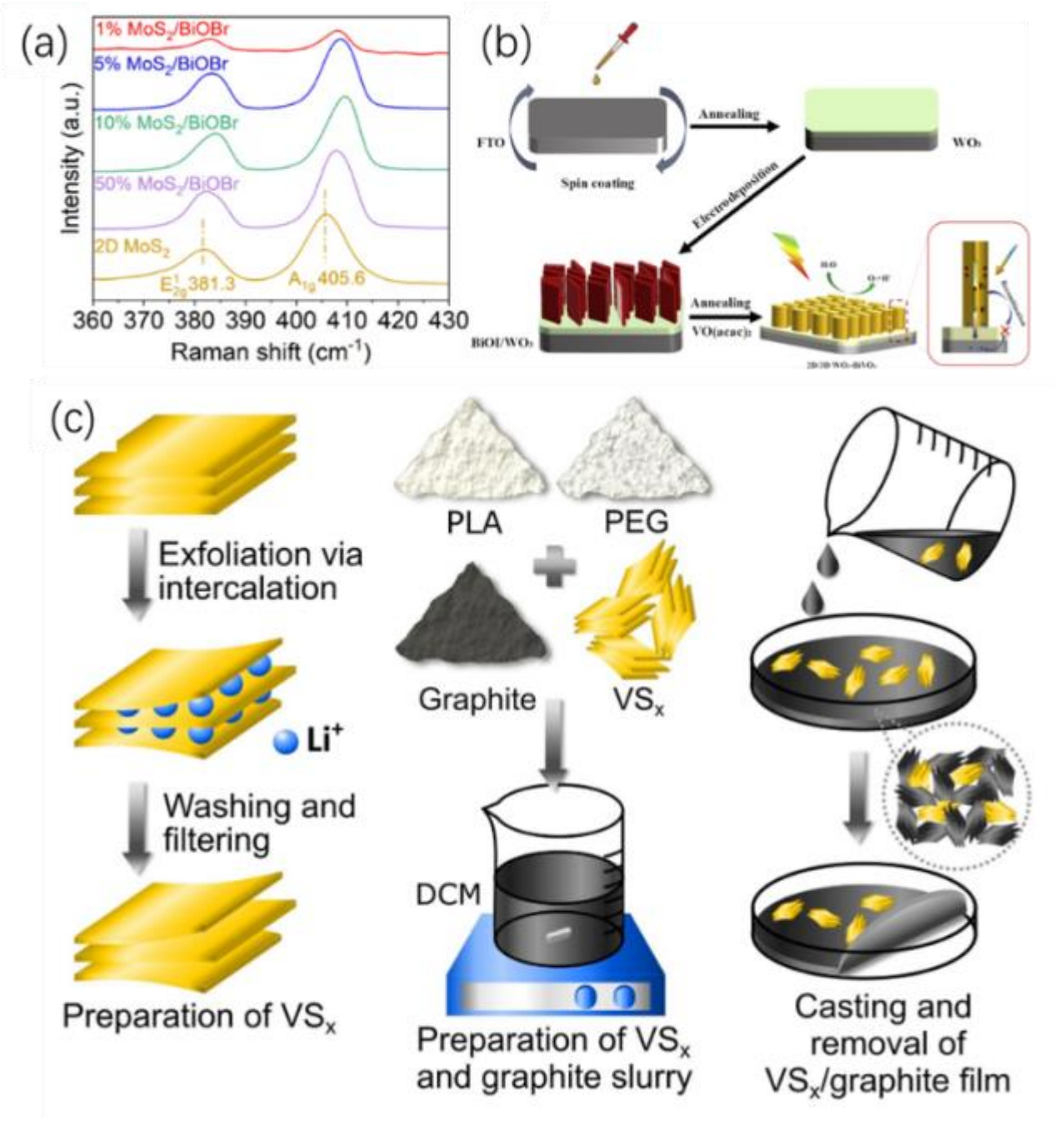

**Figure 9.** (a) Raman spectra of the HJs and 2D $MoS_2$ in the range of 360–430 $cm^{-1}$.[37] © The Royal Society of Chemistry 2023 (b) Schematic for the preparation of 2D/3D $WO_3/BiVO_4$ photoanode.**[70]** © 2021 Hydrogen Energy Publications LLC. Published by Elsevier Ltd. (c) Schematic illustration of the preparation of exfoliated $VS_x$ and a flexible graphite film that contains the exfoliated $VS_x$.**[80]** © 2022 Elsevier B.V.

Spin-coating deposition is a common method for ex-situ synthesis of HJs. To fabricate uniform HJs, the suspension of each component should have the desired stability and viscosity. For instance, to fabricate $TiO_2/In_2S_3$/GO HJ, GO powder was firstly obtained by Hummer's method, then dispersed in ethanol with acetic acid by ultrasonication to form a GO suspension. The suspension is eventually coated on the $TiO_2/In_2S_3$ structure.[52]

Electrochemical deposition can be used to fabricate photoelectrodes with HJs. For instance, $rGO/CeO_2/TiO_2$ HJ was prepared by electrochemical deposition of rGO on $CeO_2/TiO_2$, using $CeO_2/TiO_2$ as a working electrode and GO dispersion as electrolyte.[49] $WO_3/BiVO_4$ electrode was fabricated by electrodeposition of $WO_3$, in which $WO_3$ is used as a working electrode (Figure 9b).[70]

The *ex-situ* synthesis of HJs can be achieved by simply mixing and depositing the components as well. For example, to prepare $VS_x$/graphite film, Ng et al. mixed $VS_x$ and graphite with other polymers including poly(lactic acid) and poly(ethylene glycol) to make a homogeneous slurry. The HJ film is then obtained by casting and drying on a Petri dish (Figure 9c).[80] $WO_3/WS_2$ HJ can be synthesized by a simple drop casting step.[46] Zhang et al. synthesized carbon quantum dots/g-$C_3N_4$ via a simple ultrasonic dispersion self-assembly method.[58] Samsudin et al. managed to decorate rGO on g-$C_3N_4/BiVO_4$ by a simple wet-impregnation method.[47]

Compared to the *in-situ* synthesis strategy, the *ex-situ* synthesis is not used frequently. However, in the *ex-situ* synthesis, each component of the heterojunction is synthesized independently before combining them. This allows for precise control over the properties of each material, such as size, shape, composition, and crystallinity. Processing conditions for each material can be optimized independently. This can simplify the fabrication process and enhance the reproducibility of the heterojunction, as the synthesis of each component is not constrained by the presence of other materials. Researchers can optimize the properties of individual components separately, ensuring that each material is tailored to contribute specific characteristics to the heterojunction. This can lead to improved performance in the final application.

## 4. Computational methods

Theoretical modeling involves the study of materials at the atomic scale using the concepts of quantum mechanics (QM) or molecular mechanics (MM) and relying on appropriate algorithms and codes running on powerful supercomputers.[110] Theoretical calculations provide a versatile method that allow us to help understanding the experimental data by providing the necessary atomistic detail into the studied problems, as well as provide the additional insight which could be hardly accessible by experiments. Thus, they shall be regarded as an indispensable part of research, together with experiments, to elucidate the mechanism behind PEC. Their advantages include the ability to isolate a particular issue from a complex problem, and conduct cost-effective material screening, making it a suitable method for discovering suitable photoelectrocatalysts (Figure 10). In the following section, we first introduce the most common methodologies and then we show how they can be used to provide insight into different PEC reactions on 2D materials and heterostructures, thus completing the experimental data.

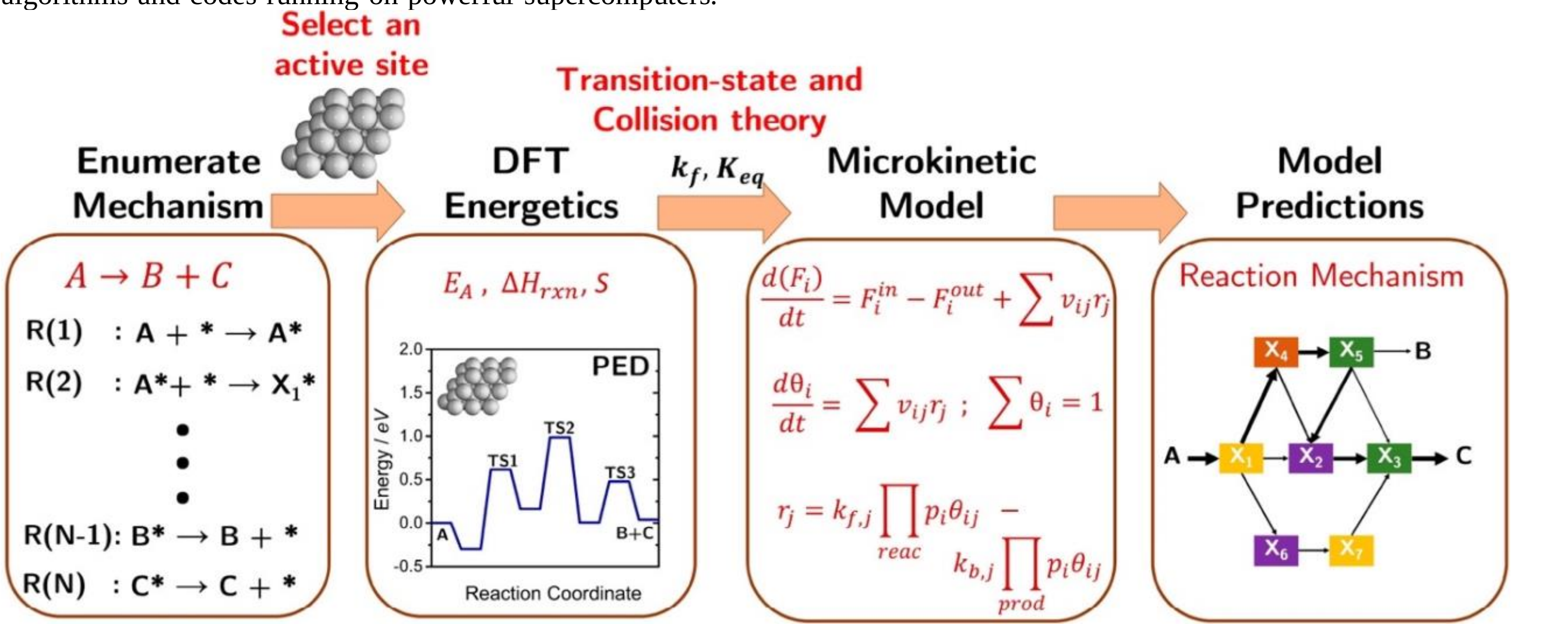

**Figure 10.** Schematic representation of the typical workflow adopted in a microkinetic-modeling (MKM)-based analysis for heterogeneous catalysis. A, B/C, and Xi refer to reactant, products, and reaction intermediates, respectively. PED refers to potential energy diagram. Reactor balance equations (system of differential-algebraic equations) solved in a MKM are shown in the third panel. Reprinted with permission from ref. [111]. Copyright 2020 American Chemical Society.

### 4.1 Interface modelling

First-principles density functional theory (DFT)[112,113] calculations have overcome a challenge to explicitly understand both the surface configuration and the structure-dependent electronic structure, charge carrier mobility and photoelectrocatalytic properties of interfaces.[114] It is essential to build an interface model of the 2D-layered HJ as close as possible to the real situation, described via employing experimental characterization techniques.[114–117] As an example, certain facets of the surface may be more reactive than others in a catalytic material,[118] or the presence of defects can constitute the most active catalytic sites due to their chemical surroundings or unsaturation.[119] Therefore, careful understanding of the experimentally measured data from different spectroscopy and microscopy techniques, etc. is vital to model the correct HJ structure. In addition, surface reconstruction events may occur after creating heterostructures by combining different dimensional systems.[95] These effects are easily captured by optimization steps in DFT calculations, and if the DFT functional inherently includes dispersion (optB88-vdW[120], vdW-DF2[121]) or dispersion corrected functionals[122,123] are selected , then the DFT shall provide correct structures.

Once the models of catalytic interfaces are built, DFT calculations allow to study their interactions which are typically described by the calculation of binding/adsorption energies and evaluation of the Gibbs energy profile to assess the catalytic mechanism. The Sabatier principle[124] for heterogeneous catalysis guides this approach. It states that the optimal catalyst binds adsorbates and intermediates with an appropriate binding energy: if intermediates are bound too strongly, they block active sites, and if they are too weakly adsorbed, they are not sufficiently present on the surface to allow for high reaction rates. The Sabatier principle is most often represented as a volcano plot.[125,126]

In addition to the static DFT calculations, molecular dynamics (MD) simulations can generate the time evolution of all the particles (atoms, ions, molecules) of a model HJ using a classical, Newtonian MD approach. However, in classical MD breaking bonds is not allowed. Thus, other types of MD simulations such as *ab-initio* MD (AIMD),[127] MD simulations with (reactive) force fields, e.g., ReaxFF,[128,129] AIMD-based metadynamics,[130,131] constrained molecular dynamics (CMD),[132] or density functional theory in classical explicit solvent (DFT-CES) simulation[133] can be employed. AIMD is a powerful tool that can quantify the interaction potentials and describe the physical–chemical nature of active sites of the catalyst. This method is very popular when explicit molecules of solvent are considered to be vital in the theoretical description.[134,135] ReaxFF based classical MD simulations allows bond breaking and forming but decreasing the computational cost as electrons are not computed. One of the most important features of ReaxFF is the use of charge equilibration (QEq) approach for handling the electrostatic interactions.[136] QEq calculates the actual charge distribution for every step of the MD trajectory and expands the electrostatic energy as a Taylor series where the linear term is identified as electronegativity of the atom and the quadratic term as electrostatic potential and self-energy.[137] DFT-CES offers an accurate description of the electrified interface at a balanced computational cost, by mean-field coupling of a QM description on the catalyst interface with a molecular dynamics description on the liquid structure of the electrolyte phase.[138] Compared with the AIMD simulation, the DFT-CES enables to investigate electrolyte phase dynamics with many more atoms over a more extended time-scale.

Apart from the information about interaction energetics between the adsorbent and the 2D-HJ, the electronic structure of the 2D-nanostructure before and after the adsorption of the targeted molecule is necessary to give insight onto the PEC process. This can be also obtained by DFT calculations.[139,140] DFT calculations with local density approximation (LDA)[141] and generalized gradient approximation (GGA)[142] functionals usually underestimate the bandgaps of semiconductors,[143] while the Heyd–Scuseria–Ernzerhof (HSE) hybrid functional[144] leads to more accurate results of bandgaps with respect to experimental data due to the long-range correction present in its functional form. A different strategy is to use wave-function based methods, such as many-body perturbation theory (MBPT) with the Green functions (G) and screened Coulomb potential (W).[145] Calculation of the energy alignment of VB and CB for periodic models and of frontier orbitals (highest occupied molecular orbital, lowest unoccupied molecular orbital) for molecular systems is a computationally simple approach used to determine the adequacy of different photoelectrocatalytic HJs for PEC e.g., in the case of water splitting, by comparing these energies with reduction potential energy, etc.[146] Additional descriptors commonly used to assess the catalytic activity of HJs are band-structure, total or partial density of states, charge difference distributions, Bader[147,148] charge analysis.

### 4.2 PEC calculations

The computational description of the full PEC process is a very challenging task, and, up to date, there are no available computational methods which can describe it in their entirety. This is mainly related to the difficulties in combining exciton dynamics with catalysis, since it requires the solution of the time-dependent Schrödinger equation for the many-body system, which has so far only been achieved for small molecules and is not yet applicable for HJs.

Involving nonradiative transitions between several electronic states using nonadiabatic MD (NAMD) is relevant in PEC, nevertheless, the methodology is still very rare in this field due to its enormous computational cost. The issue is not intrinsic of the methodologies used to properly describe the excited state (ES) potential energy surface, but relates to the hundreds of trajectories needed to be run in order to achieve reliable statistics.[149,150] The combined use of time-dependent density functional theory (TDDFT),[151,152] efficient in solving electronic problems, with NAMD can track the pathway of ES carrier energy over time, describing trajectories that evolve or jump between coupled (electronic) potential energy surfaces. The main energy loss involved in PEC at HJs is the nonadiabatic coupling between electrons and atoms. This nonadiabatic problem can be investigated by mean-field classical molecular dynamics,[153] using the so-called Ehrenfest dynamics,[154,155] or Tully's fewest switches trajectory surface hopping[156] which is self-consistently coupled with TDDFT for the description of the electronic degrees of freedom.[157] The hybrid quantum and classical dynamics approach evolves the electron wavefunction by solving the electron Schrodinger equation and describing the motion of the nucleus in Newtonian mechanics[115]. However, this technique is highly expensive in term of computational cost,[150] even though a few different codes (Hefei-NAMD, PYXAID) able to investigate the interfacial charge carrier dynamics, the electron–hole recombination dynamics, the exciting spin-polarized hole dynamics, have been developped.[158–160]

To solve this problem, the commonly used approach is to consider the PEC as two separate processes: (i) light harvesting and CT and (ii) electrocatalytic calculations. In this way, it is possible to compute a large part of the PEC process, although with the assumption that the catalysis is performed in the ground state (GS).

### 4.3 Light harvesting and charge transfer in 2D heterojunctions

QM approach based on TDDFT is the most used methodology for the description of ESs, as it enables modelling large systems and interfaces with high accuracy. However, since TDDFT is based on one-electron picture of the electronic GS, approximation must be considered.[146] The most common approximations in modelling the photogenerated holes and electrons include (i) study the positively (negatively) charged system by removing (adding) one electron, which inherently neglects the existence of the exciton and of the hole‒electron interaction and is typically used for finite systems; (ii) adding electron traps or donors, to simulate hole or electron existence by, *e.g.*, adding –OH or –H surface groups onto the periodic models, which is limited by the precise knowledge of the positioning and surface coverage of the functional groups;[161] (iii) first-principles MDs describing the GS of the system upon addition or removal of electrons, avoiding the costly ES calculations;[162] (iv) open-shell DFT calculations of the singlet photoexcited state, where the two unpaired electrons are coupled to a triplet spin state, allowing the structural optimization of the system in an excited-state potential energy surface which is supposed to be close to that of the singlet state.

For calculations of periodic systems, in addition to the prediction of the fundamental band gap, excitonic effects must be included in the determination of the optical band gap of a 2D material.[145] Most 3D semiconductors exhibit very small exciton binding energies, and hence, the distinction between optical and electronic bandgaps is frequently ignored. However, in 2D materials and for some organic semiconductors, the difference can be significant.[150] The optical absorption function is given by the imaginary part of the dielectric function and can be computed using DFT or the by applying GW quasiparticle approximation, including excitonic effects by solving the Bethe–Salpeter equation (BSE).[150,163] Despite the higher accuracy, this method is more computationally expensive compared to TDDFT, scaling as $N^4$ with the system size,[164] making TDDFT the best trade-off between accuracy and quality.[165]

TDDFT calculations can also provide information on the location of the photogenerated holes and electrons at the HJ. By analyzing the charge distribution, it is possible to discriminate between CT or energy transfer processes that can take place at the interface and observe how the generated charge carriers are transported to the adsorbed molecules at the catalytic site to induce redox reactions.

Despite its benefits, the use of TDDFT needs to be considered with caution, as it (i) is incapable to properly describe systems with polyradical character of the GS wavefunction, (ii) fails to include doubly ESs due to the adiabatic approximation of the time-dependent correlation-exchange potential,[166] (iii) sometimes incorrectly orders the ESs based on the poor choice of the exchange-correlation functional,[167,168] and (iv) fails to properly describe CT states.[169]

For 2D-HJs, it is important to consider the lowest ES providing that the photogenerated charges are easily separated and transported at the catalytic sites. Thus, it could be assumed that the system evolves on the energy surface of its lowest ES before partaking in the PEC reaction. An exact description requires to obtain the optimized geometry of the first ES, and it can be computationally demanding. To avoid this bottleneck, the classical path approximation, in which the ES structure of each intermediate step in the reaction path is approximated by the GS structures,[160,170] ignoring the response of the nuclear system to the excited electrons, can be adopted. For HJs, such approximation holds true since they have many electrons, and the excitation of a single electron has a limited influence on the system. Moreover, the exited electron lifetimes are very short, and the nuclear system has no time to respond to these short-lived excited carriers. Unfortunately, such calculations are barely performed in this field, even employing these simplifications.

### 4.4 Electrocatalysis calculations

Electrocatalysis is driven by applying a voltage across the reaction cell, providing a sensitive control of the rate. Therefore, for electrocatalysis calculations, the DFT procedure needs to be corrected to represent constant-potential conditions to make direct comparison with electrochemical available experiments.[171–173] The two most common methodologies to estimate the catalytic performance for electrocatalysis are the computational hydrogen electrode (CHE) model,[174–176] developed by Nørskov et al., and the grand canonical potential kinetics (GCP-K) method,[177–180] developed by Goddard III et al. Both these methods can calculate the thermochemistry of the entire process by including additional factors such as pH and applied potential in the Gibbs free energy formulation for each reaction step (including transition states), however, while the CHE consider constant charges, the GCP-K allows for changes in charges during the catalytic steps.

#### 4.4.1 Computational hydrogen electrode (CHE)

CHE prominently opened the door of EC to theoretical description. Thanks to its simplicity and computational efficiency, the CHE approach enabled studies of reaction pathways and large-scale computational screening studies with the purpose of finding the optimal electrocatalytic material for different reduction reactions.[181] It enables the computation of the relative Gibbs free energies of intermediates of a series of proton-coupled electron transfer (PCET) steps along an electrochemical reaction pathway. Each elementary step is formulated as a removal or addition of the electroactive species, (proton + electrons). For simplicity, we introduce CHE for the Volmer step of the HER reaction in acidic aqueous solution $H^+_{aq}$ + $e^-$ + * → H*, where * represents a catalyst surface, H* hydrogen adsorbed on the catalyst and $H^+_{aq}$ proton in the aqueous solution. For this reaction, the Gibbs free energy $\Delta G$ shall be calculated from equation 1:

$$\Delta G = G(H^*) - G(*) - G(H^+_{aq}) - \mu_e \qquad (1),$$

where $G(H^+_{aq})$ depends on the pH of the solution (2):

$$G(H^+_{aq}(pH)) = G(H^+_{aq}(pH = 0)) - 0.059pH \qquad (2).$$

The chemical potential of electrons $\mu_e$ is the change in free energy when the electron is added or removed from an (infinitely large) system and depends on the electrode potential $U$ which is referenced to the standard hydrogen electrode (SHE). Thus, equation 2 can be rewritten as:

$$\Delta G = G(H^*) - G(*) - G(H^+_{aq}(pH = 0)) - \mu_e(0_{SHE}) + 0.059pH + |e|U_{SHE} \qquad (3)$$

The $G$(H*), $G$(*), and $G(H^+_{aq})$ terms can be calculated from first-principles methods, typically DFT in the field of catalysis of 2D HJs, either in gas, but preferably in the implicit solvation as most of the experiments are done in some aqueous electrolyte. Nevertheless, it is still difficult to determine the $G(H^+_{aq}(pH=0))$ from first-principles, which not only involves the strong solvation but also the proton concentration.[182] This bottleneck is overcome by the assumption that energies of other species can be easily calculated based on the existence of thermodynamic equilibrium in the SHE (equation 4):

$$G(H^+_{aq}(pH = 0)) + \mu_e(0_{SHE}) = G(H_2)/2 \qquad (4).$$

As a result, the free energy can be calculated with DFT calculations by substituting equation 4 to equation 3 (equation 5), where all terms on the right side of the equation are known.

$$\Delta G = G(H^*) - G(*) - G(H_2)/2 + 0.059pH + |e|U_{SHE} \qquad (5)$$

Overall, the essence of the CHE is that the free energy of solvated ions, which is difficult to obtain from DFT calculations, can be deduced from the equilibrium reaction where the energies of other species can be easily calculated or have been experimentally measured.[182]

Due to its simplicity, the CHE method has been widely used for investigations of low-dimensional systems catalysis. This method allows to find an onset potential (minimum overpotential), which drives all the reaction steps towards the formation of the final product. Nevertheless, CHE still poses several drawbacks. First, it is not well adapted to study explicit pH effects and implicitly assumes that the neutral surface state computed by DFT is representative of 0 V vs reversible hydrogen electrode (RHE).[183] In other words, the electronic energies are only measured for electroneutral entities independently from the electrochemical potential as the electron transfer reaction is always coupled, even though in real systems electrons are exchanged between catalyst and electrode to reach equilibrium, where Fermi level of the catalyst and electrode potential shall be equal. Moreover, the effect of surface charge/constant potential can be significant in heterogeneous electrochemistry, especially for low-dimensional HJs, where the change of charge affecting the electronic structure is not confined to the catalytic center. Secondly, the solvation effects are assumed to be negligible, and only computations in the gas-phase are necessary to evaluate electrocatalytic reaction pathways. Even though simple approaches to predict reaction barriers have been derived,[184,185] CHE model neglects the activation energies of all PCET steps and therefore is not reliable for accurate kinetics.

Hence, different methods have been derived and implemented to reach more realistic theoretical description. Methodologies such as GCP-K, or constant potential-hybrid solvation-dynamic model (CP-HS-DM) appear to be useful to describe the kinetics at solid−water interface[186] and for decoupled electron transfer/proton transfer.[187,188] CP-HS-DM methodology overcomes the oversimplification of CHE by (i) considering that the configurations and interactions of explicit water molecules can evolve along the reaction coordinate; (ii) not assuming that the system has zero (or constant) net electronic charges and does not neglect the surface charge effects. The key feature of the CP-HS-DM is the implementation of the constant-potential condition for electrons to the slow-growth AIMD. Practically, it uses several layers of explicit water molecules in conjunction with an implicit solution to solve the reaction species. The electrons are coupled with a fictitious potentiostat so that the Fermi level of the system fluctuates around a constant and the number of electrons evolves following the grand-canonical distribution at the preset electrode potential. The constrained AIMD simulations are performed with reaction coordinate gradually changing from the initial to the final state. Eventually, the mean force with respect to the reaction coordinate is integrated, giving the free energy profile, from which the activation energy can be identified.[186]

### 4.4.2 Grand canonical potential kinetics formulation (GCP-K)

The recently developed GCP-K QM formulation based on thermodynamics calculations provides the understanding of heterogeneous electrochemical reactions by describing reaction kinetics as a function of an applied potential (*U*). In traditional QM, the number of electrons is fixed. In the GCP-K method, the number of electrons is automatically adjusted to the fixed applied potential, leading to a quadratic dependence of the potential on *U*. This provides physical quantities which describe differential capacitance during CT from the electrode to reacting species.[189] GCP-K method is elegantly capable of predicting electrochemical kinetics, thus, calculate current density and turnover frequency as a function of applied potential rather than for fixed charges as in standard QM. It describes how the energy barrier of transition state and its structure change with a fixed applied potential, which leads directly to current versus potential relation (Tafel slope). It has been demonstrated that GCP-K accurately describes the PCET process during different electrocatalytic reactions and it has already been successfully employed in a few theoretical works where a good agreement with experiments has been found.[177,189,190]

In more details, the voltage-dependent GCP for surface states can be derived from traditional fixed-electron based free energies by using a Legendre transformation.[177] The GCP-K formulation arises from minimizing the free energy for each value of n in Equation (7) using a Legendre transform relating the net charge of the system and the applied voltage. Performing this macroscopic transformation explicitly allows to make the connection of GCP-K to the traditional Butler−Volmer kinetics (Figure 11),

$$G(n;U) = F(n) - ne\,(U_{\mathrm{SHE}} - U) \qquad (7),$$

where $G$ is the grand canonical free energy, which depends on the applied potential $U_{\mathrm{SHE}}$ vs SHE, $n$ is the number of electrons, $e$ the unit electronvolt in energy, $F$ is the free energy obtained as a function of $n$ and $U_{\mathrm{SHE}} = \mu_{e,\mathrm{SHE}}/e$ is the electronic energy at SHE conditions. The Fermi level can be shifted by changing the reference electrode based on the pH of the solution. To use $G(n;U)$ as a thermodynamic potential, the number of electrons needs to be equilibrated with the applied potential. The Fermi level shift with applied potential is achieved by changing the number of electrons of a system (changing the occupation of the electronic bands) during QM self-consistent calculations.[189] GCP is then obtained by minimizing the free energy following Equation (8):

$$GCP(U) = \min_n G(n;U) = \min_n F(n) - ne(U_{\mathrm{SHE}} - U) \qquad (9).$$

The minimization of free energy as a function of $n$ shall be at least quadratic to describe the minimization of $GCP(n;U)$. That is why free energy $F(n)$ as a function of number of electrons is expanded as a quadratic form (Equation (9)):

$$F(n) = \mathrm{a}(n - n_0)^2 + b(n - n_0) + c \qquad (10),$$

where $a$, $b$, $c$ are the fitting parameters. This leads to Equation (11)

$$GCP(U) = -\frac{1}{4a}(b - \mu_{e,\mathrm{SHE}} + eU)^2 + c - n_0\mu_{e,\mathrm{SHE}} + n_0 eU \qquad (11),$$

where the final form of GCP becomes

$$n = n_0 - \frac{C_{diff}}{e} - (U - U_{PZC}) \qquad (12)$$

$$GCP(U) = -\frac{C_{diff}}{2}(U - U_{PZC})^2 + F_0 - n_0\mu_{e,\mathrm{SHE}} + n_0 eU \qquad (13),$$

$$F(n) = \frac{e^2}{2C_{diff}}(n - n_0)^2 + (\mu_{e,\mathrm{SHE}} - eU_{PZC})(n - n_0) + F_0 \qquad (14).$$

where $U_{PZC}$ is the potential at zero charge; $\mu_{e,\mathrm{SHE}}$ is the chemical potential of an electron at SHE; $F_0$ is the free energy at zero net charge; $n_0$ is the number of electrons at zero net charge.

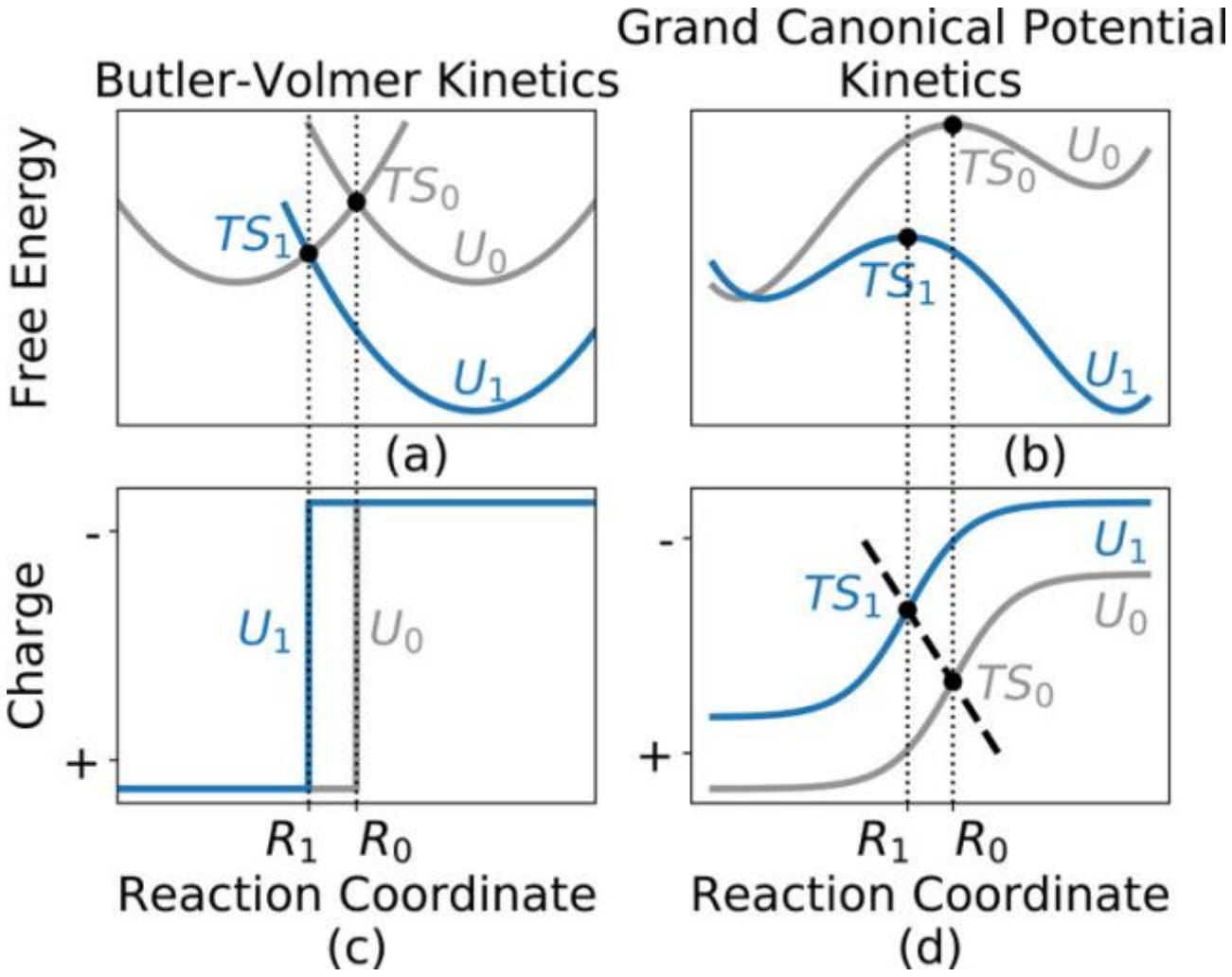

**Figure 11.** Schematics showing how voltage dependent electrochemical reactions are described by GCP-K. (b) and (d) differ from the more standard view of Butler−Volmer kinetics (a) and (c). As the voltage is changed from $U_0$ to $U_1$, the energy profiles shift as in (a) and (b), while the relevant reaction coordinate changes from $R_0$ to $R_1$. The Butler−Volmer picture in (c) can be considered as a special case of the GCP-K scheme (d) in which the electron transfers instantaneously. Reprinted with permission from ref. [177]. Copyright 2018 American Chemical Society.

For neutral system, $F_o(n = n_o) = c$ (from Equation 10); $b = \mu_{e,\mathrm{SHE}} - eU_{PZC}$ (from Equation 4). Because of this formulation, the differential capacitance $C_{diff}$ across the electrode and electrolyte can be calculated $C_{diff} = -e\frac{\partial n}{\partial U} = \frac{e^2}{2a}$; where $a = \frac{e^2}{2C_{diff}}$ (from Equation 11). This all allows a continuous description of the evolution of the reaction intermediates and transition states.

Despite its versatility, the computational cost of the GCP-K approach is higher than the CHE one, and its application to low dimensional materials just started to be relevant.[191] Nevertheless, its high accuracy, implementation, and ability to provide physical observables, make this method a much reliable choice for EC reaction studies compared to CHE.

### 4.5 Microkinetic modelling

First-principles kinetics can simulate EC processes in real time, considering rare reaction events, and the method may be parametrized by data, including adsorption energies and kinetic barriers, obtained from DFT calculations and transition state theories. First-principles kinetics is then followed by kinetic modeling such as (i) microkinetic modeling or (ii) kinetic Monte Carlo (kMC) simulations. In microkinetic modeling, a set of differential equations is simultaneously solved to calculate the overall reaction rates and species concentrations, describing the catalyst surface, gaseous phases, and, if relevant, liquid phase and interfaces (Figure 10).[111] Usually, competitive adsorption, desorption, and surface transformations are included.[192,193] Microkinetic modeling needs all relevant rate constants of the reaction as an input, calculated from all the energies of the relevant reaction intermediates and transition states as functions of applied potential using the quadratic grand canonical potential, and then obtained using the Eyring equation. For periodic systems, transition states are commonly computed using the climbing image nudged elastic band calculations.[194,195] Another effective way to obtain the kinetic barrier is the Brønsted–Evans–Polanyi (BEP) relationship.[196,197] It relates the kinetic barrier to the corresponding reaction energy for a class of materials: $\Delta G^{\ddagger} = \beta\Delta G + \alpha$, where $\Delta G^{\ddagger}$, $\beta$, $\Delta G$, and $\alpha$ are the kinetic barrier, BEP coefficient, reaction energy, and a constant, respectively. BEP relationship allows one to estimate the kinetic barrier by simply calculating the reaction energy, leading to a significant reduction in computational costs and permitting high-throughput screening of materials.[198,199] For EC, its main approximation is that the transition state for the nonelectrochemical step is equivalent to that of the electrochemical step at a specific potential $U$ (Equation (15)):

$$\Delta G^{\ddagger}(U) = \alpha(U)(\Delta G(U) + neU) + \beta(U) \qquad (15),$$

where $\alpha(U)$ and $\beta(U)$ represent the best fit slope and intercept of the BEP scaling relation that vary with potential.[191]

Microkinetic modeling is not computationally demanding, however, it deals with averages, hence it lacks the capabilities to probe the catalyst surface on the atomistic level, such as tracing the effects of adsorbate relative positions, *i.e.*, lateral interactions and cooperative effects.[192]

On the other hand, kMC[200,201] is a relatively cheap method, which can be applied for EC in estimating material properties, charge transfers at the interface or reaction paths.[110202] The combined DFT and kMC approach could provide a detailed picture of the interplay between thermodynamics and kinetics, as well as precise predictions of the current and voltages for electroreduction reactions as a function of the applied electrochemical potential.[203] kMC simulations are also capable to assess the charge transport and electron dynamics processes at the interface in the framework of a hopping regime,[204] as well as to cast down the contributions from the different constituents of the system and helping in avoiding possible causes of charge recombination.[205]

## 5. Photoelectrochemical applications of 2D materials

### 5.1 PEC water splitting

The PEC water splitting reaction occurs at the interface of the photoelectrode and the electrolyte solution. The photo-generated electrons reduce water to form hydrogen gas while the holes generated in the photoelectrode oxidize water to form oxygen gas. The overall reaction is represented as:

$2H_2O$ + energy (light/electricity) → $2H_2 + O_2$

The whole reaction can be divided into two half reactions, namely HER and OER, which take place on photocathode and photoanode, respectively:

$2H^+ + 2e^-$ → $H_2$ (HER)

$2H_2O$ → $O_2 + 4H^+ + 4e^-$ (OER)

2D material HJs have shown great potential in photoelectrocatalytic water splitting because of their special properties. One of the most promising 2D materials used for PEC water splitting is graphene, acting mostly as CT mediator or supporting matrix because of its electrical conductivity, excellent mechanical strength, large surface area and stability in a wide range of environments.

Other 2D materials that have been investigated for photoelectrocatalytic water splitting include TMDCs, which have tunable bandgap and strong light-matter interactions, thus showing excellent active charge carrier production and high catalytic activity.[33] The formation of HJs composed of TMDCs is a promising strategy to improve the PEC activity, and there are some HJs designed for catalysing water splitting, like a triple W/$WO_3$/$WS_2$ HJ.[46] As the photoanode, the $WO_3$/$WS_2$ is able to separate charges, use the photogenerated holes for OER, while accumulating and transferring electrons to the W substrate and the counter electrode for HER. This HJ showed a photocurrent density of 5.6 mA/$cm^2$ at 1.23 V vs. Ag/AgCl, which was 7.2 times higher than pure 2D $WO_3$. Moreover, the $WO_3$/$WS_2$ HJ enhances the incident photocurrent efficiency (IPCE) by 55%, and the applied-bias photon-to-current conversion efficiency of the $WO_3$/$WS_2$ films was approximately 2.26% at 0.75 V (vs. Ag/AgCl), which is 9 times higher than pure $WO_3$.

In some cases, PEC water splitting can combine with other techniques to improve the reaction efficiency. For example, HJ photoelectrocatalysts can be used with ion membranes to facilitate the production of $H_2$ and $O_2$.[28] A ZnCr-LDH/g-$C_3N_4$ HJ was applied as a catalyst with an interlayer bipolar membrane to split water for HER and OER near each Pt electrode, while the remaining unreacted $H^+$ and $OH^-$ combine with $Cl^-$ and $Na^+$, respectively, and the cation-exchange and anion exchange membrane is permselective to specific ions (Figure 12a).

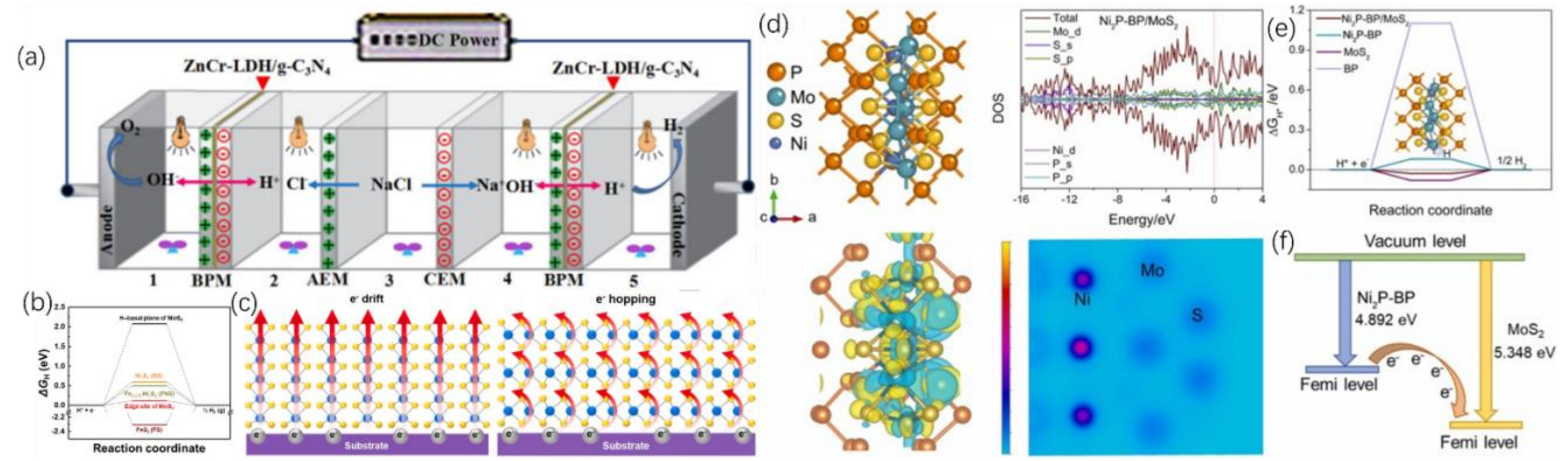

**Figure 12.** (a) Schematic diagram of photoelectrocatalytic water dissociation performance tests.[28] © 2017 Elsevier B.V. (b) Free-energy diagram of H adsorption on the site of each catalyst from DFT calculation. Schematics of (c) pristine $MoS_2$ and vertically aligned MoS2.**[86]** © 2021 Elsevier B.V. (d) DFT calculation results: model, Bader charge distribution, density of states (DOS), and local electron state density distribution of the $Ni_2P$-BP/$MoS_2$ heterointerface, (e) $\Delta GH^*$ of BP, $MoS_2$, $Ni_2P$-BP, and the $Ni_2P$-BP/$MoS_2$ heterointerface. (f) Difference in the work functions of Ni2P-BP and $MoS_2$. Reproduced with permission from ref. [206]

There are several reports focusing separately on catalysts for either HER or OER, since most photoelectrocatalysts are either oxidative or reductive. However, there are also catalysts which are active for both HER and OER. To design a proper catalyst for the whole water splitting process, one of the strategies is to build a *p-n* junction, to separate excited electrons and holes more efficiently and transfer them on different sites for HER and OER, separately. As mentioned before in Section 2.4.1, the 2D/1D BiOI/g-$C_3N_4$ HJ is able to gather holes on 2D BiOI and electrons on g-$C_3N_4$ nanotubes.[55] Thus, both the photogenerated charge carriers can be exploited, resulting in an enhanced PEC water splitting.

DFT calculations on CT efficiency in type II $Cu_2O$/g-$C_3N_4$ HJ[95] showed an enhanced hydrogen generation via photocatalytic and electrocatalytic water splitting reaction, which was induced by the transfer of photogenerated electrons from p-type $Cu_2O$ to n-type g-$C_3N_4$. These DFT calculations focused on changes in the electronic structure upon adsorption of $Cu_2O$ nanoparticle on g-$C_3N_4$ nanosheets. Although the $Cu_2O$ (111) facet is stable, a surface reconstruction of outermost layer of $Cu_2O$ was revealed using AIMD simulation. Moreover, two adsorbed g-$C_3N_4$ sheets suffered significant structural distortions because of the surface-bound bonding interactions between $sp^2$ hybridized N and Cu atoms. This structural deformation of g-$C_3N_4$ nanosheets is thought to be a crucial component in the enhanced catalytic activity of the HJs. Importantly, Bader electron charge analysis carried out after the HJ formation revealed the net electron transfer from the $Cu_2O$ surface related to the adsorption and limited electron tunnelling toward g-$C_3N_4$ sheets without direct contact with the $Cu_2O$ surface.

DFT calculations with the HSE06 functional (on the PBE optimized geometries) were used to analyze the evolution of optoelectronic properties and high-frequency dielectric constant profiles of various 2D $MoO_{3-x}S_x$/$MoS_2$ HJs modified by chemical and physical approaches.[207] This $MoO_3$/$MoS_2$ HJ is a type III (broken gap) HJ associated with a metallic character. Nevertheless, the DFT calculations proposed a strategy to tune the transitions from type III to type II band alignment of materials made of the 2D $MoO_{3-x}$Sx/$MoS_2$ HJ by exchanging the terminal oxo atoms of the $MoO_{3-x}S_x$ single layer with sulfur enables shifting its CB position above the VB position of the $MoS_2$ single layer. Interestingly, at medium and high S-doping (>5%), the bandgap and CT of the $MoS_2$/$MoO_{3-x}S_x$ HJ evolve continuously with the S concentration in the $MoO_{3-x}S_x$. At medium S-doping (9–10%), the 2D $MoS_{0.28}O_{2.72}$/$MoS_2$ HJ was identified as a potential PEC material for water splitting since it may generate a direct Z-scheme system due to the proximity of the CB of S-doped $MoO_3$ to the VB of $MoS_2$ (bandgap ~0.4 eV).

### 5.1.1 Hydrogen evolution reaction

HER in PEC involves the use of a photocathode to catalyze the reduction of $H_2O$ into $H_2$ in the presence of light and bias. Theoretically, $H^+$ needs to combine with electrons and change to $H_2$. Therefore, a photoelectrocatalyst with proper CB edge can provide sufficient electrons for the reaction. Meanwhile, the bandgap of the catalyst should be as small as possible to absorb as wide a range of the solar spectrum as possible.

Different kinds of MCs have been reported as photoelectrocatalysts. Among them, $MoS_2$ has caught most of the attention as a component of HJs since its band position is suitable to convert UV and visible light to active charges and provide reaction active sites for HER.[208] It is known that in the PEC of HER, the reaction active sites where the intermediates are adsorbed and reactions happens play a crucial role. The active sites of 2D layered materials in PEC of HER are typically located at the edges and defects of the material. This is because the edges and defects of 2D materials have higher energy states than the flat surface, which can provide sites for the adsorption and activation of reactant molecules. For instance, Paulraj et al. wrapped 2D $MoS_2$ on silicon nanowires, the photoelectrocatalytic results evidence the significant enhancement in performance of $MoS_2$/Si NWs based hybrid photocathode with ~300 mV vs. RHE, and the current density of –26.5 mA/cm$^2$ was achieved at the applied bias of 0 V vs. RHE.[106] This performance is also comparable with stable Pt/Si NWs photoelectrode. GaTe/ZnO HJ is designed similarly to ensure as many active sites are exposed to the reactants as possible.[64] In this HJ, the ZnO nanowires acted as the supporting framework for the favour layer-over-layer stacked growth of GaTe, which provided high surface area and high density of exposure edges for HER. The dark and photocathodic current reached −2 mA/cm$^2$ and −2.5 mA/cm$^2$ at −0.39 V vs RHE. High stability was indicated by the 20% decreased photocurrent density in 2 h. The hydrogen evolution rate was about 1.5 μmol/cm$^2$/h, which is much higher than that obtained at the pristine ZnO nanowires.

Furthermore, the activation of basal plane of $MoS_2$ is another strategy to intensify the photoelectrocatalytic properties. Roy et al. report basal plane activation of $MoS_2$ by heterostructuring with 2D $MoSe_2$ for enhanced PEC of HER. $MoS_2$/$MoSe_2$ HJ grown on silicon nanowire (SiNW) array shows 1.2 times higher photocurrent density and 1.36 times higher incident IPCE than pristine $MoS_2$ grown on SiNW array along with 4.44 times higher $H_2$ evolution rate compared to pristine SiNW photocathode.[209] Here, DFT calculations of heterostructure reveal that CT from the $MoSe_2$ layer to the basal plane of $MoS_2$ increases overall electron density resulting in its increased affinity towards proton reduction, which supports the experimental findings.

HJ containing vertically aligned $MoS_2$ was designed by introducing $Ni_3S_2$ and Fe doped $Ni_3S_2$ on the edge-terminated surface of $MoS_2$ as well. The onset potentials of this photocathode markedly shifted toward the anodic direction to 280 mV, with photocurrent density at 0 V of −25.4 mA/cm$^2$. The significant decrease in Gibbs free energy at the interface between $MoS_2$ and $Ni_3S_2$ or Fe-doped $Ni_3S_2$ promotes robust proton reduction at the interface (Figure 12b). Moreover, the resistance is reduced dramatically at the HJs with vertically aligned MoS2, indicating that the electrode-to-electrolyte shuttling of electrons on this HJ was the fastest among all the samples. This is because the movement of the electrons inside the 2D layers of MoS2 is much faster than through the H-basal plane, as shown in Figure 12c.

However, recent reports find that the crystal defects can behave as electron traps to hinder the reaction efficiency, as the defects are the recombination center of the photogenerated charge carriers. By passivating the edge defects, the charge recombination is reduced, and the cathodic photocurrent increased dramatically; while the photocurrent increases even more after the sample was treated by pre-annealing and surface passivation to eliminate the internal and edge defects, respectively. These contradictions in the reaction mechanism encourage scientists to invest more attention on these directions to develop more efficient photoelectrocatalysts for HER.

For some 2D HJs, the photoelectrocatalytic property is optimized by tuning the amount ratio of each component. For instance, the 2D/2D $BiVO_4$/$SnS_2$ HJ with a 1:3 molar ratio exhibited dramatically higher current densities of 0.21 mA/cm$^2$ than the HJ with a 3:1 ratio (0.01 mA/cm$^2$).[62] However, the mechanism governing the components' ratio is still unknown and requires further investigations. A new 2D MC, VSx, has been reported as a photoelectrocatalyst for HER as well.[80] VSx/graphite HJ forms a flexible electrode functionalized as a photoelectrocatalyst for enhanced HER by visible and near-infrared light irradiation (overpotential ≈500 mV at the current density of −10 mA/cm$^2$), in which graphite is a charge separator and mediator.

Transition metals such as Fe, Co, Ni and Mn, and their compounds have been proved active for electrocatalytic and photoelectrocatalytic HER. This is due to their partially filled d orbitals, allowing them to readily donate or accept electrons during the catalytic processes. The NiMn-LDH/g-$C_3N_4$ HJ can improve the HER performance by reducing the potential at −60 mA/cm$^2$ to −126 mV and Tafel slope to 50 mV/dec, thus boosting the PEC of HER dramatically. Similarly, using Co instead of Ni in the HJ also show promising PEC activity for HER, and the performance are listed in Table 2.[26]

**Table 2.** Comparison of PEC HER activities for some state-of-the-art 2D-HJ related materials.

| Catalyst | electrolyte | Light source | Current density | Tafel slope (mV/dec) | H2 production | Ref. |
|---|---|---|---|---|---|---|
| **Nd-doped g-$C_3N_4$/BiOI** | 0.5 M $Na_2SO_4$ | Air Mass, AM 1.5 G | 15.5 mA/cm$^2$ at −1.23 V vs. RHE | / | 288 μmol/h/cm$^2$ | 53 |
| **Pt/$MoS_2$ /$TiO_2$** | 0.5 M $H_2SO_4$ | 250 W Xenon lamp | −10 mA/cm$^2$ at −74 mV vs RHE | 30 | / | 83 |
| **BiOI/g-$C_3N_4$** | 1.0 M KOH | 300 W Xe lamp | −0.23 mA/cm$^2$ at −1 V vs. Ag/AgCl | / | / | 55 |
| **GaTe/ZnO** | 0.1 M $Na_2$SO4 | 300 W Xe lamp (λ > 420 nm), 100 mW cm–2 | −2 mA/cm$^2$ at −0.39 V vs. RHE | / | 1.5 μmol/h/cm$^2$ | 64 |
| **$MoSe_2$/Si** | 0.1 M$H_2SO_4$ | white LED | 4.5 mA/cm$^2$ at −1.0 V vs Ag/AgCl | / | / | 210 |
| **SmV/g-$C_3N_4$** | 0.5 M $H_2SO_4$ | 400 W Xe light | −10 mA/cm$^2$ at −200 mV vs RHE | 63 | / | 43 |
| **1T/2H $MoS_2$** | 0.5 M $Na_2SO_4$ | visible light | −1.4 mA/cm$^2$ at -0.6 V vs. Ag/AgCl | | / | 18 |
| **$MoS_2$ nanoribbons** | 0.5 M $H_2SO_4$ | 405 nm LED, 140 μW/cm2 | −10 mA/cm$^2$ at −280 mV vs RHE | 124 | / | 211 |
| **$MoS_2$/Si** | 0.5 M $Na_2SO_4$ | He-Ne lamp 100 mW/cm2 | -33 mA/cm$^2$ at -0.3 V vs. RHE | 34 | / | 106 |
| **$BiVO_4$/GQD/g-$C_3N_4$** | Lake water | 500 W halogen lamp | 19.2 mA/cm$^2$ at 1.0 V vs Ag/AgCl | / | 84.9 mmol/h/cm$^2$ | 56 |
| **$BiVO_4$/$SnS_2$** | 0.5 M Na2SO4 | solar simulator, 100 mW/cm2 | 0.21 mA/cm$^2$ at 1.23 V vs RHE | / | ~21 µmol/sec/cm$^2$ | 62 |
| **RGO/g-$C_3N_4$/BiVO** | 0.5 M Na2SO4 | 500 W halogen lamp, 100 mW.cm-2 | 14.44 mA/cm$^2$ at 1.0 V vs Ag/AgCl | / | 63.5 mmol/h/cm$^2$ | 47 |
| **P/g-$C_3N_4$** | 0.1 M Na2SO4 | | 202 µA/cm$^2$ at −1 V vs. Ag/AgCl | / | / | 57 |
| **$MoS_2$/$TiO_2$** | 0.5 M H2SO4 | 300 W Xenon lamp | 10 mA/cm$^2$ at −320 mV | 135 | / | 90 |

The decoration of small number of noble metals has become a strategy to improve catalytic properties of 2D HJs. For instance, the $MoS_2$/$TiO_2$ HJ can be used as photoelectrocatalyst for HER and the onset potential is −215 mV, and the Tafel slope is 92 mV/dec.[83] The performance can be strongly improved by decorating Pt nanoparticles on the surface of $MoS_2$ to form more effective reaction active sites. Specifically, the onset potential is reduced to −9 mV vs. RHE, which is dramatically lower than the benchmark Pt/C (−48 mV vs. RHE under the same experimental conditions). Meanwhile, the Tafel slope decreased from 92 mV/dec to 30 mV/dec with the addition of Pt, revealing faster kinetics by using Pt as catalyst and $MoS_2$/$TiO_2$ HJ as charge mediator.

In some reports, the HJs are not directly used as photocathode for HER, but to separate and migrate the charge, while the counter electrode collects the electrons for HER. In the work of Samsudin et al., the $BiVO_4$/rGO/g-$C_3N_4$ is the photoanode to separate and migrate photogenerated electrons to the cathode for HER, while the production of $H_2$ is on the Pt counter electrode.[56] To quantify the yield of $H_2$, the gas analyzer was applied. The high photoelectrochemical hydrogen production of 63.5 mmol/h demonstrated by this HJ exceeds the pure $BiVO_4$ (9.5 mmol/h) and g-$C_3N_4$ (11.9 mmol/h), implying that the rGO serves as an excellent electron mediator for the reduction process which lowers the photocatalytic overpotential and minimizes the photocharge carrier recombination. Similarly, h-$MoO_3$/1T-$MoS_2$ HJ is used as photoanode to generate active electrons for HER as well.[88] There are more other HJs reported as an indirect photoelectrocatalysts, which are listed in Table 2.

Although pristine MXenes show unsatisfactory catalytic activity, phosphorus and oxygen modified $Mo_2CT_x$ MXenes appear to be different class of promising HER photocatalysts.[212] Using the CHE method, Qu et al. observed that MXenes doped with non-metals possess a metallic band structure and an optimal value of $\Delta G_{H*}$, which lead to improved conductivity and electrocatalytic kinetics, respectively.[212] Interestingly, both O and P needed to be present in the model, otherwise the resulting $\Delta G_{H*}$ value would be too negative, restraining the adsorbed hydrogen escaping from the surface. Doping of $Ti_3C_2T_x$ MXenes by nitrogen atoms was also found advantageous for HER.[213] $Ti_3C_2T_x$ MXene used as an efficient solid support to host a N and S coordinated ruthenium single atom (RuSA) catalyst was demonstrated to be display superior activity toward the HER.[214] The model of RuSA/N/S/$Ti_3C_2T_x$ catalyst was constructed based on XPS and XAFS results, and DFT calculations employing CHE method. partial density of states and total density of states analyses revealed that the coordination interaction of RuSA altered the electronic structure of the $Ti_3C_2T_x$ support while achieving an optimal $\Delta G_H^*$ close to zero.

An enhancement in HER performance under UV illumination was found for pristine 2D BiOBr and the 1% $MoS_2$/BiOBr vdWs HJ, with the latter showing better catalytic performance and stability than the former.[139] DFT calculations revealed that excellent properties of 1% $MoS_2$/BiOBr may be attributed to an optimal distribution of $MoS_2$ nanosheets among 2D BiOBr layers, as was revealed by total and partial density of state (DOS) analysis. The CT analysis demonstrated that an almost constant number of ∼0.35 electrons was transferred from BiOBr to $MoS_2$. DFT calculations also proved that the holes photogenerated in BiOBr can remain in BiOBr or be partially transferred on the surface of $MoS_2$. Overall, it was concluded that the system can be considered a Type-I HJ, in which a formal energy transfer process takes place from BiOBr to $MoS_2$ following photoexcitation of the former through a charge exchange mechanism.

Another novel dual in-plane/out-of-plane $Ni_2P$-black phosphorus (BP)/$MoS_2$ HJ (Figure 12d-f) was found to be a promising catalyst for HER.[206] Here, the DFT calculations with PBE functional and vdWs corrections verified that the work function difference at this Mott-Schottky interface caused an electron transfer from $Ni_2P$-BP to the outer-surface $MoS_2$, which boosts the HER process. This was demonstrated by Bader charge and DOS analysis. As such, the $Ni_2P$-BP/$MoS_2$ heterostructure presents almost zero $\Delta G_{H^*}$ (−0.03 eV), indicating superior HER activity beyond the counterparts ($MoS_2$, BP, and $Ni_2P$-BP).[206] Hydrogen was found to be preferentially adsorbed on the S atom of the $Ni_2P$-BP/$MoS_2$ HJ.

Moreover, ultrasmall NiSe/$WSe_2$@NC HJ exhibited an enhanced alkaline HER.[215] All calculations were carried out by spin-polarized DFT. Changes in Gibbs free energy were calculated by the CHE model. The calculated work function of $WSe_2$ of 5.67 eV was higher than that of NiSe (4.84 eV). Consequently, the contact potential at the NiSe/$WSe_2$ interface could induce a built-in electric field, resulting in electron transfer from NiSe to $WSe_2$. In accordance, the charge density difference analysis showed that 1.83 $e^-$ are transferred from NiSe to $WSe_2$, resulting in the electron accumulation on $WSe_2$, favoring the water dissociation energies, and electron depletion on NiSe, being beneficial for the release of $H_2$. Moreover, the calculated adsorption energy of water on NiSe/$WSe_2$ was presented to be suitable adsorption energy at W sites (−1.07 eV), and the $\Delta G_{H^*}$ at Ni sites is as low as 0.37 eV, which are superior to those of both single NiSe and $WSe_2$ surfaces.

2D HJs InSe nanosheets modified with Au nanoparticles, where Au atoms are adsorbed on the hollow site of the InSe nanosheet, also achieved the theoretical enhancement of the HER performance comparing to the not-modified InSe nanosheets.[216] Here, the DFT calculation of the band structure revealed that Au nanoparticles can improve the conductivity of InSe nanosheets after the Au adsorption, as the indirect band semiconductor, with a gap of approximately 1.36 eV, is changed to metal after the formation of the heterojunction. Moreover, charge density distribution analysis demonstrated that the adsorption of Au induces the formation of Au–Se CT channels with ionic bond characteristics. This feature can not only inject free electrons on the surface of Au ions into the InSe nanosheet to improve its conductivity but also serves as a CT channel for the HER process. Additionally, the calculated $\Delta G_{H^*}$ of the HER process changed from 1.72 eV for InSe nanosheets to −0.59 eV for InSe/Au HJ, confirming that the formation of the HJ optimizes the hydrogen adsorption/desorption kinetics process, effectively improving the HER performance.

### 5.1.2 OER

OER is the other half reaction of water splitting. It occurs on the photoanode, which requires the active holes to oxidize hydroxyl species and produce $O_2$. Thus, a photoelectrocatalyst which can accumulate and release holes to the reactants is required.

g-$C_3N_4$ 2D HJs are commonly used as a photoelectrocatalyst for OER. Among them, the $TiO_2$/g-$C_3N_4$ HJ is the most reported because of its flexibility of synthesis, good energy conversion ability and the excellent stability of $TiO_2$. g-$C_3N_4$ is the catalyst for OER considering the band alignment of the two components. For example, Murugan et al. combine g-$C_3N_4$ with $TiO_2$ to prepare a type II HJ, in which g-$C_3N_4$ absorbs the incident photons and excites the electrons between the frontier orbitals, while $TiO_2$ attracts the excited electrons to hinder the charge recombination.[39] As such, the HJ with 20 wt% of g-$C_3N_4$ exhibits 1.4 and 595 folds of OER efficiency than the bare components, respectively. Rajaitha et al. obtained a similar HJ with a simple wet chemical synthesis method and found that this photoanode yields a maximum efficiency of ~0.072%, which is twofold higher than bare $TiO_2$ (0.035%).[38] Moreover, g-$C_3N_4$/BiOI photoanode was able to generate a photocurrent density as high as 0.70 mA/$cm^2$. This 50% increase in photoresponse over the pristine, low bandgap BiOI films indicates the superior charge separation ability of the HJ.[42] Besides, it displayed almost no loss in photoelectrochemical performance and did not suffer damage to their structure and optical properties due to reuse, which bodes well for their long-term operational stability and durability.

HJs containing $MoS_2$ are widely used in PEC of OER as well. The VB edges of monolayer $MoS_2$ are more positive than the oxidation potential of water, which means 2D $MoS_2$ could be used as catalyst for PEC of OER.[217] However, the pure $MoS_2$ suffers from fast charge recombination which inhibits its usage for OER. Therefore, a variety of $MoS_2$ HJs are fabricated for efficient charge separation and enhanced OER performance. For instance, the combination of $MoS_2$ and $\alpha$-$Fe_2O_3$ can change the band structure of the catalyst and separate the photo-generated charges more efficiently.[44] As shown in Figure 13a, the band bending achieved equilibrium of the Fermi level when $MoS_2$ is in contact with $\alpha$-$Fe_2O_3$ in the p-n junction. Following visible light illumination, the photo-generated electrons were transferred from the CB of $MoS_2$ to the CB of $\alpha$-$Fe_2O_3$. The holes were simultaneously driven by the electrostatic field from the VB of $\alpha$-$Fe_2O_3$ to the VB of $MoS_2$, and they reacted with •OH to generate $O_2$. Vertical $MoS_2$ NSs-decorated $WO_3$ nanorods possessed an enlarged surface area, improved light absorption performance, and appropriately organized the staggered HJ of $MoS_2$/$WO_3$ for promoting the PEC reaction, showing a 72% enhancement in PEC performance compared to pristine $WO_3$. The photoanode with 3D $BiVO_4$ grown on 2D $WO_3$ underlayer displays a photocurrent of 2.55 mA/$cm^2$ at 1.23 V vs. RHE as well.[70] Moreover, the onset potential of the 2D/3D $WO_3$/$BiVO_4$ photoanode is shifted to 300 mV comparing to the pristine $BiVO_4$, ensuring lower voltage requirements for PEC of OER. $MoS_2$/$WS_2$ monolayers can form a type II HJ with matching crystal lattice parameters, allowing for ultrafast CT and separation between the discrete materials. Sherrel et al. synthesized the $MoS_2$/$WS_2$ HJ on Au substrate, reaching an IPCE of ∼1.6% and a visible-light-driven photocurrent density of 1.7 mA/$cm^2$ (at 1.19 V vs RHE) for OER (Figure 13b).

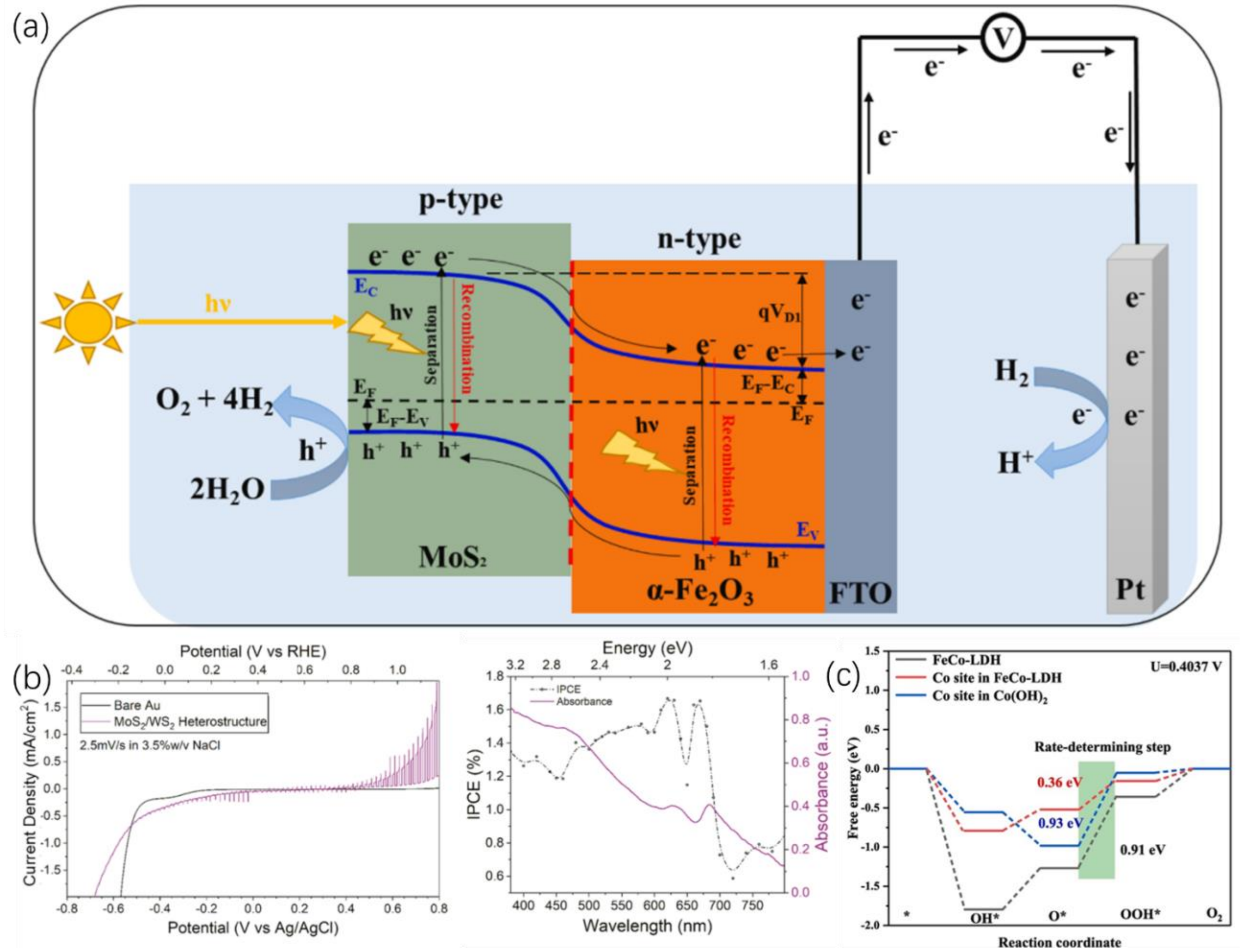

**Figure 13.** (a) Mechanism for the p–n HJ of α-$Fe_2O_3$/$MoS_2$.[44] © 2020 The Author(s). Published by Elsevier B.V. (b) Linear sweep voltammetry at 2.5 mV/s from +0.8 to −0.8 V vs Ag/AgCl, demonstrating the complete protection of the Au surface as minimal noble-metal hydrogen evolution is observed and a negligible contribution of the substrate to either dark or photocurrent in the water oxidation region occurred. Photocurrent dependence on monochromatic light at +0.75 V vs Ag/AgCl and corresponding absorbance spectra of the heterostructures dispersed via sonication in ethanol.[77] © 2019 American Chemical Society. (c) Adsorption free energy of FeCo-LDH and FeCo-LDH@Co$(OH)_2$ (with Co site in Co(OH)2 and FeCo-LDH) electrocatalysts for OER.[218] © 2023 Elsevier B.V.

The construction of HJs with $MoS_2$ and materials with metallic properties like $Sn_xMo_{1-x}S_2$ is helpful for the enhancement of the charge separation and transfer.[81] The $Sn_xMo_{1-x}S_2$/$MoS_2$ interface can effectively eliminate the Fermi level pinning effect and minimize the contact resistance, therefore ensuring electron transfer through $Sn_xMo_{1-x}S_2$ to the counter electrode, while the photogenerated holes participate in OER on the photoanode, improving the photoelectrocatalytic performance by 2.5 times.

Besides, the thin layer of 2D $MoS_2$ on the surface of $BiVO_4$ dramatically increases the photocurrent density in the range of 0.6–1.8 V vs. RHE[219]. This is mainly because $MoS_2$ provides the reaction active sites from OER, as the pure $BiVO_4$ shows negligible photocurrent in the range of 0.6–1.8 V. vs. RHE, since it is probably a charge mediator in the PEC.

LDHs containing Fe, Co and Ni are largely studied as electrocatalysts for OER because of the intrinsic properties of Fe, Co and Ni. Therefore, these compounds are combined with other photoelectrocatalysts for more efficient OER. For instance, α-$Fe_2O_3$/$Ni_{0.5}Fe_{0.5}$-LDH photoelectrode exhibits a negative shift of onset potential compared with pure α-$Fe_2O_3$, suggesting a boost in the HJs kinetics for OER.[19] CoMn-LDH shows photocurrent for OER as well, and the catalytic property can be enhanced by adding 10 wt% of g-$C_3N_4$.[26] A comparison of the photoelectrocatalytic properties of different photoelectrocatalysts is listed in Table 3.

**Table 3.** Comparison of PEC OER activities for some state-of-the-art 2D-HJ related materials.

| Catalyst | electrolyte | Light source | Current density | Efficiency | Ref. |
|---|---|---|---|---|---|
| **$MoS_2$/$WO_3$** | | AM 1.5 G illumination | 1.12 mA/cm$^2$ at 0.8 V vs. Hg2/Hg2Cl2 | photoconversion efficiency at 0.8 V: 0.52% | 107 |
| **$Sn_xMo_{1-x}S_2/MoS_2$** | 0.5M Na2SO4 | 300 W Xe lamp (100 mW cm−2) | > 0.8 mA/cm$^2$ at 1.23 V vs RHE | / | |
| **BiOI/g-$C_3N_4$** | 1.0 M KOH | 300 W Xe lamp | 81.5 μA/cm$^2$ at −1 V vs. Ag/AgCl | / | 55 |
| **g-$C_3N_4$/$TiO_2$** | 1 M KOH | 100 mW cm-2 (AM 1.5 G) | 72.3 μA/cm$^2$ at 1.23 V vs RHE | applied bias photon-to-current efficiency ~0.028% | 39 |

PH-dependent MkM was used to evaluate the relative performance of 12 different transition metal catalysts embedded in fourfold N-substituted double carbon vacancies in graphene for OER.[220] It is revealed that the used methodology to calculate the reaction pathways led to the results with enhanced catalytic activity when compared to purely thermodynamics-based predictions, which underestimated an onset potentials. The reason is that the catalytic reaction pathways performed on graphene-based catalysts are naturally also influenced by features such as metal site coverage, etc.

Ultra-thin FeCo-LDH@Co$(OH)_2$ HJ is found to have many surface exposed active sites, which resulted in enhanced catalytic activity for the OER, and the spin-polarized DFT calculations were helpful to reveal the reaction rate determining step (RDS).[218] The absorption Gibbs free energies of OH*, O*, and OOH* were calculated on different Co active site. The RDS of both FeCo-LDH and the FeCo-LDH@Co$(OH)_2$ HJ (different Co sites) was found to be the third step of the process in which O* generates OOH* (Figure 13c). These calculations also indicated that the interfacial cooperation between the electrons from FeCo-LDH and Co$(OH)_2$ effectively regulated the adsorption strength of intermediates on the HJ and reduced the barrier of RDS in the OER process. Interestingly, the real active site of was described to be the Co site.

### 5.2 Organic detoxification

Organic pollutants can be harmful to the environment, particularly if they enter waterways and affect aquatic life. PEC organic detoxification is a useful process for removing organic pollutants from wastewater and other industrial effluents. By applying light and electricity as the reaction power source, PEC can be a cost-effective method compared to the traditional methods like chemical oxidation or biological treatment. It can also be combined with other treatment methods to further increase the efficiency of the overall treatment process. As a versatile method for wastewater treatment, PEC organic degradation can be used to treat a wide range of organics, including pharmaceuticals, dyes and pesticides. As shown in Figure 14a, both the photogenerated electrons and holes can be used indirectly or directly for organic degradation, since the organic pollutants are mainly decomposed into non-toxic species by oxidation reaction. The excited electrons can react with $O_2$ and form $•O_2^-$, which is a powerful oxidant to degrade organic molecules. Meanwhile the holes can be directly used to oxidize $OH^-$ and $H_2O$ into •OH, which can be used for the removal of pollutants. These free radicals can oxidize most of the organic pollutants regardless of their nature. Usually, organic pollutants can be degraded totally into inorganic non-toxic molecules or changed into other non-toxic or value-added commodity organic molecules, depending on the types of the active species and specific reactions.

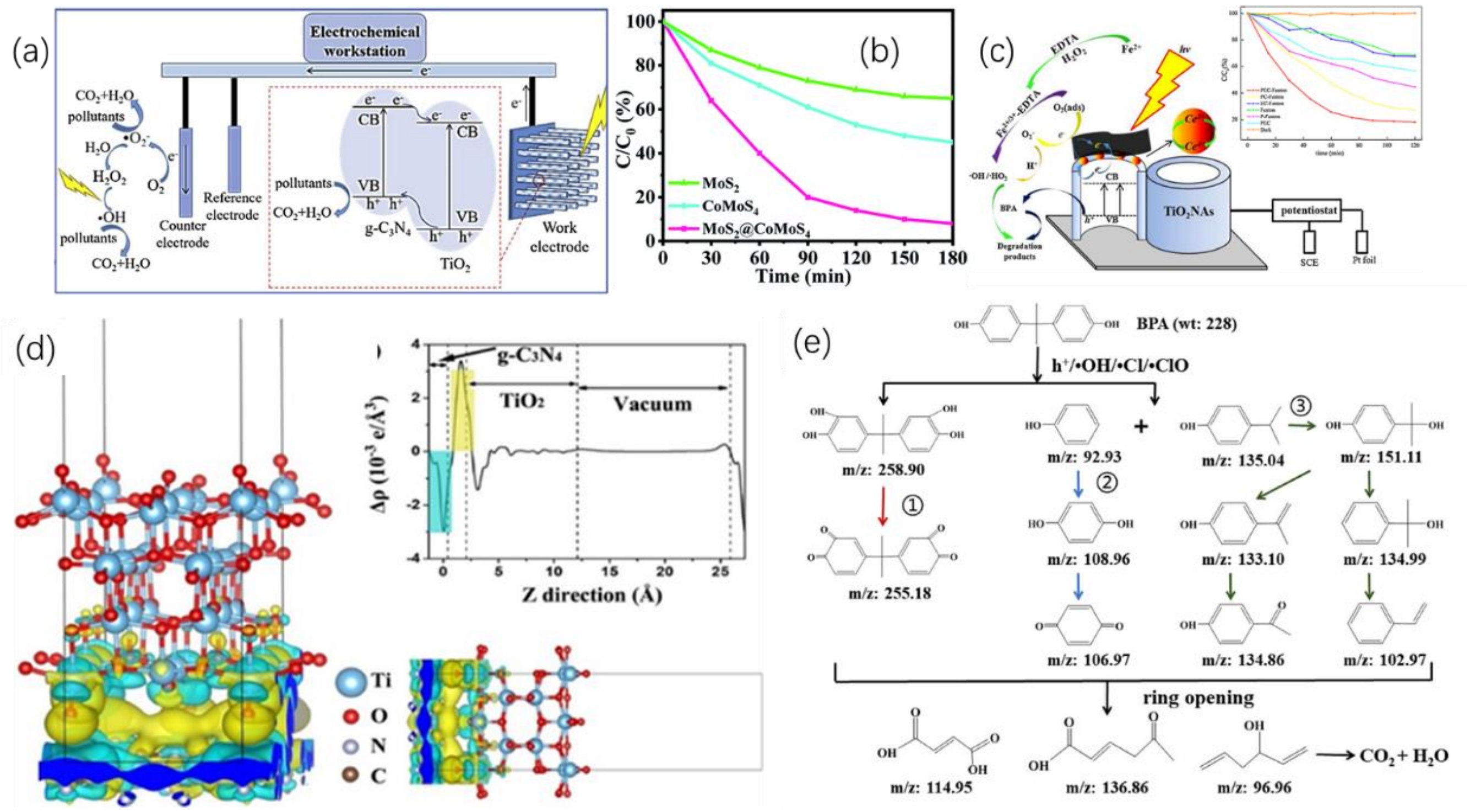

**Figure 14**. (a) Schematic of photoelectrocatalytic degradation of organic pollutants.[40] © 2019 Elsevier B.V. (b) the PEC activities of $MoS_2$, $CoMoS_4$ and $MoS_2$/$CoMoS_4$.[74] © The Royal Society of Chemistry 2020. (c) Schematic diagrams for PEC degradation towards Bisphenol A (BPA) using rGO/$CeO_2$/$TiO_2$.[49] © 2016 Elsevier Ltd. (d) The side view of the charge density difference for 2D $TiO_2$/g-

$C_3N_4$; Planar-averaged electron density difference $\Delta\rho(z)$ for 2D $TiO_2$/g-$C_3N_4$. The cyan and yellow areas indicate electron depletion and accumulation, respectively.[41] © 2019 Elsevier Inc. (e) Proposed pathway of BPA photoelectrocatalytic degradation over $MoS_2$/$BiVO_4$ photoanode under visible light illumination.[108] © 2020 Elsevier Ltd.

Tetracycline (TC), lomefloxacin (LOM), and p-nitrophenol (PNP) are common antibacterial agents or precursors for medicines and can influence the microorganisms once released to nature. Therefore, there is plenty of work focused on the detoxification of these organics by PEC. In most cases, the photocatalytic oxidation is incomplete, and the final products are still organic molecules presenting lower toxicity. For instance, TC can be photoelectrocatalytic oxidized by rGO/AgCl[48] and BiOI/$BiPO_4$,[45] though the final product is not $CO_2$ and $H_2O$. The ratio of the HJ components has an optimized value, meaning that either materials can inhibits the PEC degradation efficiency.[45] The comparison of performance of these catalysts are listed in Table 4. Considering the complexity of the final products from PEC degradation of pharmaceuticals, it is more reasonable to evaluate the toxicity of all the products than quantify all the products after PEC.

**Table 4.** Comparison of photoelectrocatalytic degradation of organics for some state-of-the-art 2D-HJ related materials.

| Catalyst | Pollutants | Conditions | Degradation efficiency | Ref. |
|---|---|---|---|---|
| **g-$C_3N_4$/ $Ag_3PO_4$** | RhB | [RhB] 10 ppm<br>[NaCl] 0.01 M<br>J =10 mA cm− 2 | 99% in 30 min | 54 |
| **GO/$Ag_3PO_4$/Ni** | RhB | [RhB] =5 ppm<br>0.1 M $Na_2SO_4$<br>Xenon lamp<br>E = 0.6 V | 97 % in 18 min | 14 |
| **rGO/AgCl** | TC | [TC] = 20 mg/L<br>0.5 M $Na_2SO_4$<br>E=1.5 V (vs Ag/AgCl).<br>Visible light | 85.2% in 120 min | 48 |
| **$TiO_2$/g-$C_3N_4$** | TC | 10 mg/L<br>0.1 M $Na_2SO_4$<br>Xenon lamp<br>1 V vs. Ag/AgCl | 100% in 120 min | 40 |
| **$Co_3O_4$/g-$C_3N_4$** | Reactive brilliant blue KN-R | C = 60 mg/L<br>0.1 mol/L Na2SO4<br>Xenon lamp<br>J =0.01 µA cm−2 | 91.4% in 120 min | 13 |
| **CQD/g-$C_3N_4$** | Methylene blue (MB) | 5 ppm<br>0.1 M $Na_2SO_4$<br>xenon lamp<br>E=1.5 V | 97.21% in 3 h | 58 |
| **$Ti_3C_2T_x$/$Bi_{12}TiO_{20}$** | MB | 10 mg/L<br>0.5 M $Na_2SO_4$<br>xenon lamp<br>E=1V | 85.4% in 120 min | 65 |
| **BiOBr/$TiO_2$** | RhB | xenon lamp<br>10 mg/L Rh.B<br>mol/L NaCl<br>E = 0.3 V | 60% in 100 min | 69 |
| **$Al_2O_3$/$SnS_2$** | MB | 10 mg/l<br>xenon lamp<br>0.5 M $Na_2SO_4$<br>E= 0.49 V (vs. Ag/AgCl) | 85.9% in 3 h | 76 |
| **Pt/g-$C_3N_4$/CdS** | MB | 5 mg L−1)<br>E = 0.6 V<br>Xe arc lamp | 60% in 2 h | 16 |
| **C/g-$C_3N_4$** | MB | $10^{-6}$ g/mL<br>0.1 M KOH<br>white light LED<br>E = 1.5 V vs RHE | 90% in 25 min | 59 |
| **CQD/g-$C_3N_4$** | phenol | 5 ppm<br>150 mg/mL $Na_2SO_4$<br>xenon lamp<br>E=1.5 V | 51.6% in 3 h | 58 |
| **BiOI/$BiPO_4$** | TC | 0.1 mol/L $Na_2SO_4$<br>10 ppm<br>Xenon lamp<br>E=1.2 V | 77% in 250 min | 45 |
| **$MoS_2$/$BiVO_4$** | BPA | 0.1 M of NaCl<br>10 ppm<br>Xenon lamp<br>E=1.5 V | 100% in 75 min | 108 |
| **CQDs/g-$C_3N_4$** | MB | 0.1 M $Na_2SO_4$<br>MB: 5 ppm<br>Xe lamp<br>E=1 V | 100% in 180 min | 58 |
| **Au-$WS_2$** | phenol | 50 mg L− 1 phenol<br>1 M $Na_2SO_4$<br>Xenon lamp<br>Cyclic voltammetry (CV): 0-1.0 V | 100% in 70 min | 63 |
| **Ni /NiFe-LDH/$Co_3O_4$** | BPA | 10 mg/L<br>0.1 M $Na_2SO_4$<br>Xenon lamp<br>E = 0.7 V | 100% in 120 min | 32 |
| **g-$C_3N_4$/$TiO_2$** | TC | C0 = 10 mg/L,<br>0.1 M $Na_2SO_4$<br>Xe lamp<br>E=1.0 V vs. Ag/AgCl | 100% in 120 min | 40 |
| **g-$C_3N_4$/α-$Fe_2O_3$** | PNP | 10 mg/L<br>0.2 M Na2SO4<br>Xe lamp<br>E = 1.5 V | 83.5% in 150 min | 61 |
| **RGO-$CeO_2$-$TiO_2$** | BPA | 10 mg L-1<br>0.05 M Na2SO4<br>Xe lamp | 80% in 120 min | 49 |

Zhang et al. use the $MoS_2$/$CoMoS_4$ HJ to perform LOM degradation.[74] The electrons accumulated on the CB of MoS react with $O_2$ to form $•O_2^-$, which is a strong oxidant for LOM degradation. Simultaneously, the photogenerated holes are left on the VB of $CoMoS_4$ to enhance the PEC degradation efficiency of LOM. The reaction equations can be proposed as follows:

$$•O_2^- + LOM \rightarrow CO_2 + H_2O$$

$$h^+ + LOM \rightarrow CO_2 + H_2O + \text{harmless ions}$$

Eventually, 79% of LOM was degraded in 180 min with $MoS_2$/$MoMoS_4$ as photoanode (Figure 14b).

As to the PEC of PNP degradation, the photogenerated electrons can combine with toxic PNP and convert it to less toxic p-aminophenol (PAP), which is an organic intermediate to produce various medicines, dyes, antioxidants and oil additives.

Total degradation of the PNP into $N_2$, NO, $CO_2$ and $H_2O$ is performed by g-$C_3N_4$/α-$Fe_2O_3$ HJ along both oxidation and reduction pathways.[61] The presence of the HJ improves the CT efficiency, as proved by the photocurrent density and the resistance. At the optimum experiment conditions at PAP initial concentration of 10 mg/L, pH value of 6.1, electrolyte concentration of 0.2 mol/L, and bias voltage of 1.5 V, the optimized photoanode displayed the highest PNP removal rate, which results from the synergistic effect of photocatalysis and electrocatalysis.

As a group of typical organic pollutants, dyes are widely used in the textile, paper, and leather industries, and present in wastewater from these industries. They are often resistant to conventional treatment methods and can persist in the environment for a long time, leading to environmental pollution and health hazards. To eliminate them efficiently, methods including PEC have been developed. The mechanism of dye degradation is like the degradation of other organics. Theoretically, both the active electrons and holes can join the PEC of dye by reacting with water and oxygen to produce reactive oxygen species (ROS). Different kinds of dyes are used for PEC degradation by 2D HJs, such as RhB, reactive brilliant blue KN-R, MB, phenol et al. Table 4 reports the photoelectrocatalytic activities for some state-of-the-art catalysts.

Bisphenol A (BPA) is a common chemical compound primarily used in the manufacturing of various plastics. It has been linked to a variety of health problems, including reproductive and developmental disorders, diabetes, cardiovascular disease, and certain cancers. BPA pollution can occur through a variety of sources, including industrial wastewater discharges, leaching from plastic products and packaging, and landfill leachate. It can also be found in food and drinking water due to its use in the production of food packaging, can linings, and water bottles. Therefore, this compound has become one of the targets to be removed by PEC. $TiO_2$ is commonly used as catalyst for this reaction. However, its catalytic efficiency is low due to its poor conductivity, unexpected recombination of photoinduced charges, and broad bandgap (3.2 eV). Therefore, 2D materials/$TiO_2$ HJs are designed to reinforce the PEC efficiency. For example, $TiO_2$/g-$C_3N_4$ is prepared for the photodegradation of BPA.[41] Its composition reduces internal resistance and enhances the electrocatalytic conductivity compared to the pristine components. The $TiO_2$/g-$C_3N_4$/carbon fiber electrode exhibits a higher PEC activity, with reaction rates of 1.7, 2.5, and 3 times faster than that of g-$C_3N_4$, $TiO_2$, and commercial P25, respectively. DFT calculations with PBE functional suggested that the generated interface HJ of 2D $TiO_2$/g-$C_3N_4$ can provide quick charge separation and transfer via both Ti–N and C–O bridges, resulting in a prepared catalyst that can facilitate the effective separation and transportation of photoinduced electron-hole pairs.[41] The redistribution of charges mostly took place at the 2D $TiO_2$/g-$C_3N_4$ interface region, largely because of the weak vdW interaction between $TiO_2$ and g-$C_3N_4$ (Figure 14d). Additionally, the results indicated that electrons move from g-$C_3N_4$ to the $TiO_2$ slab at the 2D $TiO_2$-g-$C_3N_4$ interface.

rGO/$CeO_2$/$TiO_2$ HJ is used for the PEC degradation of BPA as well, while the efficiency rate is only 40%. Fenton oxidation is reported helpful in increasing the amount of ROS, and the following formulas explain how the Fenton oxidation generate ROS continuously:

$Fe^{2+} + H_2O_2 + H+ \rightarrow Fe^{3+} + \cdot OH + H_2O$

$Fe^{2+}\text{-EDTA} + H_2O_2 \rightarrow Fe^{3+}\text{-EDTA} + \cdot OH + OH\cdot$

$Fe^{3+}\text{-EDTA} + H_2O_2 \rightarrow Fe^{2+}\text{-EDTA} + HO_2\cdot + H^+$

As shown in figure 14c, with the combination of PEC and Fenton oxidation, the degradation rate reaches 81.8% in 120 min, meanwhile the kinetics is improved from 0.0045 $min^{-1}$ to 0.0146 $min^{-1}$, indicating a synergetic effect between Fenton and PEC process (Figure 14e).

This type of electrolyte can influence the photoelectrocatalytic degradation of BPA as well. By using 100 mM of NaCl as the supporting electrolyte, 10 ppm of BPA could be completely degraded in 75 min by $MoS_2$/$BiVO_4$ photoanode, with a bias of 1.5 V vs. Ag/AgCl. This is because chloride anions can be activated by photogenerated holes to form chlorine oxide radical (•ClO), which plays a dominant role in the degradation of BPA. NiFe-LDH/$Co_3O_4$ HJ is reported to exhibit 100% removal rate of BPA of 10 mg/L in 120 min under visible light illumination.[32]

### 4.3 $CO_2$ reduction reaction

Valorization of $CO_2$ by converting it into value add-products and fuels is one of the leading scientific challenges to face the effects of global warming. PEC is one of the most promising techniques, along with photocatalysis, electrocatalysis, or thermocatalysis, capable of transforming carbon dioxide into useful products.[221] PEC for $CO_2$ reduction is usually performed in organic media due to the good solubility of $CO_2$. However, in order to have an easier scale-up application and lower cost, aqueous media is still the first choice for PEC of $CO_2$RR.[3] However, in water it is important to take into account the HER competitive process. Half reduction reactions in Table 5 below show the main reduction of products that can be obtained from $CO_2$RR.[3]

**Table 5.** Electrochemical reduction potentials evaluated in an aqueous solution at pH = 7 vs. standard hydrogen electrode (SHE).

| Reaction | $E_0$ vs SHE (V) |
|---|---|
| $CO_2 + e^- \rightarrow CO_2\bullet^-$ | –1.850 |
| $CO_2 + 2H^+ + 2e^- \rightarrow CO + H_2O$ | –0.665 |
| $CO_2 + 2H^+ + 2e^- \rightarrow HCO_2H$ | –0.521 |
| $CO_2 + 4H^+ + 4e^- \rightarrow HCHO + H_2O$ | –0.485 |
| $CO_2 + 6H^+ + 6e^- \rightarrow CH_3OH + H_2O$ | –0.399 |
| $CO_2 + 8H^+ + 8e^- \rightarrow CH_4 + 2H_2O$ | –0.246 |
| $2CO_2 + 12H^+ + 12e^- \rightarrow C_2H_4 + 4H_2O$ | -0.349 |
| $2CO_2 + 12H^+ + 12e^- \rightarrow C_2H_5OH + 3H_2O$ | -0.329 |
| $2CO_2 + 14H^+ + 14e^- \rightarrow C_2H_6 + 4H_2O$ | -0.270 |
| $3CO_2 + 18H^+ + 18e^- \rightarrow C_3H_7OH + 5H_2O$ | -0.310 |
| $2H^+ + 2e^- \rightarrow H_2$ | –0.414 |

Usually, CO and formate are the main products since they require 2 electrons in the process. However, since thermodynamically the reduction potentials have similar values, product selectivity has to be controlled as best as possible by tuning the system properly.[222] Moreover, $CO_2$ adsorption on the material surface is a key step in the general $CO_2$ conversion efficiency. In this regard, thanks to the numerous surface active sites and the tunable electronic properties, 2D HJs have been studied with increasing effort.[223] Currently, the production of C2 compounds such as alcohols and ethylene has become an important direction in the field of $CO_2$RR, but the study of 2D HJs photoelectocatalysts is still poor both theoretically and experimentally.

One of the most promising 2D materials for $CO_2$RR is g-$C_3N_4$, Zhu et al. demonstrated how $CO_2$ molecules have better adsorption on 2D monolayer compared to the bulk g-$C_3N_4$. Thanks to DFT calculation it was possible also to show how the adsorption is favored on two coordinated N atoms.[224] In order to increase the catalytic performance of g-$C_3N_4$, Wang et al. inserted S atoms into the g-$C_3N_4$ structure, showing a reduction of the bandgap and an increased photogeneration of charge carriers thanks to the presence of the impurity states.[225] Sagara et al. combined B atoms with g-C3N4, showing a 5 times increase in the photocurrent compared to the undoped of g-$C_3N_4$ (Figure 15a).[91] In the same work, the addition of a co-catalyst on the surface of the semiconductor in the form of gold nanoparticles, further increased the selectivity toward a specific product (ethanol) and the catalytic activity (Figure 15b). More HJs for PEC of $CO_2$RR have been reported as a theoretical model. For instance, BiOCl/g-$C_3N_4$ is constructed for DFT calculations of Gibbs free energy (Figure 15c).[96] The comparison of the calculated Gibbs free energy for the $CO_2$RR to CO on the surface of ns-CN, p-BiOCl/ns-CN, OVs-BiOCl/ns-CN, and OVs-BiOCl/ns-CN reveal that for all systems, $*CO_2 + H^+ + e^- \rightarrow *COOH$ is a RDS. Importantly, it was described that O vacancies and N vacancies in the BiOCl and ns-CN, respectively, in the HJ led to the lowest energy barrier. Although the formation of the heterojunction from BiOCl nanosheets with 2D g-$C_3N_4$ with vacancies seems to be promising for PEC, the adsorbed CO molecules have difficulty forming the final products involved with more than two-electron processes as the free energy change of *CO is negative, hence *CO is desorbed from the catalyst surface as the final product.

Another class of 2D materials studied for $CO_2$RR is the TMDCs. Hong et al. worked on a simple system using $MoS_2$ and $MoSe_2$ thin film on silica substrates showing an increase of current density shining light on the photoelectrodes, where $MoSe_2$ has the best performance (Figure 15d).[226] Hu et al. investigated on $MoS_2$ nanosheets loaded on $SnO_2$ nanoparticles as a photocathode.[227] It was discovered that a 5% loading of MoS2 is the best for increasing the catalytic activity in terms of faradic efficiency and current density, with respect to the bare $SnO_2$.

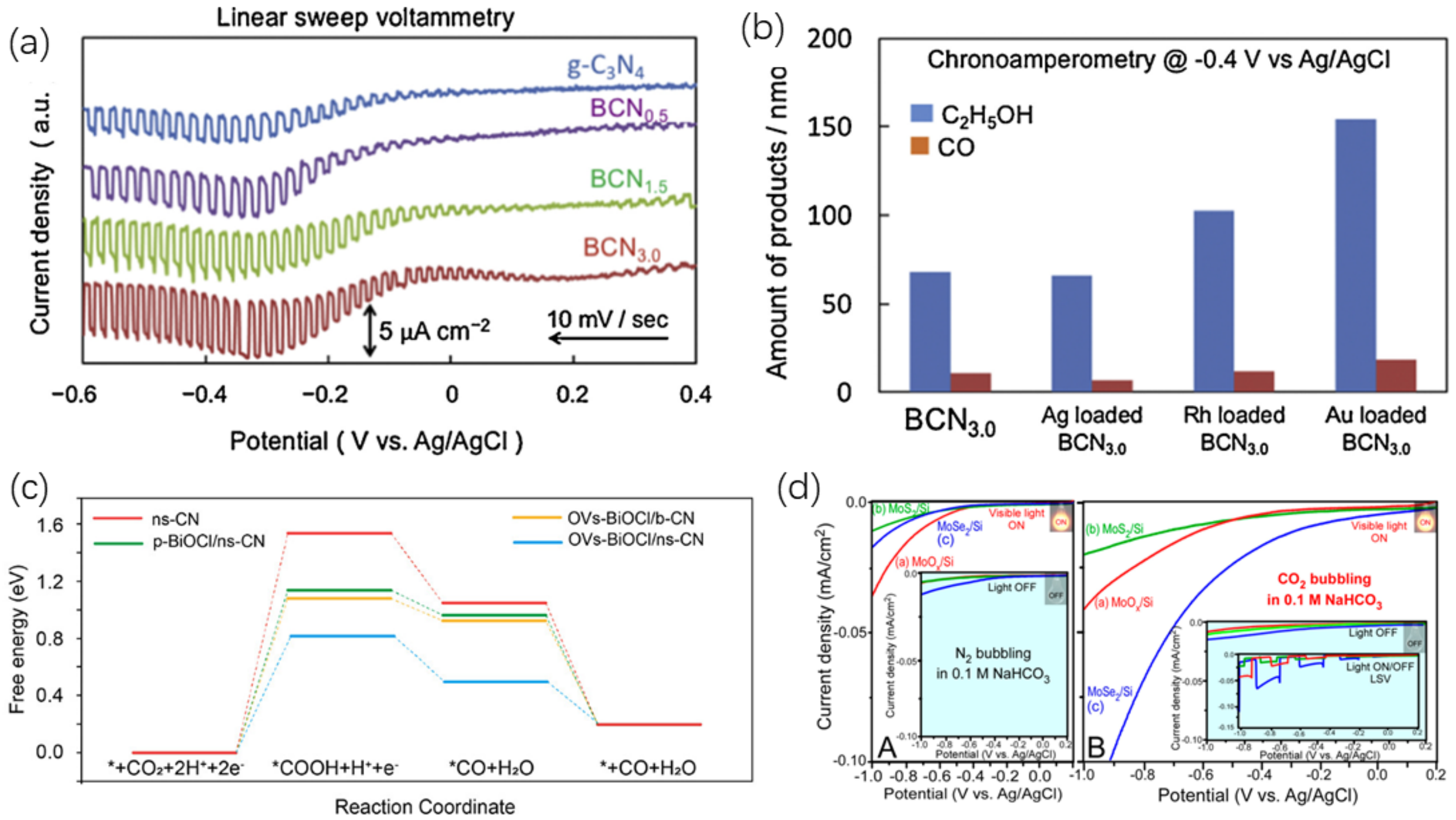

**Figure 15.** (a) Linear sweep voltammetry of the g-$C_3N_4$ and B/g-$C_3N_4$ electrodes. (b) Products analyses of PEC of CO2RR over co-catalyst loaded B/g-$C_3N_4$ electrodes.[91] © 2016 Elsevier B.V. (c) Gibbs free energy calculations.[96] © 2023 American Chemical Society. (d) Linear sweep voltammetry curves (voltage range: +0.2~1.0 V) in N2- (A) and CO2-purged (B) 0.1 M $NaHCO_3$ electrolyte at a scan rate of 10 mV/sec under dark (in the corresponding inset Figure) and the visible light exposure condition for $MoO_x$/Si, $MoS_2$/Si and $MoSe_2$/Si samples. The inset (B) shows the light ON-and-OFF linear sweep voltammetry curves.[210] © 2019 by the authors. Licensee MDPI, Basel, Switzerland.

### 5.4 $N_2$ reduction reaction (NRR)

Ammonia is one of the most important chemicals, and its production, exclusively carried out through the Haber-Bosch process, accounts for about 1.3% of total $CO_2$ global emissions.[228] Since the production requires high temperatures (300–500 °C) and pressures (150–300 atm), processes that work under mild conditions are highly desirable.

The overall reaction (below) has an equilibrium potential E° equal to +0.55 V vs NHE.[229]

$$N_2 + 6H^+ + 6e^- \rightleftharpoons 2NH_3(g)$$

It is well known how the $N_2$ double bond activation is particularly challenging due to the high cleavage energy (945 kJ $mol^{-1}$) and the addition of the first H atom to form $N_2H^+$ has a quite negative potential (−3.2 V vs RHE).[230] Moreover, as well as for NRR, HER is still a competitive process that makes the selectivity toward ammonia even more difficult. PEC of NRR is still a new field since most of the research so far has been more active on the photochemical and electrochemical processes for ammonia production.[231]

Simple DFT calculations were performed to study the electronic structures of $MoS_2$ and map out the energy profile of NRR on $MoS_2$ (Figure 16a and 16b).[232] Ideal scenario for NRR catalysis requires a catalyst with a strong activation (small activation energy, Ea) to $N_2$, but a relatively weak binding (small adsorption energy, $\Delta E$) for the intermediate species. The calculations revealed that the basal plane was found to be inert and $N_2$ molecules even cannot be effectively adsorbed. Moreover, these calculations suggested that the positively charged Mo-edge plays the key role to polarize and activate the $N_2$ molecules. The computed free-energy profile also demonstrated that the RDS is the reductive protonation of adsorbed $N_2$, with a barrier of 0.68 eV without external potential.

As for NRR, 2D HJs can improve the catalytic activity by reducing the charge recombination. Mushtaq et al. reported a $MoSe_2$/g-$C_3N_4$ HJ used in NRR with 28.9% of faradic efficiency at –0.3 V vs RHE (Figure 16c).[79] By optimizing the quantity of g-$C_3N_4$ (7%) inside the HJ, the photocurrent obtained is higher than the individual semiconductor. In another work by Ye et al., a $MoS_2$/$TiO_2$ HJ optimized with a molar ratio to 1:2 was capable of reaching 65.5 % of faradic efficiency at –0.2 V vs RHE.[233]

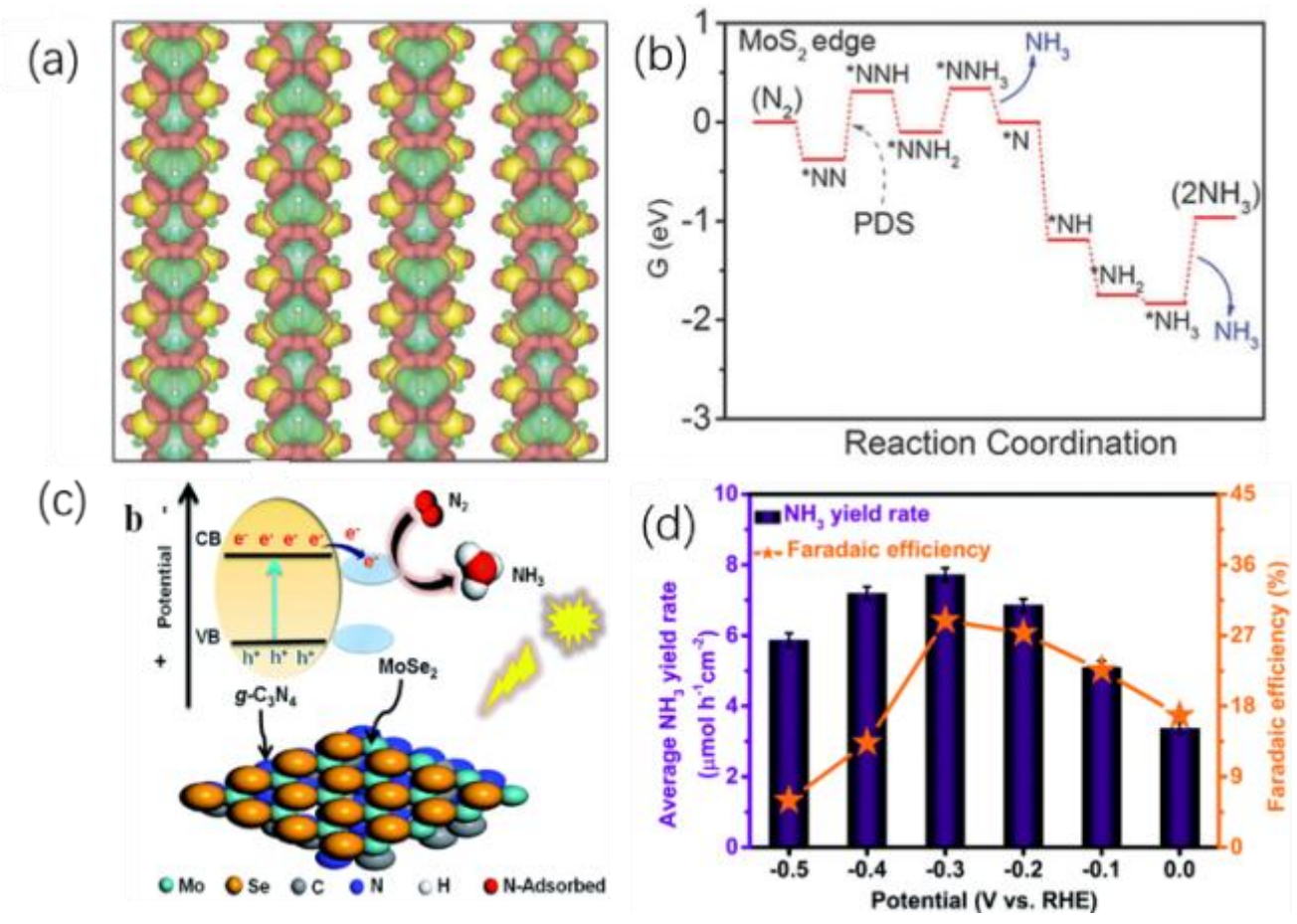

**Figure 16.** (a) The isosurface of deformation charge density from the top view. Red and green represent charge accumulation and loss, respectively. Isosurface is 0.0025 a.u. (b) Free-energy profile for NRR at MoS2 edge site. An asterisk (*) denotes as the adsorption site.[232] © 2018 WILEY-VCH Verlag GmbH & Co. KGaA, Weinheim (c) PEC illustration of $MoSe_2$@g-$C_3N_4$ HJ. (d) Equivalent $NH_3$ yield rates and FE at the potential from 0 to −0.5 V *vs.* RHE of $MoSe_2$@g-$C_3N_4$ heterojunctions. Reproduced with permission from ref [79]. © 2020 Royal Society of Chemistry.

DFT calculations further predicted that 2D MXene compounds, specifically $Nb_3C_2$, can function as efficient NRR catalysts.[234] The mechanism for the electrochemical conversion of $N_2$ into $NH_3$ catalysed by $d^2$ (Ti, Zr, and Hf), $d^3$ (V, Nb, and Ta), and $d^4$ (Cr and Mo) 2D transition metal carbides, or MXenes, with formulae $M_{n+1}C_n$ (n = 2, $M_3C_2$ unit-cell), has been studied by means of DFT with PBE functional. The calculated results suggested that $V_3C_2$ and $Nb_3C_2$ exhibited a strong $N_2$ fixation ability and had a stronger adsorption energy of $N_2$ than that of CO and $H_2O$ molecules. Once $N_2$ was spontaneously captured by $V_3C_2$ and $Nb_3C_2$, limiting barriers of just 0.64 and 0.90 V, respectively, were demanded for its electrochemical conversion into $NH_3$.

### 5.5 Alcohol oxidation

Alcohol oxidation is the anodic reaction in fuel cells which converts chemical energy into electrical energy. Although Pt is currently used as the most effective and stable catalyst for it, the attempt to reduce the amount of Pt by spreading Pt nanoparticles on cheaper and stable substrates with optoelectronic properties is a trend in this field. Till now, there are several 2D catalysts applied as substrates in this direction. Perovskites are a big group of semiconductors used in all the optoelectronic field, and the decoration of Pt on 2D perovskite $La_2Ti_2O_7$ was performed by Hu et al.[73] This system is able to make use of the light energy efficiently, with the catalytic methanol oxidation activities 10.6 times higher compared to activities in the dark condition.

g-$C_3N_4$ can be used as promising photoactivated support for Pt towards the alcohol oxidation reaction as well.[15] To perform the PEC of methanol oxidation in a 1.0 M $CH_3OH$ + 1.0 M KOH solution, it is observed that the forward peak current intensity of Pt/g-$C_3N_4$ nanosheets can reach 520.4 mA $mg^{-1}$, which is higher than that of pure Pt nanoparticles modified electrode (104.8 mA $mg^{-1}$) (Figure 17a). The introduced ultrathin g-$C_3N_4$ as a support will not only prevent the aggregation of Pt during the synthesis process, resulting in small size Pt nanoparticles, but also in favor of adsorption of target molecules owing to 2D structures. On the other hand, when the Pt/g-$C_3N_4$ electrode was upon visible light (>400 nm) irradiation, the g-$C_3N_4$ can be excited and generated electrons in the CB and holes in the VB. Therein, the holes have oxidative ability and can react with surface adsorbed $OH^-$/$H_2O$ to form strong oxidative hydroxyl radicals (•OHs). The adsorbed methanol molecules on the surface of catalysts can be also oxidized upon these •OHs, leading to a photoelectrocatalytic alcohol oxidation process (Figure 17b). Usually, these electron-hole pairs will quickly recombine and only a fraction of holes can be used. On the other hand, the photoexcited electrons transfer to Pt firstly in the Pt/g-$C_3N_4$ composite, and then flow to the circuit under an external electric field, thus preventing the charges recombination. With the combination of g-$C_3N_4$/CdS, the Pt nanoparticles shows 7.4 times enhanced methanol oxidation under visible light irradiation compared to the test in the dark environment.[16] Beside the catalytic activity of methanol oxidation, the stability of corresponding electrode is also significantly improved with assistance of visible light irradiation.

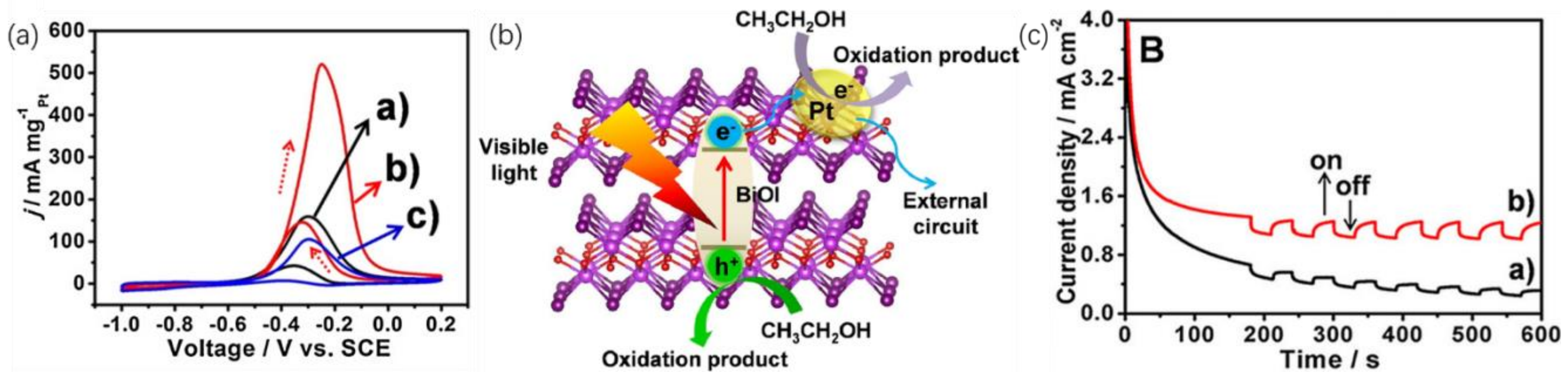

**Figure 17.** (a) CVs of the Pt/g-$C_3N_4$ under dark a) and visible light illumination b), and pure Pt nanoparticles c) in 1.0 M $CH_3OH$ + 1.0 M KOH solution at a scan rate of 50 mV $s^{-1}$. © 2016 Elsevier B.V. (b) Schematic illustration for oxidation ethanol process on Pt/BiOI electrode under visible light illumination. © 2017 Elsevier B.V. (c) Chronoamperometric curves for a) Pt/CdS and b) Pt/CdS/rGO electrodes with and without visible-light illumination every 30 s at –0.3 V. © 2017 Wiley-VCH Verlag GmbH & Co. KGaA, Weinheim.

Additionally, there are more than two components in the catalyst system, in which the 2D materials with metallic properties are helpful for charge separation and transfer. For example, Pt/CdS/rGO nanocomposites were reported to have an enhanced catalytic property in methanol oxidation, mainly because the synergetic effect of each composite[51]. Specifically, Pt provide the reactive site, CdS oversees light absorption and conversion, rGO mainly favors CT and the stability of the catalyst, as the photoelectric current density remains at around 1.2 mA/$cm^2$ at –0.3 V, while the Pt/CdS has a smaller photocurrent and continuously decreasing current density during 600 s of test (Figure 17c).

BiOI has been used as the support for Pt and charge separator, as the pure BiOI does not show any catalytic activity on ethanol oxidation.[34] For Pt/BiOI, the CV curves display two typical ethanol electro-oxidation features in the range of −0.8 ~ 0.2 V with forward (ca. −0.20 V) and backward (ca. −0.30 V) peaks in alkaline media. When the Pt/BiOI electrode was upon visible light irradiation, the current density was 874.4 mA $mg^{-1}$ Pt, which is about 3 times higher than the current in dark. This proved that the photogenerated charges from BiOI can be converted to Pt and used for PEC of ethanol oxidation eventually.

Deep into the mechanism of MOR process, GCP-K DFT calculations were able to find the difference in methylation using chloromethane on 1T $MoS_2$ and 1T $WS_2$ surfaces.[235] Thermodynamics and kinetics calculations were performed for the first methyl addition on the experimentally observed coverages. These calculations at constant potential elucidated that (i) the methylation reaction became more favorable as the potential decreased and (ii) the thermodynamics and kinetics of 1T $MoS_2$ methylation were more favorable than the equivalent reaction on 1T $WS_2$ at constant potential. The reason behind this was described as differences in the charge per $WS_2$/$MoS_2$ unit at constant potential. Methodologically interesting observation was that differences in reactivity decreased, and even reversed, when these reactions were calculated at constant charge instead of constant potential.

### 5.6 $H_2O_2$ production

The oxygen reduction reaction is a crucial PEC process that occurs at the cathode. This reaction involves the reduction of $O_2$ to $H_2O$ or other oxygen-containing species like $H_2O_2$. The PEC of $H_2O_2$ production reaction can be described as follows:

2e$^-$ one-step: $O_2 + 2H^+ + 2e^- \rightarrow H_2O_2$

1e$^-$ two-step: (i) $e^- + O_2 \rightarrow \bullet O_2^-$; (ii) $\bullet O2^- + e^- + 2H^+ \rightarrow H_2O_2$

Under PEC environment, the active electrons and holes can also lead to $H_2O_2$ decomposition:

(i) $H_2O_2 + H^+ + e^- \rightarrow H_2O + OH^*$, $H_2O_2 + 2H^+ + 2e^- \rightarrow H_2O$;

(ii) $H_2O_2 + h^+ \rightarrow H^+ + HO_2^*$; $H_2O_2 + 2h^+ \rightarrow 2H^+ + O_2$,

where $H^+/e^-$ is the main cause of $H_2O_2$ decomposition at lower pH and light-induced holes are the main cause of $H_2O_2$ decomposition at higher $H_2O_2$ concentration.[74] Therefore, the amount of $H_2O_2$ produced gradually increased with time, and the formation rate gradually slowed down and the yield of $H_2O_2$ tended to stabilize. $MoS_2$/$CoMoS_4$ is a perfect example of this situation. During the PEC $H_2O_2$ production, the amount of $H_2O_2$ increases fast from 0 to ~180 μM in the first 60 min under irradiation, while reaches about 205 μM after another 60 min. Therefore, the separation of the product from the reaction cell could be designed to improve the reaction, such as the application of H-cell as the reaction cell.

The puzzling question why many catalysts exhibit high selectivity for $H_2O_2$ during ORR despite the strong thermodynamic preference for the O−OH breaking leading to the formation of water was studied by first-principles calculations of the electrochemical reaction kinetics at the solid−water interface of single Co-graphene or single vacancy graphene.[186] These authors considered water movement and net electronic charges, using the CP-HS-DM method. The obtained reaction mechanism demonstrated that using these catalysts, breaking the O−OH bond has a higher energy barrier than breaking *−O, due to the rigidity of the O−OH bond. Moreover, the potential and pH affected the selectivity towards formation of $H_2O$ or $H_2O_2$. For single Co-graphene decreasing potential promoted proton adsorption to the O absorbed onto Co, thus increasing the $H_2O_2$ selectivity. On the contrary, for single vacancy graphene, the proton preferred to adsorb onto the latter O, resulting in a lower $H_2O_2$ selectivity in acid condition.

### 5.7 Metal cations detoxification

Photoelectrocatalysts can be used to detoxify poisonous metal ions by changing their valence state. For example, the oxidation of As(III) to As(V) is performed by the Au/$WS_2$ photoanode with an efficiency of 95% under visible light illumination[63]. Removal of uranium from uranium-containing wastewater is of great significance for uranium contamination remediation and development of nuclear energy. To solve this problem, Dai et al designed a photoelectrocatalytic method to reduce the U(VI) cations to U(IV), which can be deposited on the cathode of the cell. By using the g-$C_3N_4$/$Sn_3O_4$/Ni photoanode as a S-scheme catalyst to produce and transfer active electrons to the Pt cathode, the deposition rate of U(IV) on the Pt cathode reached 94.28 % at pH of 5.0 for photoelectrocatalytic approach, while the removal rates of U(VI) only amounted to 10.56 % and 36.65 % for the photocatalytic and electrochemical extraction approaches, respectively. The reduction of Cr(VI) to Cr(III) is reported as well.[32] By the combination of the active electrons generated from NiFe-LDH/$Co_3O_4$ with $Cr_2O_7^{2-}$ ions in the waste water, the toxic $Cr_2O_7^{2-}$ ions are able to be reduced to $Cr^{3+}$.

## 6. Perspective and conclusion

In summary, we have thoroughly reviewed the various 2D layered materials which are used in PEC. This exercise is motivated by the infant research on 2D materials as one of the most promising materials for PEC. Currently, many popular 2D materials (such as graphene and its derivatives, and mostly MXenes) suffer from their metallic property, which means they are not able to behave as an active charge generator. Therefore, they always behave as a co-catalyst in PEC and are mainly responsible for charge separation and transfer. Moreover, semiconductive 2D materials (such as TMDs, g-$C_3N_4$, and LDHs) can suffer from their unfavorable band positions, poor charge carrier mobilities and poor stabilities due to the photocorrosion or electrochemical environment. In fact, several strategies have been applied to solve these problems, including HJ, doping, defects engineering and tuning of morphology. From the literature, the efforts in developing the photoelectrocatalysts with these strategies are prominent. In principle, all these strategies aim at tuning the bandgap of the catalytic system to make it suitable for the specific reactions. The formation of HJs, impurity state, defect state and band alignment due to the 2D material size and thickness are proved to work properly in many published works. Meanwhile, some of the strategies described may help to increase the surface-active sites, which synergize the band tuning effort and facilitate the PEC efficiency. Among all these strategies for improvement, we reviewed different kind of structure alignment in the HJs (such as type I, type II, Z-scheme, and S-scheme), in which the 2D materials can behave either as charge generator, charge separator or CT pathways, depending on the natural properties of the 2D materials. In a type I HJs, the component with broader bandgap usually absorbs light and produces active charge carriers, while the component with narrower bandgap can receive the photogenerated charges and provide active sites for the photoelectrocatalytic reactions. Type II is more frequently compared to other types of HJs due to its simplicity and efficiency in charge separation and transport. Z-scheme and S-scheme are less established systems, but show notable architecture developments in CT channels. Thus, in our perspectives, deeper exploration in Z-scheme and S-scheme HJs is to be designed and pursued in the future.

Despite the design of the 2D catalysts, the synthesis methodology is the first step to realize these designs. It is clear that the development of synthesis methods is still at the laboratory level and some steps away from practical applications. The synthesis top-down and bottom-up methods were reviewed in Section 3. In general, the "top-down" methods are more scalable compared to the "bottom-up" ones, as they can leverage existing manufacturing techniques and equipment. Moreover, top-down methods can utilize existing materials, which may already possess desirable properties or characteristics. Specifically, mechanical force is the main source to destroy the weak interaction between the layers during the synthesis of 2D materials, thanks to the special 2D structure of the materials. Eventually, the synthesized products can keep the original phase with dimension decreased to 2D. In contrast, certain materials may not be suitable for top-down processing due to their inherent properties, making it difficult to achieve desired 2D structures with specific properties. Bottom-up methods allow for precise control at the nanoscale or atomic level, enabling the creation of complex structures with high precision and functionality. Different from top-down methods, bottom-up approaches can produce materials with unique properties not found in bulk materials. By controlling the assembly or synthesis process, it is possible to achieve tailored material properties. Moreover, HJs with complicated design of shapes and compositions are available by bottom-up methods. Furthermore, bottom-up methods are not limited to the phase of the precursors, enabling the creation of new materials with tailored properties and functionalities, opening up possibilities for advancements in various fields. However, there are several disadvantages of bottom-up methods as well. First, scaling up the bottom-up approach from the laboratory to large-scale production can be challenging due to issues related to reproducibility, efficiency, and cost. Second, the assembly or synthesis processes in the bottom-up approach can be complex, requiring precise control of reaction conditions, self-assembly mechanisms, or chemical reactions. The probability of impurities is quite high compared to the top-down methods. As to the industrial synthesis, scalability, reproducibility, cost-effectiveness, and quality control will be always considered. Considering both top-down and bottom-up methods have their strengths and limitations, the representative methods of both of them are developed at the industrial level, resulting in different forms of products. Typically, different kinds of exfoliation have been widely used in producing large quantities of graphene, TMDCs and BP suspensions or powders, while CVD can be used in fabricating large area of films. Although many synthetic strategies have been developed for the synthesis of high quality, large surface area ultrathin 2D materials, the controlled synthesis of the desired layer with high charge carrier mobility of the materials still needs further research attention. Furthermore, practical research is needed for designing efficient hybrid materials with intimate contact and efficient interfacial coupling. By continuously optimizing the parameters of synthesis techniques, large-scale production of different high-quality 2D material HJs for various photoelectrocatalytic reactions will be developed properly. Similarly, even though the theoretical research has reached the maturity to provide great insight into the mechanism of the specific (photo)electrocatalytic reaction using 2D HJs, there is a great space for improvement. The most urgent methodological improvement consists of the simultaneous consideration of both light and applied voltage in the calculation of catalytical processes. This will be realized by considering the catalytic process in the excited state, a holy grail for PEC. Moreover, the current state-of-the-art focuses only on the static description of light interaction with 2D HJs, which can suggest the location of hole and electrons either before or after irradiation. However, the time evolution after the photoexcitation is not usually explicitly calculated and relies only on this static picture and dynamic events are neglected. On the other hand, the description of the EC has progressed significantly as the state-of-the-art calculations are capable of considering constant potential, as it is applied in the experiments, instead of the most common constant charge approach in the DFT calculations. Here, the challenge is to apply these methodologies, e.g., GCP-K, to large systems, and at the same time reduce its computational cost.

Although 2D materials have achieved very encouraging progress in a wide range of fields from PEC, some challenges remain, which demand an urgent solution. The various challenges and opportunities over 2D material HJs are elaborately discussed in the following section. i) A thorough experimental and theoretical review regarding the activity improvement of 2D material HJs has been presented. In the PEC of water splitting and organic detoxification, there are already plenty of 2D-HJs applied for the reactions. However, some fundamental aspects still need to be addressed to achieve the highest activity result in the target reaction. ii) 2D-HJs have received eye-catching progress in different photoelectrocatalytic reactions such as N2RR, CO2RR, ORR, reduction of inorganic pollutants and alcohol oxidation. These are merging as new reactions for PEC; thus, more attention is needed and more kinds of 2D materials should be applied for these reactions in order to find the more suitable catalysts for these reactions. iii) The selection of a photocatalytic system with long-term use and durability is very challenging. Currently, the overall solar energy conversion efficiency differs from 1% to 10%, and the photo energy in the whole PEC occupies around 10%. Hence, a higher photo conversion efficiency must be achieved for the process to be commercialized. 2D semiconductive materials such as g-$C_3N_4$ and TMDs provide a wide photon absorption range and generates multiple excitons, and thus can be an ideal candidate to obtain photocatalytic HJ for future use. But once the catalysts are used under photoelectrochemical environment, the durability is decreased compared to the ones in photocatalytic system. Thus, designing 2D materials which are resistant to the electrochemical environment is another urgent problem to solve.

All these above perspectives are focused on the 2D materials which have been explored to some extent. Besides, considering that there are still a large number of materials with 2D structures not studied deeply or not applied in the field of PEC, it is promising to expand the type of 2D materials for PEC. To achieve this goal, more synthesis methods should be developed. Besides, novel HJs containing the new 2D materials can be designed as well for better efficiency and stability. Finally, the scope of PEC over 2D-HJs can be further expanded in the aspect of new materials, new methods and new applications. Meanwhile, the mechanisms can be deeply studied to modify the efficiency and durability of the catalysts.

## AUTHOR INFORMATION

### Corresponding Authors

**Mengjiao Wang** - *Department of Applied Science and Technology, Politecnico di Torino, 10129 Torino, Italy. E-mail: mengjiao.wang@polito.it*

**Teresa Gatti** - *Department of Applied Science and Technology, Politecnico di Torino, 10129 Torino, Italy. E-mail: teresa.gatti@polito.it*

**Silvio Osella** - *Chemical and Biological Systems Simulation Lab, Centre of New Technologies, University of Warsaw, 02097 Warsaw, Poland. E-mail: s.osella@cent.uw.edu.pl*

### Authors

**Roberto Altieri** - *Institute of Physical Chemistry and Center for Materials Research (LaMa), Justus Liebig University, 35392 Giessen, Germany.*

**Matteo Crisci** - *Institute of Physical Chemistry and Center for Materials Research (LaMa), Justus Liebig University, 35392 Giessen, Germany.*

**Michal Langer** - *Chemical and Biological Systems Simulation Lab, Centre of New Technologies, University of Warsaw, 02097 Warsaw, Poland.*

### Author Contributions

The manuscript was written through contributions of all authors. / All authors have given approval to the final version of the manuscript.

### Funding Sources

European Research Council, European Commission, Deutsche Forschungsgemeinschaft, Fondazione Compagnia di San Paolo, National Science Centre, Poland.

### Notes

Any additional relevant notes should be placed here.

## ACKNOWLEDGMENT

T.G. would acknowledges the support of the European Research Council for the project JANUS BI (grant agreement no. [101041229]). M.W. and T.G. also thank Fondazione Compagnia di San Paolo for financial support through the "Bando TRAPEZIO - Paving the way to research excellence and talent attraction". R.A. would like to thank the DFG for project 460609161. M.C. and T.G. thank the European Commission for the project LIGHT CAP (grant agreement no. [101017821]). S.O. is grateful to the National Science Centre, Poland (grant no. UMO/2020/39/I/ST4/01446) and the "Excellence Initiative – Research University" (IDUB) Program, Action I.3.3 – "Establishment of the Institute for Advanced Studies (IAS)" for funding (grant no. UW/IDUB/2020/25).

## ABBREVIATIONS

2D, two-dimensional; PEC, (photo)electrocatalysis; CT, charge transfer; CB, conduction band; HRTEM, High resolution transmission electron microscopy; g-C3N4, graphitic carbon nitride; MCs, metal chalcogenides; HJ, heterojunction; vdWs, van der Waals; MPCh3, metal phosphorous trichalcogenides; h-BN, hexagonal boron nitride; MOFs, metal organic frameworks; LDHs, layered double hydroxides; rGO, reduced graphene oxide; GO, graphene oxide; TMDCs, transition metal dichalcogenides; ITO, indium tin cxide; VB, valence band; IEF, internal electric field; SmV, samarium vanadate; CO2RR, CO2 reduction reaction; MOR, Methanol oxidation reaction; RhB, rhodamine B; CQDs, carbon quantum dots; OER, oxygen evolution reaction; HER, hydrogen evolution reaction; CVD, chemical vapor deposition; Ni-$B_i$, nanosized pure nickel boron oxide; CN, carbon nitride; QM, quantum mechanics; MM, molecular mechanics; MKM, microkinetic-modeling; DFT, density functional theory; MD, molecular dynamics; AIMD, *ab-initio* MD; CMD, constrained molecular dynamics; DFT-CES, density functional theory in classical explicit solvent; QEq, charge equilibration; LDA, local density approximation; GGA, generalized gradient approximation; HSE, Heyd–Scuseria–Ernzerhof; MBPT, many-body perturbation theory; NAMD, nonadiabatic MD; TDDFT, time dependent density functional theory; GS, ground state; ES, excited state; BSE, Bethe–Salpeter equation; CHE, computational hydrogen electrode; PCET, proton-coupled electron transfer; SHE, standard hydrogen electrode; RHE, reversible hydrogen electrode; CP-HS-DM, constant potential-hybrid solvation-dynamic model; GCP-K, grand canonical potential kinetics; kMC, kinetic Monte Carlo; BEP, Brønsted–Evans–Polanyi; IPCE, incident photocurrent efficiency; SiNW, silicon nanowire; RuSA, ruthenium single atom; BP, black phosphorus; NGr, nitrogen-doped graphene; RDS, rate determining step; ROS, reactive oxygen species; BPA, Bisphenol A; TC, Tetracycline; LOM, Lomefloxacin; PNP, p-nitrophenol; MB, Methylene blue; PAP, p-aminophenol; NRR, $N_2$ reduction reaction;

Insert Table of Contents artwork here

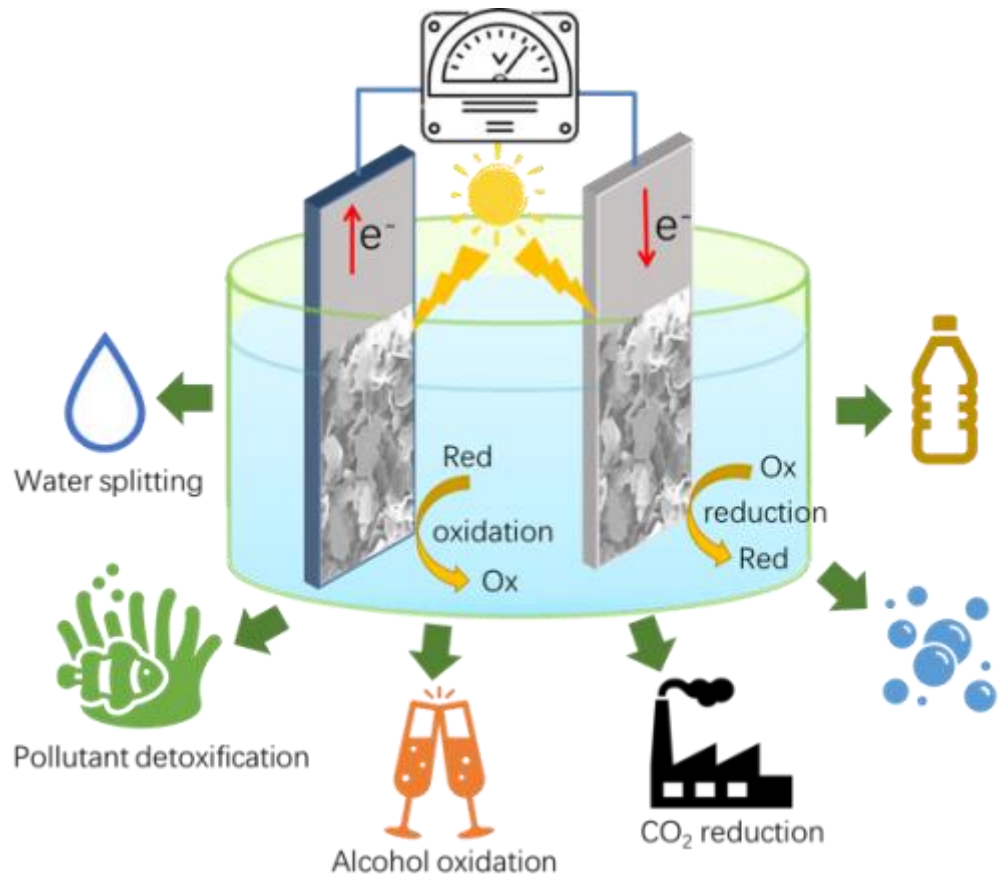